\documentclass[seceq]{ptptex}

\usepackage{graphicx}
\usepackage{wrapft}

\newcommand{\vecP}{\mbox{\boldmath $ P $}}
\newcommand{\vecW}{\mbox{\boldmath $ W $}}
\newcommand{\vecX}{\mbox{\boldmath $ X $}}
\newcommand{\vecZ}{\mbox{\boldmath $ Z $}}
\newcommand{\vecLambda}{\mbox{\boldmath $ \Lambda $}}

\title{
 Study of Quantum Decoherence in a Finite System. %
}

\subtitle{%
Three Schr\"odinger
 Cats and Crossing of Classical Orbits. %
}    

\author{
Takuji \textsc{ISHIKAWA} %
}

\inst{
Graduate School of Science and Engineering, Ibaraki University,\\
        Mito 310-8512, Ibaraki, Japan
}

\abst{\noindent\includegraphics[scale=1.0]{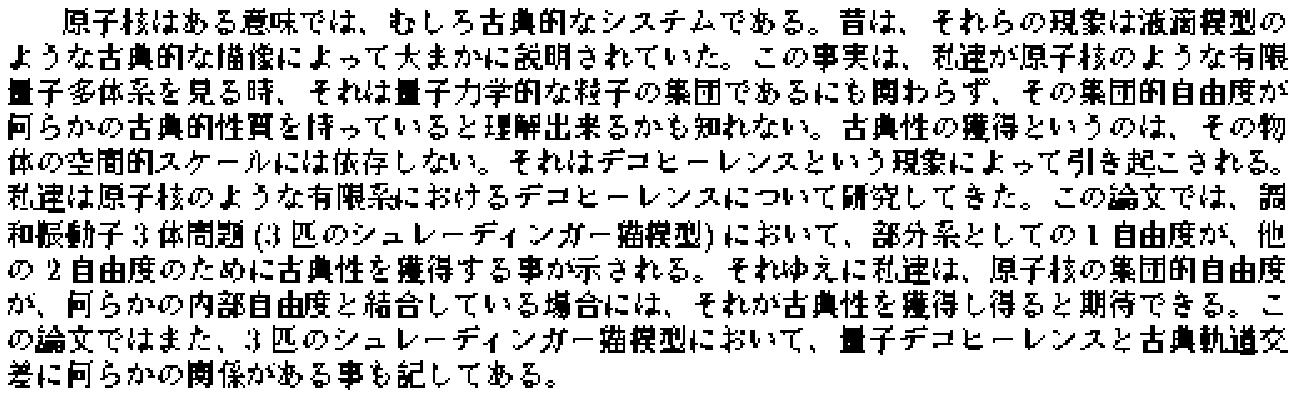}
Nuclei are rather classical systems in a sense. In the old days, their
phenomena were roughly explained in classical rules such as the liquid drop model. This fact may be understood that when
we see an finite quantum many body system like nucleus, though which is a group
of quantum mechanical particles, its any collective degree of freedom
has any classicality. Getting a classicality does not depend on 
spatial scales of objects. It is made by a phenomenon called Quantum Decoherence. We have studied
about Quantum Decoherence in a finite system as nucleus. In this
paper, at the
harmonic three body problem (3 Schr\"odinger cats), it is shown that one degree of freedom as a
sub-system would get classicality because of the other two degrees of
freedom. Therefore we can assume that a
nuclear collective degree of freedom would get classicality when it
couples with any other internal degrees of freedom. In this paper, we
also note that there is some relationship between the Quantum
Decoherence and Crossing of classical orbits in 3 Schr\"odinger cats model.
}

\begin{document}

\maketitle

\section{Introduction.}
\noindent\includegraphics[scale=1.0]{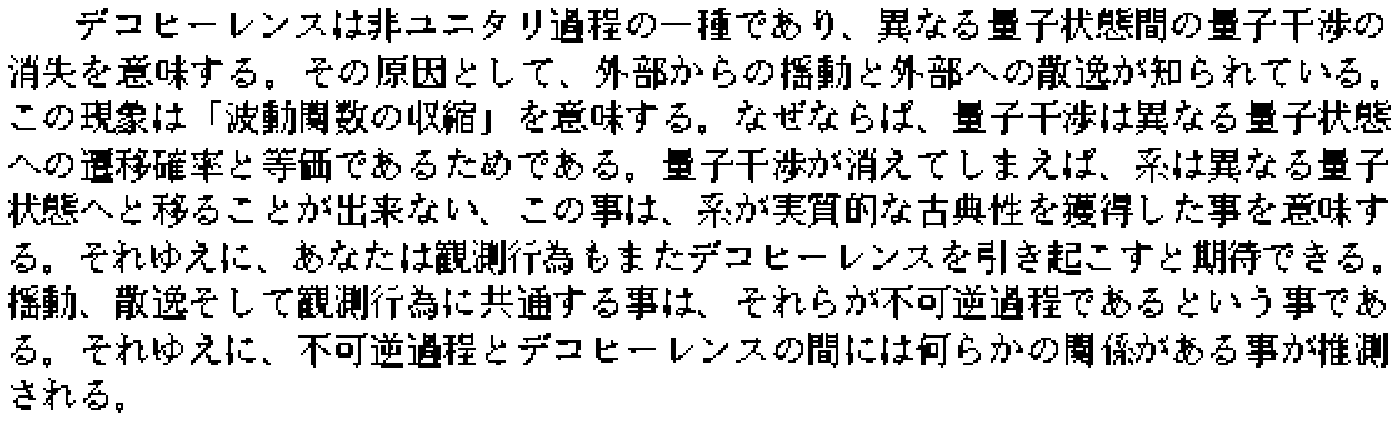}

Decoherence is a kind of non-unitary process, which is a disapperance of
quantum interferences among differrent state vectors. Its origin is known as
external fluctuations and dissipations. This phenomenon means ``the
collapse of wave function''. Because the interference is equal to the
transition probability to the other quantum states. When the
interference vanishes, the system can not be transported to any other
quantum states, it means the system gets effective classicality. So you
can expect that an observation may also cause decoherence. The
similarity among fluctuations, dissipations and observations is that
these are irreversible processes. Therefore, it is supposed that there
is any relation between irreversibility and quantum decoherence.

\noindent\includegraphics[scale=1.0]{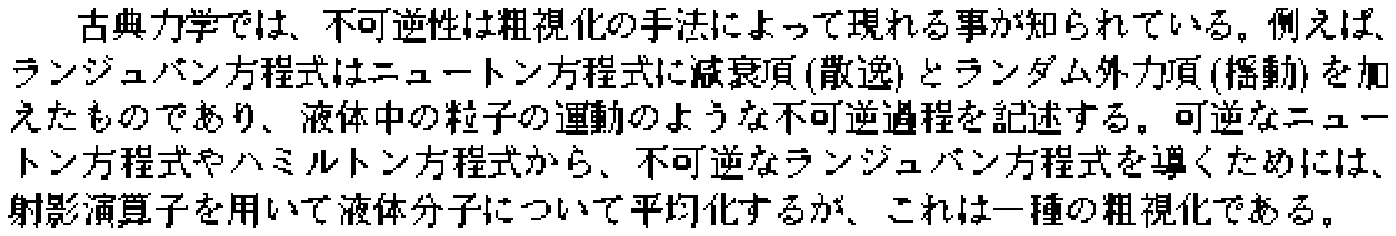}

In classical mechanics, it is known that the irreversibility is made by the coarse
graining procedure. For example, the Langevin equation is the Newton
equation with damping (dissipation) and random force (fluctuation), which describes irreversible
processes such as a motion of a particle in liquid. To derive the
irreversible Langevin equation from the reversible Newton equation or
Hamilton equation, we need to average out about degrees of freedom of
liquid molecules using the projection operators, that is a kind of
coarse graining.

\noindent\includegraphics[scale=1.0]{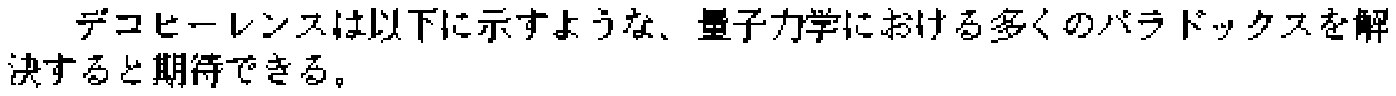}

We can expect that Decoherence would solve a lot of quantum mechanical paradoxes as follows.

 \subsection{Schr\"odinger's cat.}

\noindent\includegraphics[scale=1.0]{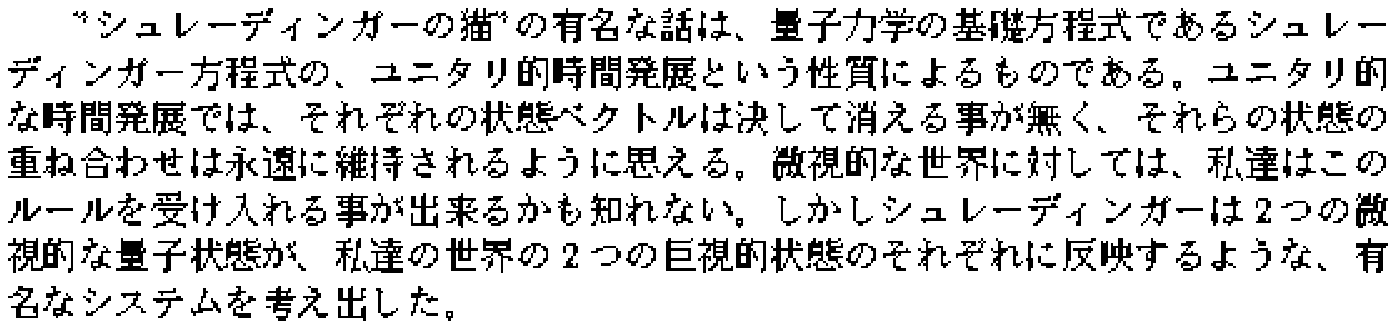}

The famous story of ``the Schr\"odinger's cat'' is due to the unitary
 evolution of the Schr\"odinger equation which is the basic equation of
 the Quantum Mechanics. By unitary evolutions, each state vector will never vanish,
 and the superposition of states will be kept forever. For microscopic
 world, we may accept this rule. But Schr\"odinger thought up a famous system in which the two
microscopic quantum states reflect the two macroscopic states in our
 world respectively.

\noindent\includegraphics[scale=1.0]{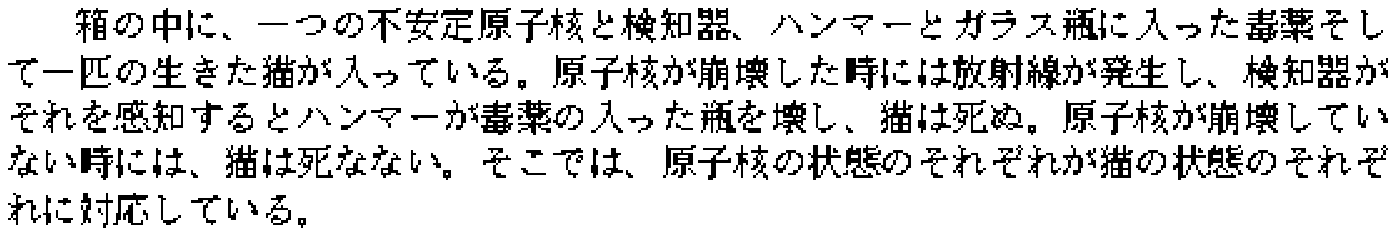}

In a box, there are an unstable nucleus, a geiger counter, a hammer,
poison in a glass bottle, and a living cat. When the nucleus decays and
a radiation occurs, the geiger counter will sense it, then the hammer
will crash the poison bottle, and the cat will die. When the nucleus
does not decay, the cat won't die. There, each of the nuclear states
corresponds to each of the cat's states respectively.

\noindent\includegraphics[scale=1.0]{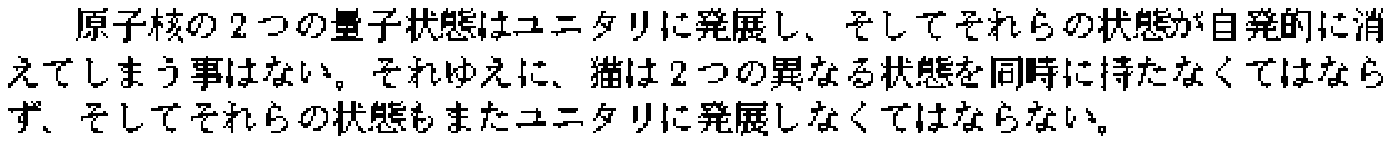}

The nuclear two quantum states evolve unitary, and the states will never
vanish spontateously, therefore the cat has to have two differrent
states simultaneously and the states have to evolve unitary,
too.
\begin{eqnarray}
 \mbox{Nuclear states} &:&
   |\psi(t)\rangle = |\psi_{notdecay}(t)\rangle + |\psi_{decay}(t)\rangle \ , \\
  \mbox{Cat's states} &:& |\phi(t) \rangle =  |\phi_{alive}(t) \rangle +
   |\phi_{dead}(t)  \rangle \
\end{eqnarray}

\noindent\includegraphics[scale=1.0]{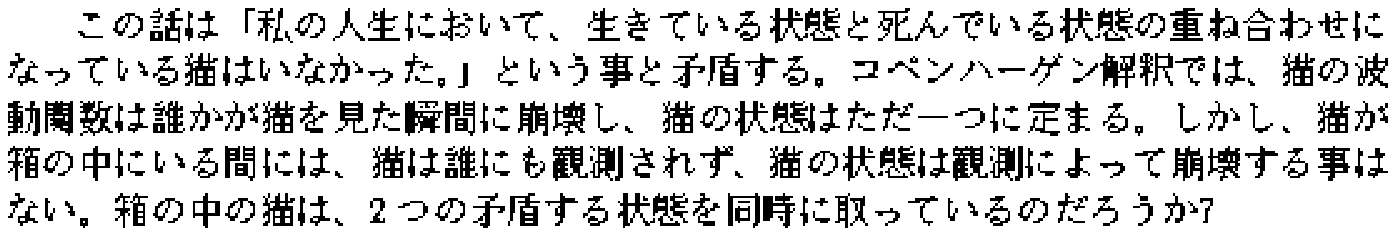}

This story conflicts with the fact that ``there has been no cat whose state is
superposition of living and dead ever in my life.'' In Copenhagen
interpretation, the cat's wave function collapses at the moment someone
watches the cat, and the state of cat defines uniquely. But while the
cat is in a box, he can't be observed by anyone and his states can't be collapsed by observations.
Is the cat in the box in the two contradictorily different state at the
same time? 

\noindent\includegraphics[scale=1.0]{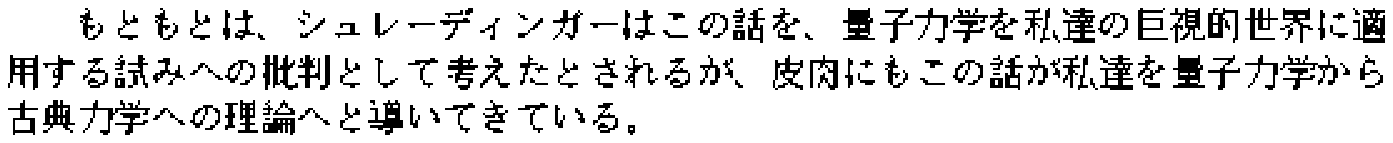}

Originally, Schr\"odinger thought this story as a criticism for attempts
to apply the quantum mechanics to our macroscopic world, but ironically
this story has led us the theory of quantum to classical.

\noindent\includegraphics[scale=1.0]{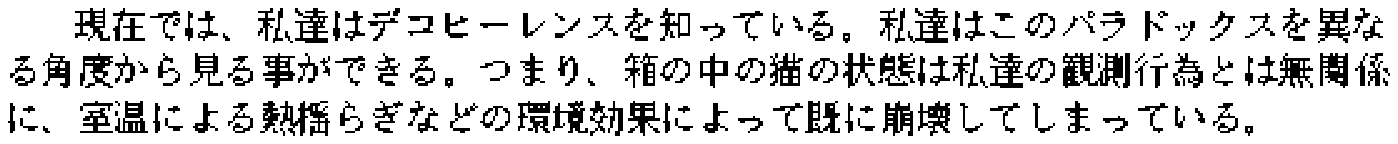}

Now, we know decoherence. We can understand this paradox another angle.
That is, the states of cat in a box had been already collapsed by
environmental effects such as thermal fluctuation by room temperature,
regardless of our observations.

 \subsection{Many Worlds interpretation.}

\noindent\includegraphics[scale=1.0]{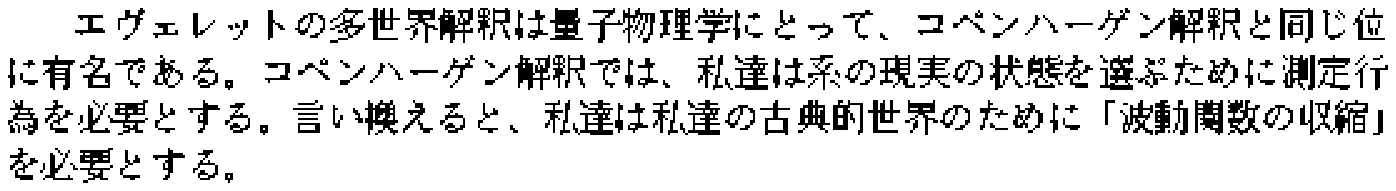}

Everret's many worlds interpretation is as famous as the Copenhagen
interpretation for quantum physics. In the Copenhagen interpretation, we
need measurements to select a real state of the system, in other words,
we need a ``Collapse of wave function'' for our classical world.   

\noindent\includegraphics[scale=1.0]{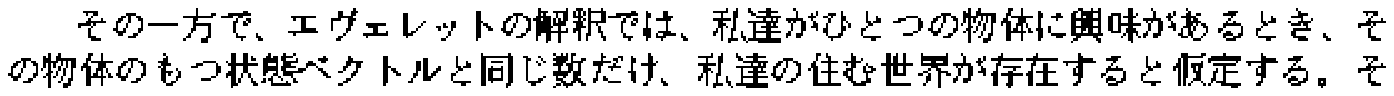}
\noindent\includegraphics[scale=1.0]{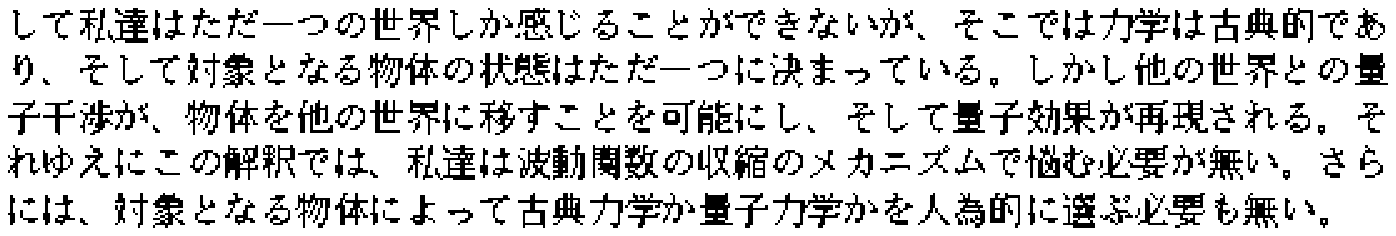}

While, in Everret's interpretation, when we interested in an object, we
assumed that there are our living worlds as many
as the number of the object's state vectors. And we can sense only one
world, where the mechanics is classical and the object's state is
defined uniformly. But the quantum interference with the other worlds
makes it possible to transfer of the object to the other worlds, then the quantum
effects are reproduced. Therefore in this interpretation, we don't need
to concern about the mechanism for collapse of wave function. Moreover,
we don't need any artificial choice of mechanics from classical or
quantum depending on the object.

\noindent\includegraphics[scale=1.0]{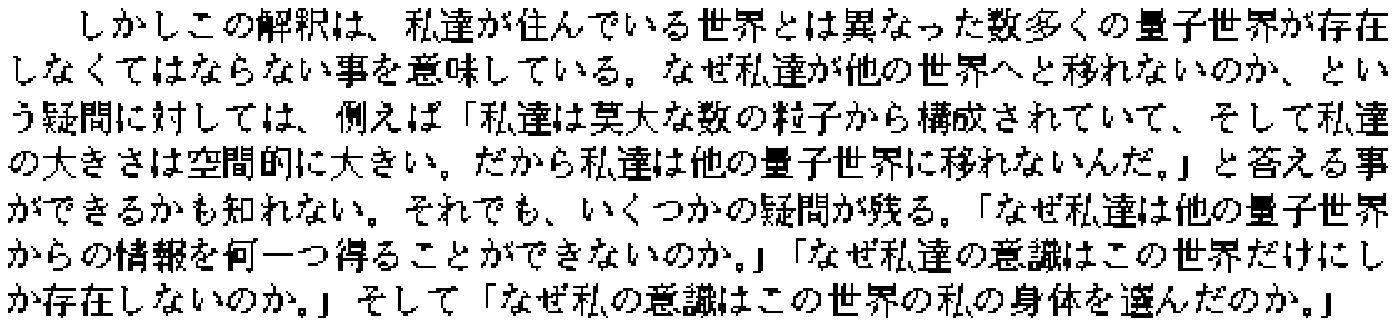}

But this interpretation means that there must be a lot of quantum worlds
different from the world we are living in. For a question why we can't
transfer to the other worlds, we may answer for example ``Because we are composed
of a huge huge numbers of particles, and our sizes are spatially
large. Therefore we can't transfer to the other quantum worlds.'' Still,
some questions remain, ``Why can't we get any information from the other
quantum worlds?'', ``Why are our consciousnesses in this world only?''
and ``Why did my consciousness sellect my body in this world?''

\noindent\includegraphics[scale=1.0]{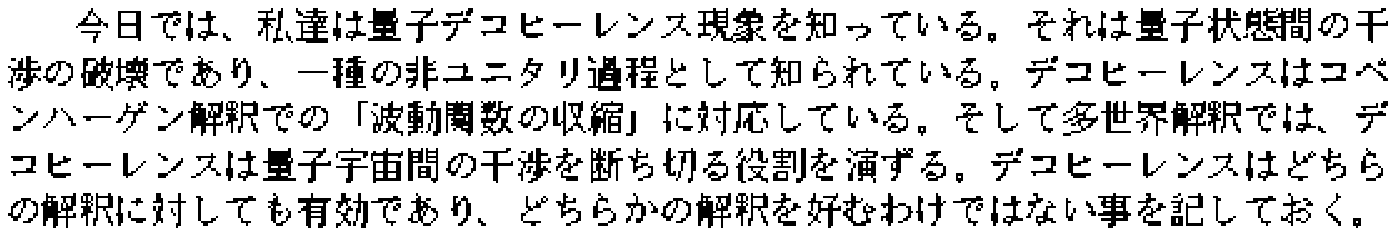}

Today, we know the quantum decoherence phenomena, which is the
destruction of interferences among quantum states, and known to be a
kind of non-unitary processes. Decoherence
corresponds to ``the collapse of wave functions'' in Copenhagen
interpretation. And in many worlds interpretation, decoherence plays a
role of the cutting of interferences among the quantum multiverses.
Note that decoherence is valid for both interpretations and does not
favor one interpretation of them over another.

\noindent\includegraphics[scale=1.0]{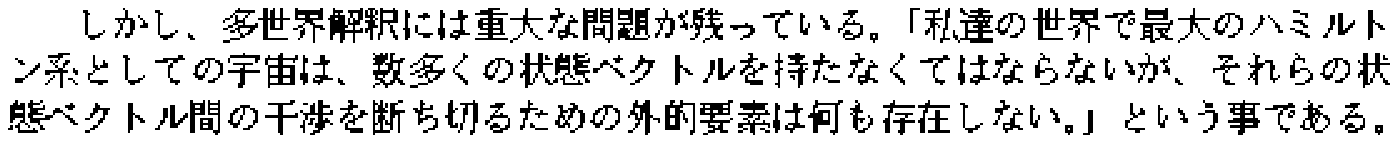}

But, the serious problem in the many worlds interpretation remains, 
``The universe, as the largest Hamiltonian system in our world, must
have so many state vectors, and there is no external elements which
destroys the interference among state vectors of the universe.''

\noindent\includegraphics[scale=0.72]{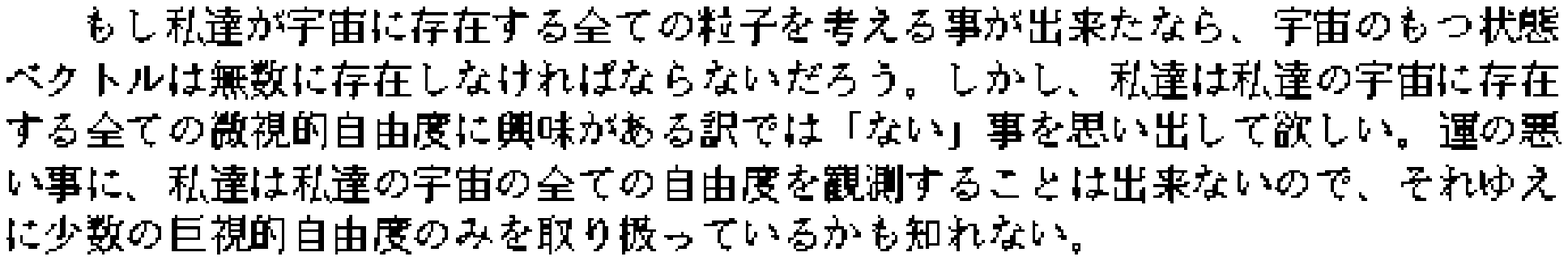}

If we could concern about a Hilbert space for whole particles in the
universe, there should be an infinite number of state vectors of the universe.
But, remember that we are {\underline {not}} interested in the whole microscopic
degrees of freedom in our universe. Unfortunately, we are not able to
observe the whole degrees of freedom in our universe, therefore we may
treat only a small numbers of macroscopic degrees of freedom.

\noindent\includegraphics[scale=0.72]{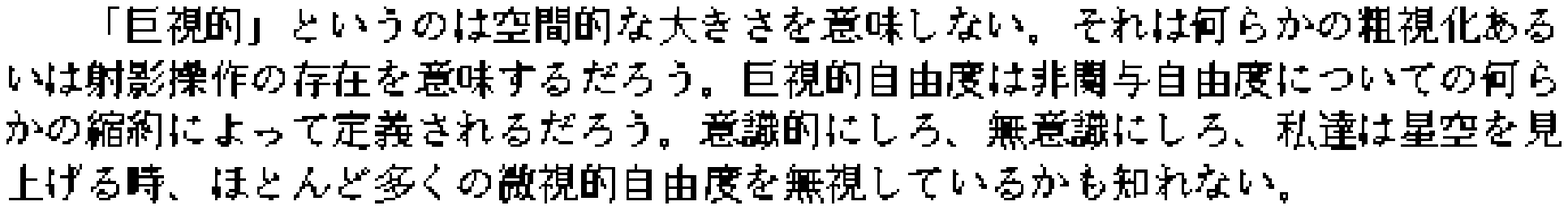}

``Macroscopic'' doesn't mean spatial scale, it would mean there is some coarse
graining or projection. The macroscopic degrees of freedom would be
defined by some reduction of irrelevant microscopic degrees of freedom.  
Whether consciously or unconsciously, we may neglect most of the microscopic
 degrees of freedom when we look up at the starry starry sky! 

\noindent\includegraphics[scale=0.72]{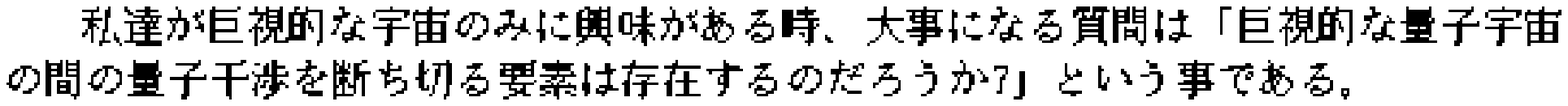}

When we interested in the macroscopic universe only, the important
question is that, ``Is there any factor to cut quantum interferences
among the macroscopic quantum universes?''\\

{ \underline {Case1. There is no such factor.}}

\noindent\includegraphics[scale=0.72]{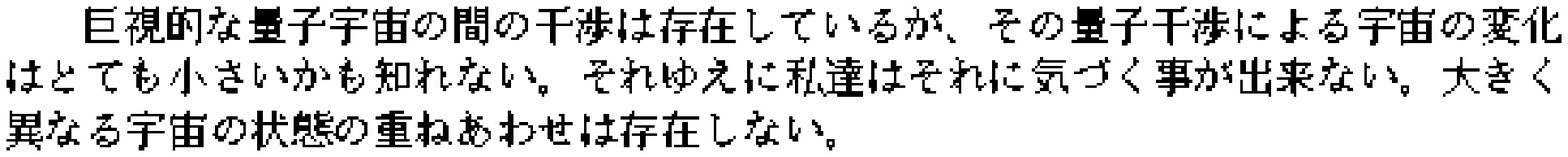}

There are interferences among the macroscopic quantum universes,
but their changes by quantum interferences may be very small. Therefore we
can't be aware of them. There is no superposition among widely different
states of universe.

{ \underline {Case2. Environmental effects.}}

\noindent\includegraphics[scale=0.72]{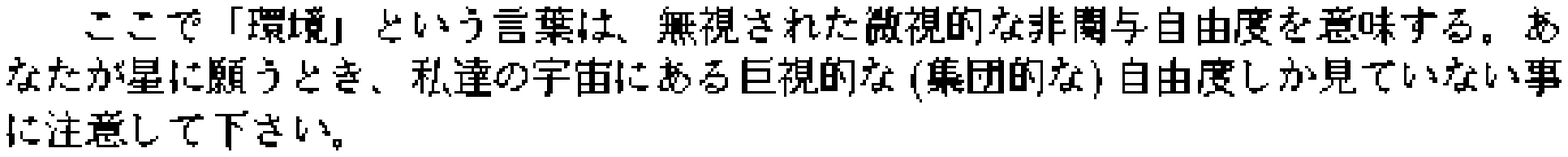}

Here, the word ``environment'' means the irrelevant microscopic degrees
of freedom neglected. Note that when you wish upon a star, you only care
about the macroscopic (collective) degrees of freedom inside our universe.

{ \underline {Case3. Spontaneous Selection.}}

\noindent\includegraphics[scale=0.72]{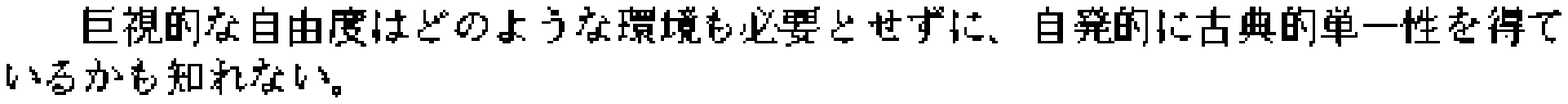}

The macroscopic degrees of freedom may get classical unity spontaneously
without any environments.\\

\noindent\includegraphics[scale=0.72]{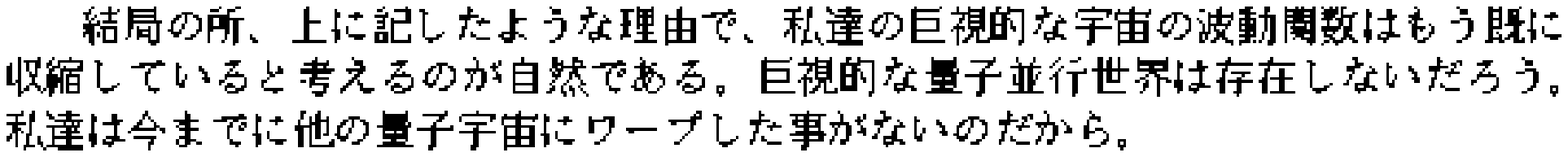}

 Eventually, it is natural that the wavefunction of our macroscopic
 universe has already collapsed for such reasons above. There wouldn't be
 any macroscopic quantum parallel worlds, because we have never been
 transported to any other quantum universes. 

\noindent\includegraphics[scale=1.0]{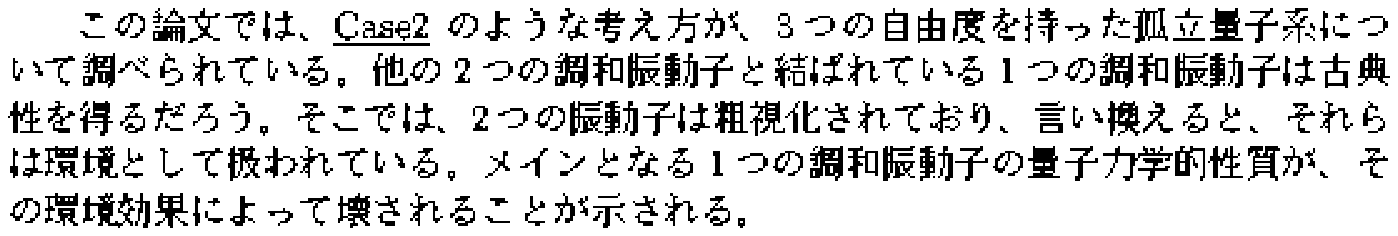}

 In this paper, an idea like \underline{Case2} is investigated for
 an isolated quantum system with 3 degrees of freedom. A harmonic
 oscillator coupled with other 2 harmonic oscillators would get
 classicality. There the latter 2 oscillators are coarse grained, in
 other words, they are treated as environments. It is showed that
 the environmental effect destroys the quantum mechanical property of the main 1 harmonic oscillator.

\subsection{Quantum Mine Sweeper and Decoherence at the nuclear fission.}

 \noindent\includegraphics[scale=1.0]{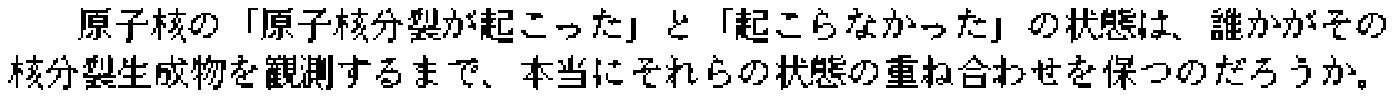}

 Do the nuclear states ``The nuclear fission has
 done.'' and ``It has not.'' really keeps their superposition until
 anyone observes the fission products? 

 \noindent\includegraphics[scale=1.0]{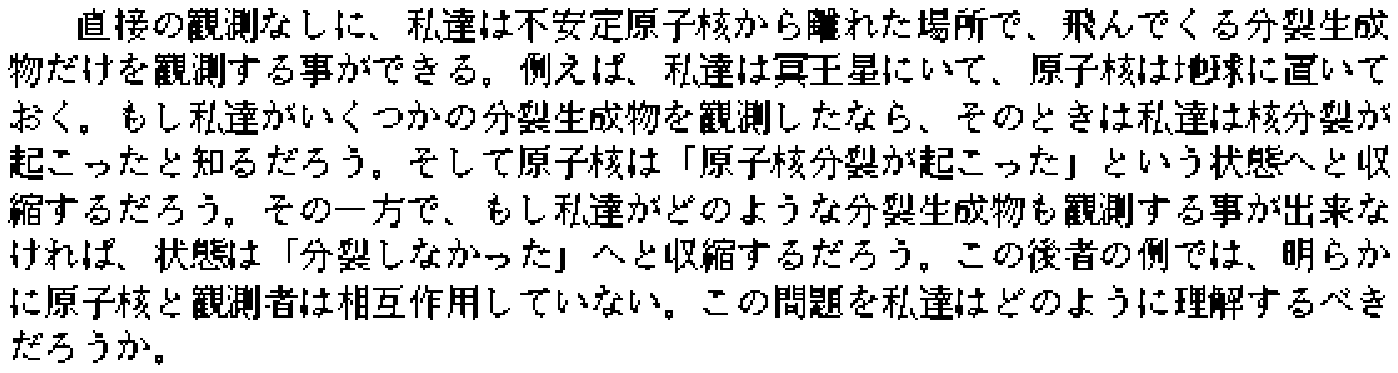}

Without a direct observation, we
 can observe only the incoming fission products at far from an unstable
 nucleus, for example, we are on the planet Pluto and the nucleus is on
 the Earth. If we observe some fission products, then we will know that the
 fission has done, and the nuclear state reduces into ``The fission has
 done.'' On the other hand, if we can not observe any fission products,
 the state would also reduce into ``It has not.'' In the latter case,
 clearly there is no interaction between the observer and the
 nucleus\(\ldots\)  How should we understand this problem?

\noindent\includegraphics[scale=1.0]{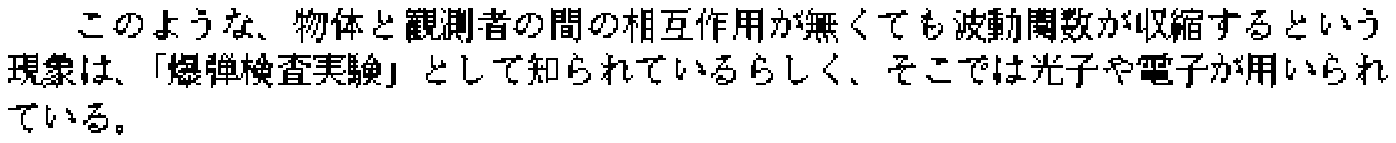}

It is said that the phenomena like this, the collapse of the wave
function without any interactions between the object and the observer,
is known as the interaction-free measurement ``Quantum Mine Sweeper'',
where photons or electorons are used. 

\noindent\includegraphics[scale=1.0]{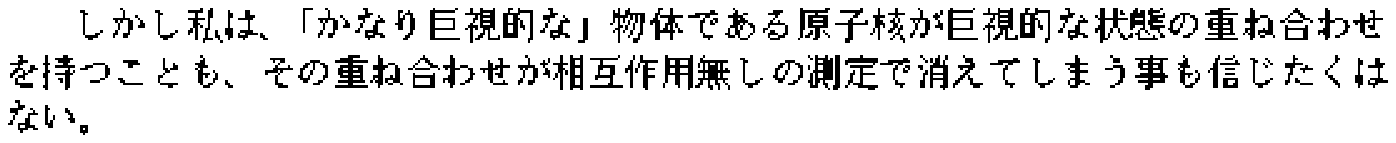}

But I don't want to think the nucleus which is ``rather
macroscopic'' object has a superposition of macroscopic states nor the
superposition vanishes with the interaction-free measurement. Will the
macroscopic different state vectors keep their superposition until
someone observes the residue of nuclear fission or fusion?

\noindent\includegraphics[scale=1.0]{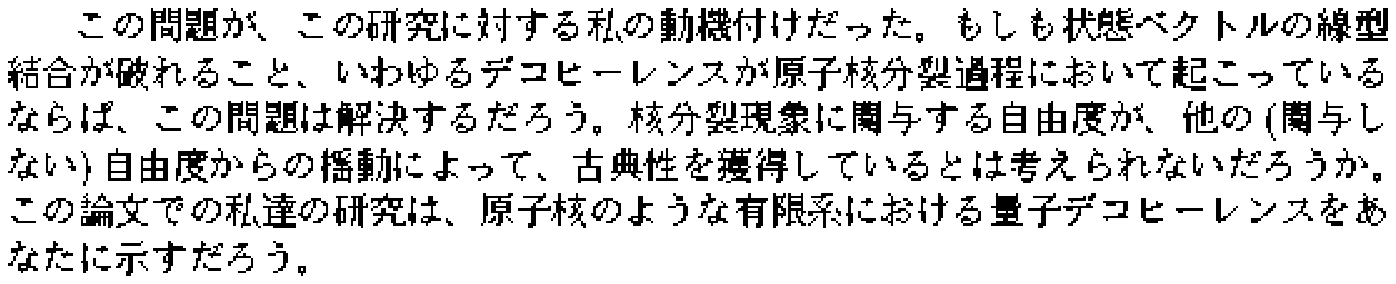}

This question was my motivation for this study. If the collapse of the linear
combination of state vectors, so called the Quantum Decoherence,
occurs in the nuclear fission process, this problem would be solved.
 Can we assume that the relevant degrees of freedom at the fission would
 get a classicality because of the fluctuation from the other
 (irrelevant) degrees of freedom? Our study in this paper will show you
 a quantum decoherence in a finite system like nucleus.
\(\diamondsuit\)

\newpage

\section{About Decoherence.}

\noindent\includegraphics[scale=1.0]{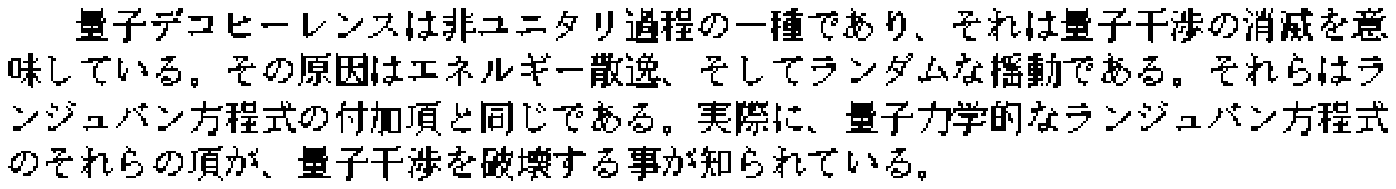}

The quantum decoherence is a kind of non-unitary processes, which means
the dissappearance of quantum interference. Its causes are dissipations
and random fluctuations. They are the same as the additional terms of 
the Langevin equation. In fact, it is known that these terms of the quantum
mechanical Langevin equation destroys quantum interferences.

\noindent\includegraphics[scale=1.0]{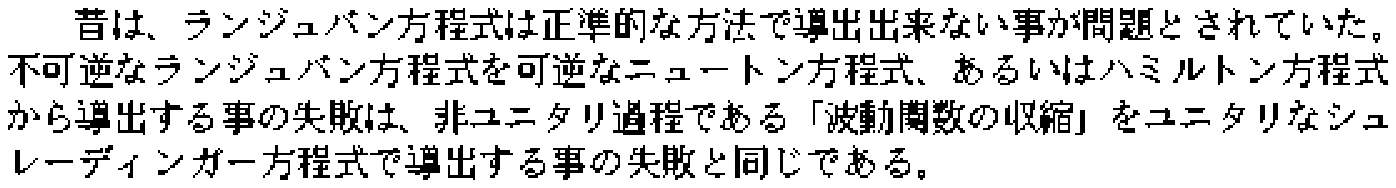}

In old days, it was a problem that the Langevin equation could not be
 derived in canonical procedures. The failure to derive the time
 irreversible Langevin equation from the time reversible Newtonian or
 Hamiltonian equation is the same as the failure to derive non-unitary
 ``The collapse of the wave function'' from the unitary Schr\"odinger
 equations.

\noindent\includegraphics[scale=1.0]{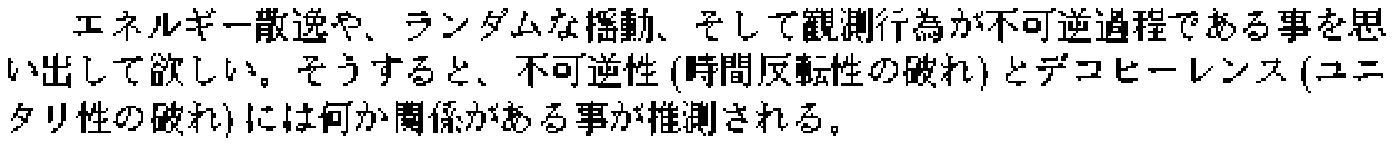}

Remember that energy dissipations, random fluctuations and observations
are irreversible processes. Then, it is assumed that there is any
relationship between the irreversibility (the break down of the time reversal symmetry), and the
quantum decoherence (the break down of unitarity) .

\noindent\includegraphics[scale=1.0]{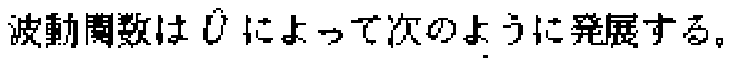}

A wave function evolves by \( \hat{U} \) as follows.
\begin{equation}
  \psi (t) = \hat{U} \psi (0) \label{eqn021}
\end{equation}
\noindent\includegraphics[scale=1.0]{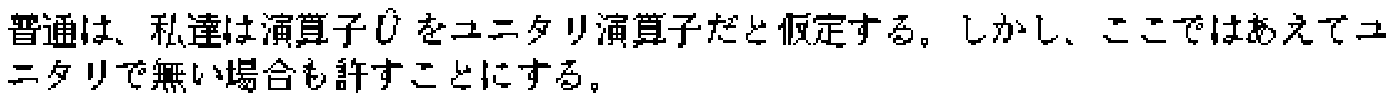}

Ordinary, we assume the operator \( \hat{U} \) as a unitary
operator. But, here we dare to allow the case \( \hat{U} \) is {\underline{not}}
unitary. 

\noindent\includegraphics[scale=1.0]{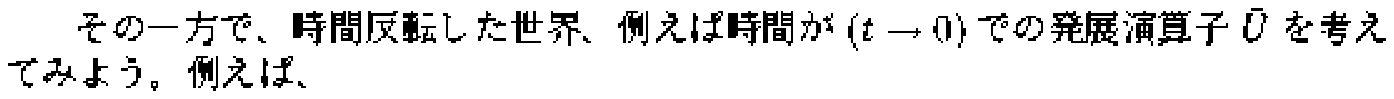}

While, please imagine the evolution operator \(\bar{U}\) for time reversal world,
such as time (\(t \rightarrow 0 \)) . For example,
\begin{equation}
  \bar{\psi} (0) = \bar{U} \psi (t)
\end{equation}

\begin{figure}
\includegraphics[scale=1.0]{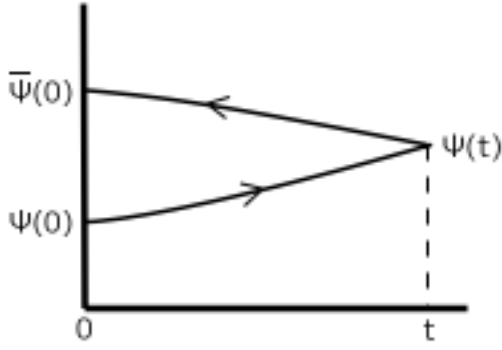}
\caption{Simple picture for irreversibility.}
\end{figure}

\noindent\includegraphics[scale=1.0]{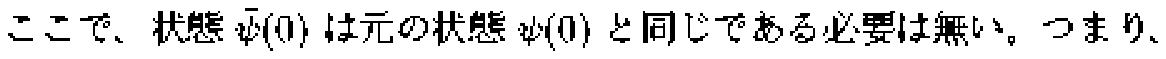}

Here, the state \( \bar{\psi} (0)\) is {\underline{not}} need to be the same as the original state
\( \psi(0) \)
, that is,
\begin{equation}
  \bar{\psi} (0) = \bar{U}\hat{U} \psi(0)
\end{equation}

\noindent\includegraphics[scale=1.0]{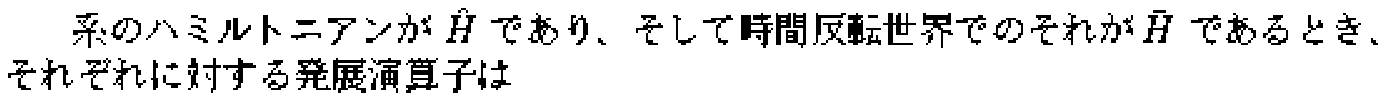}

 When the Hamiltonian of system is \(\hat{H}\), and that of the time
 reversal world is  \(\bar{H}\), evolution oparators for each are
\begin{equation}
  \hat{U} = \exp \left\{ -\frac{i}{\hbar} \hat{H} t \right\} \ , \
   \bar{U} = \exp \left\{ +\frac{i}{\hbar} \bar{H} t \right\}  \label{eqn024}
\end{equation}

\noindent\includegraphics[scale=1.0]{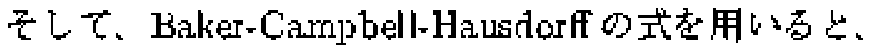}

Then, using the Baker-Campbell-Hausdorff formula,
\begin{equation}
 \bar{U}\hat{U} = \exp \left\{ -\frac{i}{\hbar}(\hat{H}-\bar{H})t
			+\frac{1}{2\hbar^2}[ \bar{H},\hat{H} ]t^2 +
			O(t^3) +\cdots \right\}
\end{equation}

\noindent\includegraphics[scale=1.0]{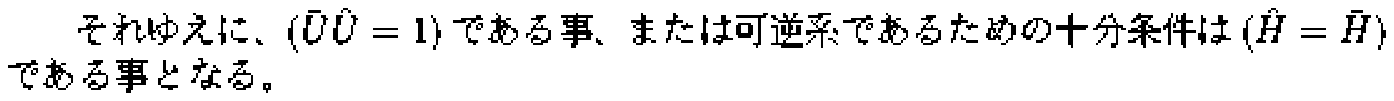}

 Therefore, the sufficient condition for being (\( \bar{U}\hat{U} = 1 \))
 or reversible system is being (\( \hat{H} = \bar{H} \)) .

\noindent\includegraphics[scale=1.0]{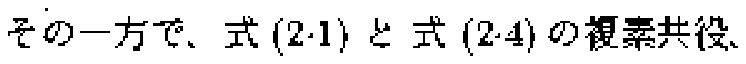}

While, the complex conjugate of eq.(\(\ref{eqn021}\)) and eq.(\(\ref{eqn024}\))
\begin{equation}
  \psi^*(t) = \hat{U}^{\dagger} \psi^*(0)  \quad , \quad
   \hat{U}^{\dagger} = \exp \left\{ +\frac{i}{\hbar} \hat{H}^{\dagger} t
			    \right\} \label{eqn026}
\end{equation}

\noindent\includegraphics[scale=1.0]{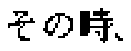}

Then
\begin{equation}
 \hat{U}^{\dagger} \hat{U} = \exp \left\{ -\frac{i}{\hbar}(\hat{H}-\hat{H}^{\dagger})t
			+\frac{1}{2\hbar^2}[ \hat{H}^{\dagger},\hat{H} ]t^2 +
			O(t^3) +\cdots \right\}
\end{equation}

\noindent\includegraphics[scale=1.0]{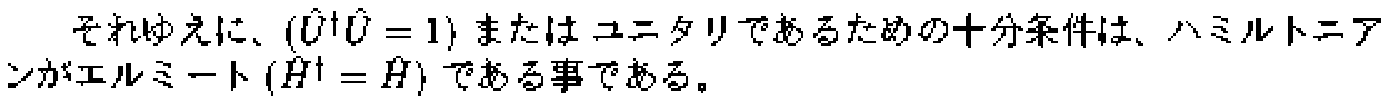}

Therefore the sufficient condition for (\( \hat{U}^{\dagger}\hat{U} = 1 \)) or  
 unitary, is that the Hamiltonian is Hermitian,
 (\(\hat{H}^{\dagger}=\hat{H}\)).

\noindent\includegraphics[scale=1.0]{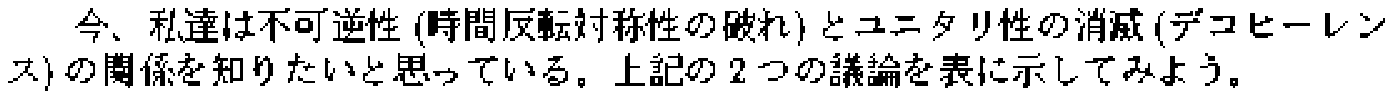}

Now we want to know the relationships between the irreversibility (the
break down of time reversal symmetry) and the disappearance of unitarity
(the quantum decoherence).
 Let us show the list of two discussions above.
\begin{equation}
  \left. \begin{array}{lcl}
    \bar{H}=\hat{H} \ \mbox{(time reversal symmetry)} & \Rightarrow & \bar{U}\hat{U}
     = 1 \ \mbox{(reversibility)}\\
    \hat{H}^{\dagger}=\hat{H} \ \mbox{(Hermitian)} & \Rightarrow &
     \hat{U}^{\dagger}\hat{U} = 1 \ \mbox{(unitarity)}
  \end{array} \right\}_{\cap} \Rightarrow \bar{U} = \hat{U}^{\dagger} \label{eqn027}
\end{equation}

\noindent\includegraphics[scale=1.0]{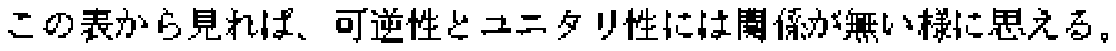}

It seems that there is no relation between the reversibility and the
unitarity from this list.

\noindent\includegraphics[scale=1.0]{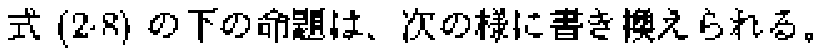}

The lower proposition of eq.(\ref{eqn027}) is rewritten as follows.
\begin{equation}
  \{ \hat{U}^{\dagger}\hat{U} = 1 \} \supset \{ \hat{H}^{\dagger} = \hat{H} \}
\end{equation}

\noindent\includegraphics[scale=1.0]{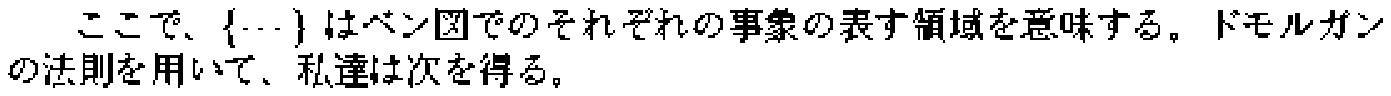}

Here, \( \{ \cdots \} \) means the region of each event in Venn
diagram. Using the De Morgan's low, we get
\begin{equation}
  \{ \hat{U}^{\dagger}\hat{U} \neq 1 \} \subset \{ \hat{H}^{\dagger} \neq \hat{H} \}
\end{equation}

\noindent\includegraphics[scale=1.0]{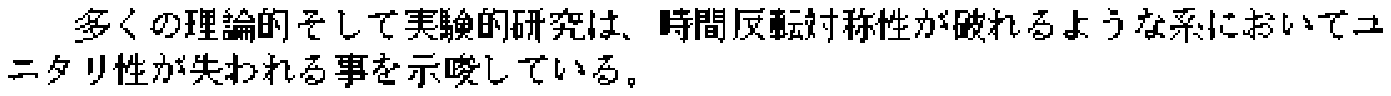}

Many theoretical and experimantal studies imply that unitarity is lost
in systems in which the time reversal symmetry breaks down. 
\begin{equation}
   \hat{U}^{\dagger}\hat{U} \neq 1  \quad \Rightarrow \quad  \bar{H}
    \neq \hat{H} \label{eqn031}
\end{equation}
\noindent\includegraphics[scale=1.0]{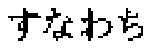}

that is
\begin{equation}
  \{ \hat{U}^{\dagger}\hat{U} \neq 1 \} \subset \{ \bar{H} \neq \hat{H} \}
\end{equation}

\noindent\includegraphics[scale=1.0]{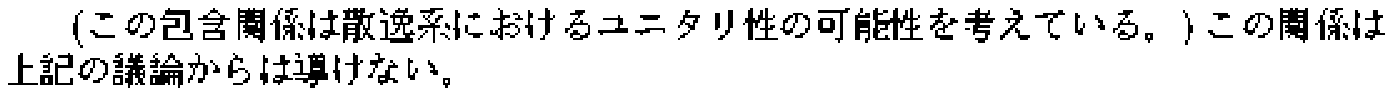}

(This inclusion relation is concerned about possibility of unitary
dissipation systems.) This relation can't be derived from
discussions above.

\noindent\includegraphics[scale=1.0]{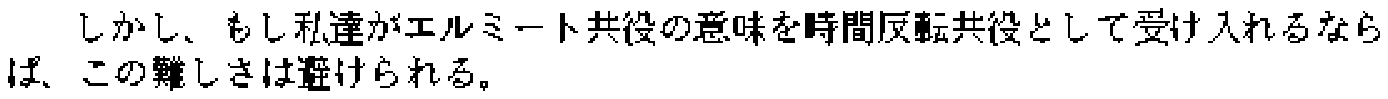}

But, If we accept a meaning for Hermitian conjugate as the time
reversal conjugate, this difficulty is avoided.
\begin{equation}
  \hat{H}^{\dagger} \equiv \bar{H} ,\quad \hat{U}^{\dagger} \equiv
   \bar{U} \label{eqn032}
\end{equation}

\noindent\includegraphics[scale=1.0]{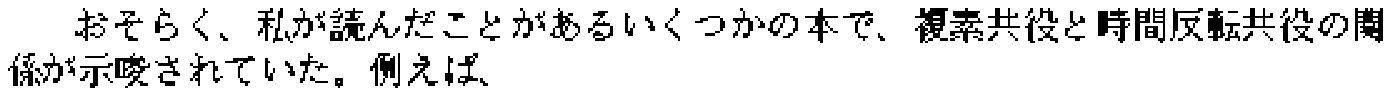}

Probably, the relation between complex conjugate and time
reversel conjugate has been implied
at some textbooks which I have ever read. For example,
\begin{equation}
  \psi(t') = \exp\left\{-\frac{i}{\hbar}\hat{H}(t'-t)\right\}\psi(t)
\end{equation}
\noindent\includegraphics[scale=1.0]{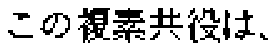}

Its complex conjugate is
\begin{equation}
  \psi^*(t')
   =\exp\left\{+\frac{i}{\hbar}\hat{H}^{\dagger}(t'-t)\right\}\psi^*(t) \label{eqn034}
\end{equation}

\noindent\includegraphics[scale=1.0]{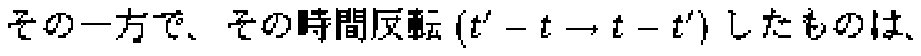}

While, the time reversal (\(t'-t \rightarrow t-t'\)) one is
\begin{equation}
  \bar{\psi}(t') =
   \exp\left\{+\frac{i}{\hbar}\bar{H}(t'-t)\right\}\bar{\psi}(t) \label{eqn035}
\end{equation}

\noindent\includegraphics[scale=1.0]{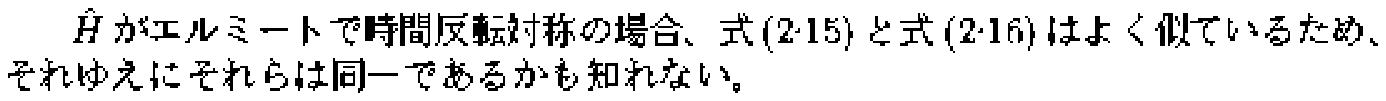}

When \(\hat{H}\) is Hermitian and time reversal
symmetric, eq.(\ref{eqn034}) and eq.(\ref{eqn035}) are very alike,
therefore they may be identical each other.
\begin{equation}
 \psi^*(t') = \bar{\psi}(t') ,  \psi^*(t) = \bar{\psi}(t)
\end{equation}
\noindent\includegraphics[scale=1.0]{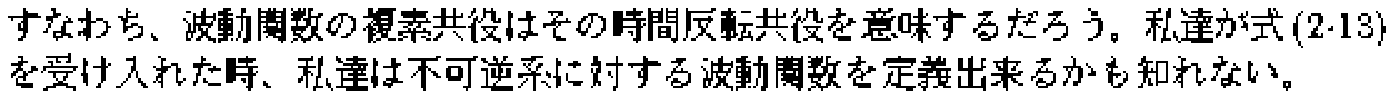}

That is, the complex conjugate of wave functions would mean its time
reversal conjugate. When we can accept eq.(\(\ref{eqn032}\)), we may define
the wave functions for irreversible systems. 

\noindent\includegraphics[scale=1.0]{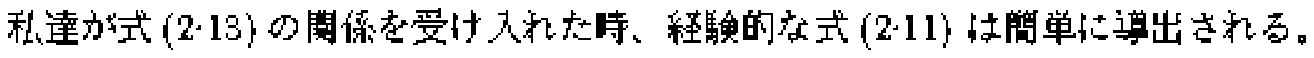}
							       
When we accept the relation eq.(\(\ref{eqn032}\)), the empirical formula
eq.(\(\ref{eqn031}\)) is easily derived.
\begin{equation}
  \hat{U}^{\dagger}\hat{U} =
   \exp\left\{+\frac{i}{\hbar}\hat{H}^{\dagger}t\right\}\exp\left\{-\frac{i}{\hbar}\hat{H}t\right\} = \exp\left\{+\frac{i}{\hbar}\bar{H}t\right\}\exp\left\{-\frac{i}{\hbar}\hat{H}t\right\}
\end{equation}

\noindent\includegraphics[scale=1.0]{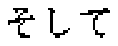}

Then,
\begin{equation}
  \bar{H} = \hat{H} \quad \Rightarrow \ \ \  \hat{U}^{\dagger}\hat{U} = 1
\end{equation}

\noindent\includegraphics[scale=1.0]{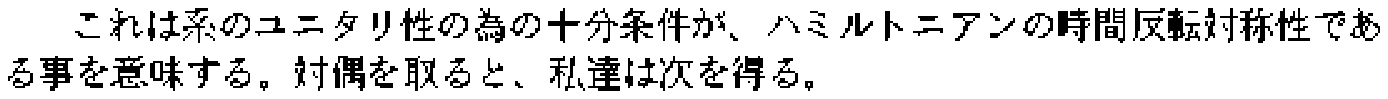}

This means that the sufficient condition for unitarity of system is the
time reversal symmetry of Hamiltonian. Taking the contraposition, we get
\begin{equation}
  \hat{U}^{\dagger}\hat{U} \neq 1  \quad \Rightarrow \ \ \    \bar{H} \neq \hat{H}
\end{equation}
\noindent\includegraphics[scale=1.0]{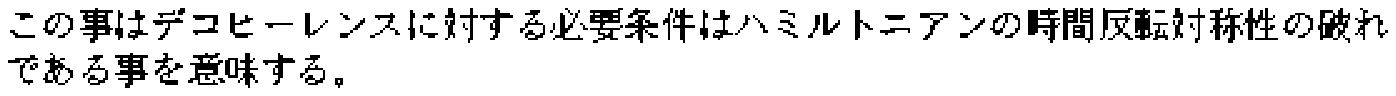}

, which means that the necessary condition for quantum decoherence is
the breaking down of time reversal symmetry of Hamiltonian.

\noindent\includegraphics[scale=1.0]{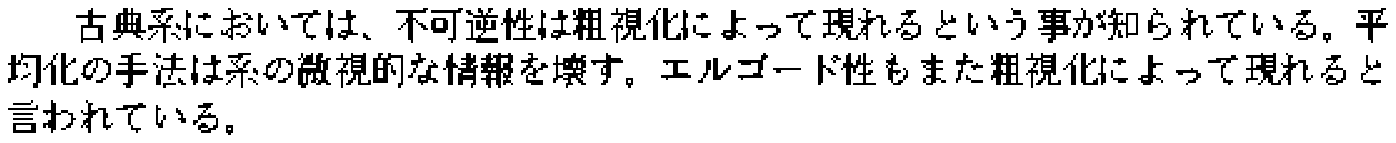}

In classical system, it is known that irreversibility appears by coarse
graining. Averaging procedures destroy microscopic informations for
system. It is said that Ergodicity also appears by coarse graining.

\noindent\includegraphics[scale=1.0]{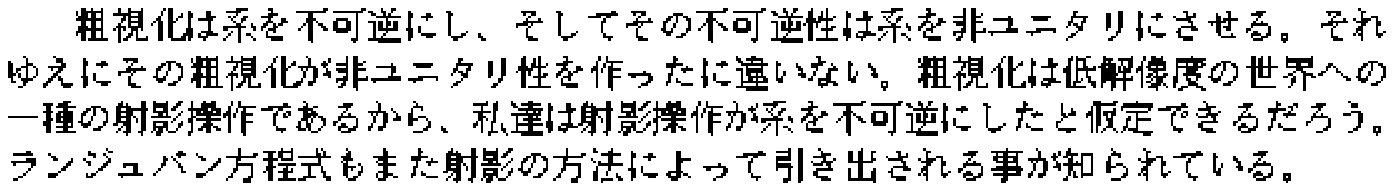}

Coarse graining makes the system irreversible, and the irreversibility
makes the system non-unitary, therefore the coarse graining must make
the non-unitarity. Coarse graining is a kind of Projection procedure
into a low resolution world, so we can assume that Projection makes the
system irreversible. It is known that the Langevin equation is also
derived by the projection method.

\noindent\includegraphics[scale=1.0]{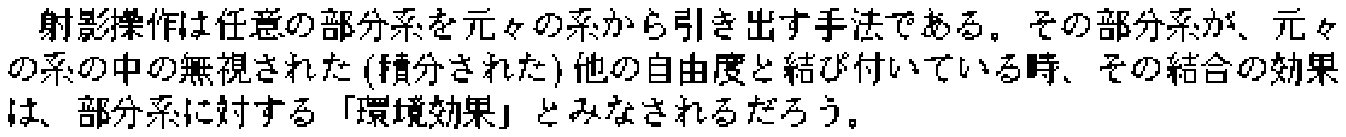}

Projection is a procedure to extract arbitrary sub system from a
original system. When the sub system couples with the other neglected
(integrated) degrees of freedom in the original system, the coupling
effects would be regarded as ``the environmental effects'' for the sub system.  \(\diamondsuit\)						    

\newpage

\section{A simple model: Asymmetric triangular
 harmonic oscillators in Schr\"odinger cat states, ``Three Schr\"odinger
 cats''.}

\noindent\includegraphics[scale=1.0]{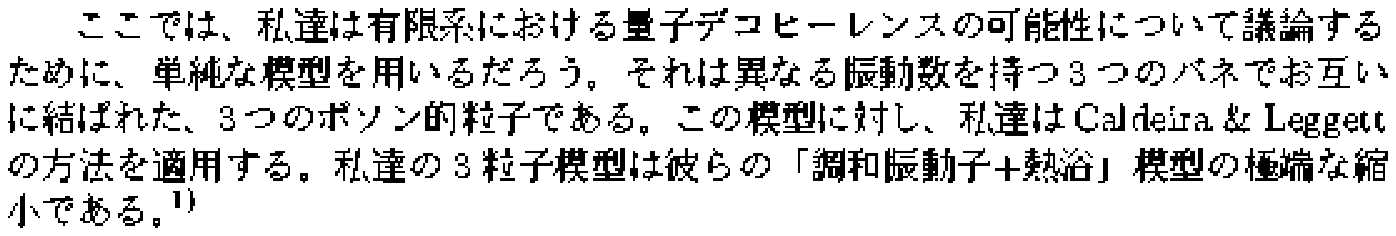}

 Here, We will use a simple model to discuss a possibility of quantum
 decoherence in a finite system. That is three bosonic particles tied up
 each other with three springs which have different frequencies. For
 this model, we will apply the Caldeira \& Leggett's technique. Our
 three particle model is a extreme reduction of their ``harmonic
 oscillator plus reservoir'' model{\cite{calleg}}.

\begin{figure}
\includegraphics[scale=.45]{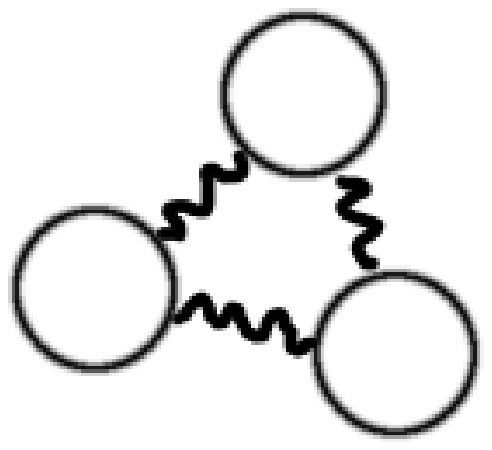}
\includegraphics[scale=.45]{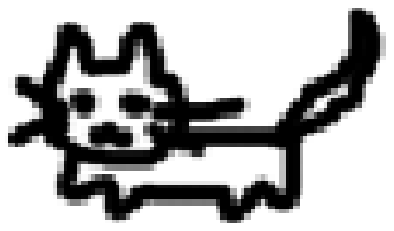}
\caption{3cats model. 3 particles are tied up with springs. And each state is Schr\"odinger cat.}
\end{figure}

\noindent\includegraphics[scale=1.0]{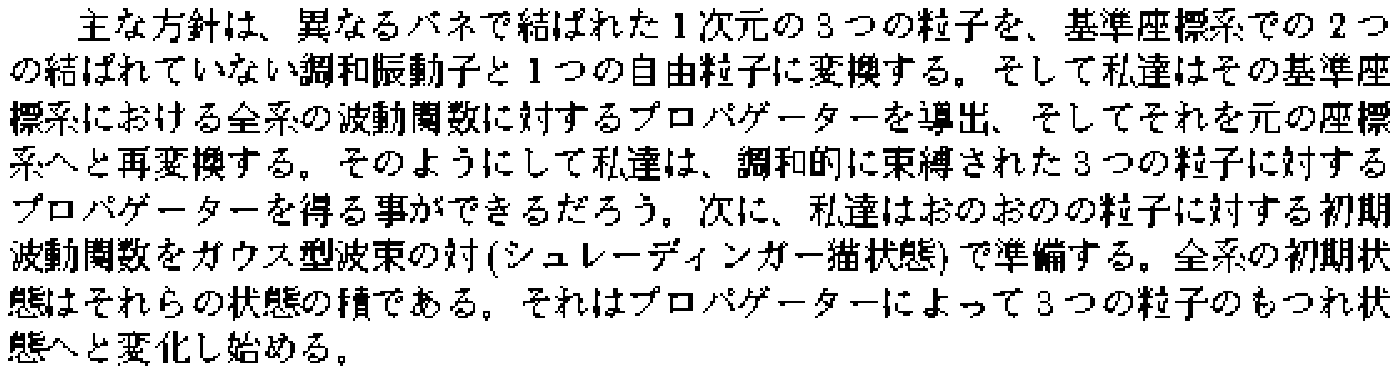}

 The main line is, to transform 1-dimensional three particles tied with different
 springs into two uncoupled harmonic oscillators and one free particle
 in normal coordinates. And we will derive a propagator for
 a total wave function of the total system in normal coordinates, 
and retranslate it into the original
 system. Then we will be able to get a propagator for harmonically
 bound three particle. Next, we will prepare the respective initial wave
 function for each particle as a pair of Gaussian wave packet (the
 Schr\"odinger cat state). The initial state of the total system is a
 product of those states. It will start to turn into a entangle
 state of three particles by propagator.

\noindent\includegraphics[scale=1.0]{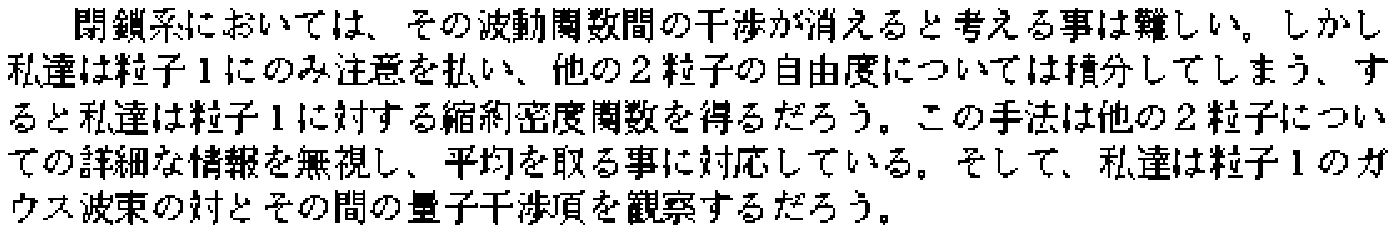}

 In a closed system, it is difficult to suppose that the interference between
 the wave functions of the closed system will vanish. But we will pay attention to the
 particle-1 only and integrate out the degrees of freedom about other two
 particles, then we will get a reduced density function for
 particle-1. This procedure corresponds to ignoring the fine
 informations about other two particles and taking an average. Then, we
 will observe changes of a pair of Gaussian wave packets of particle-1
 and the quantum interference term between them. \\

  \subsection{Classical model.}
\noindent\includegraphics[scale=1.0]{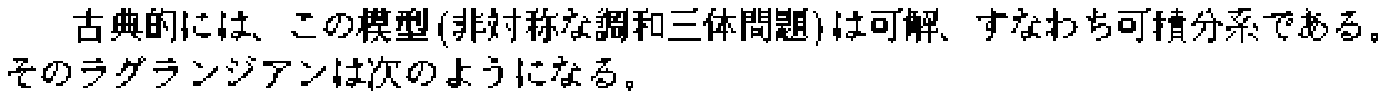}

 Classically, this model (the asymmetric harmonic three body problem) is
 soluvable, that is a integrable system. Its Lagrangean is as follows.
\begin{equation}
   L  = \frac{m}{2} \dot{x_1}^2 +\frac{m}{2} \dot{x_2}^2 + \frac{m}{2}
   \dot{x_3}^2 -\frac{m}{2} \omega_{12}^2 (x_1-x_2)^2 -\frac{m}{2}
   \omega_{13}^2 (x_1-x_3)^2 -\frac{m}{2} \omega_{23}^2 (x_2-x_3)^2 \label{eqn007}
\end{equation}
\noindent\includegraphics[scale=1.0]{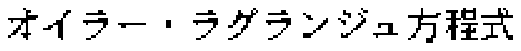}

  Applying the Euler-Lagrange equation
\begin{equation}
 \frac{\partial L}{\partial x_i} -\frac{d}{dt} \frac{\partial
  L }{\partial \dot{x_i}} = 0
\end{equation}
\noindent\includegraphics[scale=1.0]{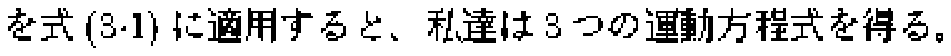}

 to eq.(\(\ref{eqn007}\)), we can get three equations of motion.
\begin{equation}
 \frac{d^2}{dt^2}
     \left( \begin{array}{c}
         x_1 \\ x_2 \\ x_3        
            \end{array} \right)
 = \left( \begin{array}{ccc}
      -(\omega_{12}^2 + \omega_{13}^2) & \omega_{12}^2 & \omega_{13}^2 \\
      \omega_{12}^2 & -(\omega_{12}^2 + \omega_{23}^2) & \omega_{23}^2 \\
      \omega_{13}^2 & \omega_{23}^2 & -(\omega_{13}^2 + \omega_{23}^2)
          \end{array} \right)
     \left( \begin{array}{c}
         x_1 \\ x_2 \\ x_3        
            \end{array} \right)
\end{equation}
\noindent\includegraphics[scale=1.0]{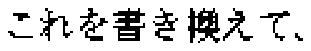}

 Rewriting this,
\begin{equation}
  \frac{d^2}{dt^2} \vecX_{(t)} = \vecW \vecX_{(t)}
\end{equation}
\noindent\includegraphics[scale=1.0]{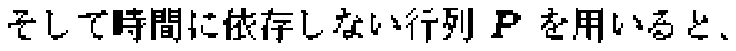}

 and using a time-independent matrix \( \vecP \) , then,
\begin{equation}
  \frac{d^2}{dt^2} ( \vecP \vecX_{(t)} ) = \vecP \vecW \vecP^{-1} (
   \vecP \vecX_{(t)} ) \label{eqn010}
\end{equation}
\noindent\includegraphics[scale=1.0]{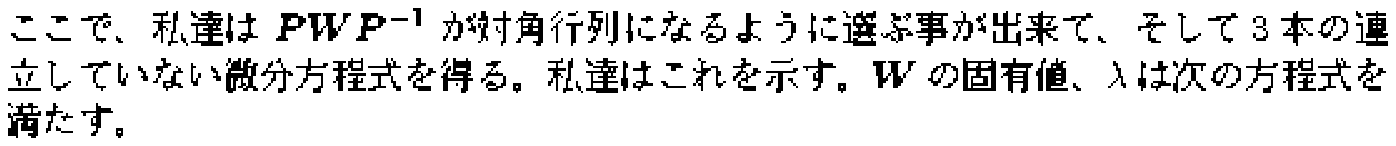}

 Here, we can set \( \vecP \) in order that \( \vecP \vecW \vecP^{-1} \)
 is diagonal, and get three uncoupled differential equations. We will
 show this procedure. Eigen values of \( \vecW \), \( \lambda \) satisfy the equation
\begin{equation}
 -\left\{ \ \lambda ^3 + 2(\omega_{12}^2 + \omega_{13}^2 + \omega_{23}^2)
   \lambda^2 + 3(\omega_{12}^2 \omega_{13}^2 +\omega_{13}^2
   \omega_{23}^2+\omega_{12}^2 \omega_{23}^2) \lambda \ \right\} \equiv
 0 \label{eqn012}
\end{equation}
\noindent\includegraphics[scale=1.0]{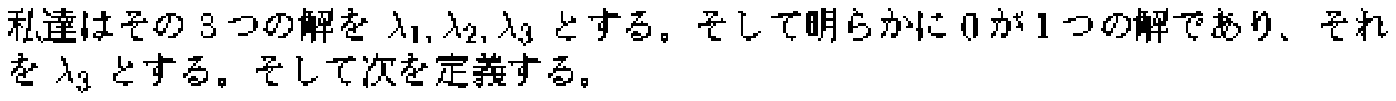}

We call its 3 solutions \( \lambda_1, \lambda_2, \lambda_3 \), and
obviously 0 is one solution, then we set it \( \lambda_3 \). And we define
\begin{equation}
  \vecP \vecW \vecP^{-1} = \left( 
     \begin{array}{ccc}     
        \lambda_1 & \ & \ \\
        \ & \lambda_2 & \ \\
        \ & \ & \lambda_3
     \end{array}  \right)
    = \left(
     \begin{array}{ccc}     
        \lambda_1 & 0 & 0 \\
        0 & \lambda_2 & 0 \\
        0 & 0 & 0
     \end{array}  \right)
    \equiv \vecLambda 
\end{equation}
\begin{equation}
  \vecP \vecX_{(t)} \equiv \vecZ_{(t)}
\end{equation}
\noindent\includegraphics[scale=1.0]{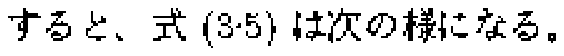}

 Then, equation (\ref{eqn010}) comes to 
\begin{equation}
  \frac{d^2}{dt^2} \vecZ _{(t)} = \vecLambda \vecZ _{(t)}
\end{equation}
\noindent\includegraphics[scale=1.0]{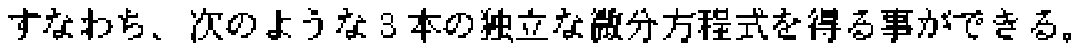}

, that is, we can get three independent differential equations as follows.
\begin{equation}
  \frac{d^2}{ d t^2} \left( 
 \begin{array}{c} z_1 \\ z_2 \\ z_3  \end{array}
  \right) = \left( 
 \begin{array}{ccc}
   \lambda_1 & \ & \ \\ \ & \lambda_2 & \ \\ \ & \ & 0
 \end{array}
  \right) \left(
 \begin{array}{c} z_1 \\ z_2 \\ z_3  \end{array}
  \right) = \left(
 \begin{array}{c} \lambda_1 z_1 \\ \lambda_2 z_2 \\ 0  \end{array}
  \right) \label{eqn017}
\end{equation}
\noindent\includegraphics[scale=1.0]{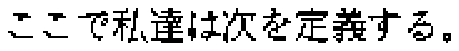}

 Here we define 
\begin{eqnarray}
  {\mit \Delta} \omega^2 &\equiv& \sqrt{ \omega_{12}^4 -\omega_{12}^2
   \omega_{13}^2 +\omega_{13}^4 -\omega_{13}^2 \omega_{23}^2 +\omega_{23}^4
   -\omega_{23}^2 \omega_{12}^2 } \nonumber \\
  &=& \sqrt{ \frac{1}{2} \left\{ (\omega_{12}^2-\omega_{13}^2)^2
   +(\omega_{13}^2-\omega_{23}^2)^2 +(\omega_{23}^2-\omega_{12}^2)^2
		       \right\} } \nonumber \\
  &=& \sqrt{(\omega_{12}^2 +\omega_{13}^2 +\omega_{23}^2 )^2
   -3\omega_{12}^2 \omega_{13}^2 -3\omega_{13}^2 \omega_{23}^2 -3\omega_{23}^2 \omega_{12}^2}
\end{eqnarray}
\noindent\includegraphics[scale=1.0]{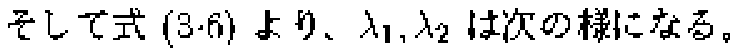}

 and from equation (\ref{eqn012}), \(\lambda_1,\lambda_2\) are 
\begin{equation}
  \left\{ \begin{array}{l}
    \lambda_1 = -\omega_{12}^2 -\omega_{13}^2 -\omega_{23}^2 +{\mit \Delta}
     \omega^2 < 0 \\
    \lambda_2 = -\omega_{12}^2 -\omega_{13}^2 -\omega_{23}^2 -{\mit \Delta}
     \omega^2 < 0
   \end{array} \right.
\end{equation}
\noindent\includegraphics[scale=1.0]{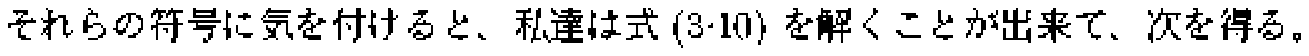}

Taking care of their signs, we can solve the equation (\ref{eqn017})
then we get 
\begin{eqnarray}
  && z_{1(t)} = A_1 \sin{\Omega_1 t} +B_1 \cos{\Omega_1 t} \quad \nonumber \\
  && z_{2(t)} = A_2 \sin{\Omega_2 t} +B_2 \cos{\Omega_2 t} \quad \nonumber \\
  && z_{3(t)} = C_1 t +C_2 \quad
\end{eqnarray}
\noindent\includegraphics[scale=1.0]{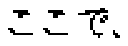}

 , where
\begin{equation}
  \Omega_1 \equiv \sqrt{-\lambda_1}, \  \Omega_2 \equiv \sqrt{-\lambda_2}
\end{equation}
\noindent\includegraphics[scale=1.0]{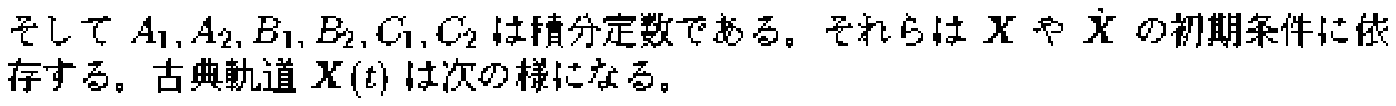}

 and \( A_1,A_2,B_1,B_2,C_1,C_2 \) are integral constants. They depend on
 the initial condition of \( \vecX \) and \( \dot{\vecX} \). The
 classical orbit \( \vecX (t) \) is 
\begin{equation}
 \vecX _{(t)} = \vecP^{-1} \vecZ_{(t)}
\end{equation}
\noindent\includegraphics[scale=1.0]{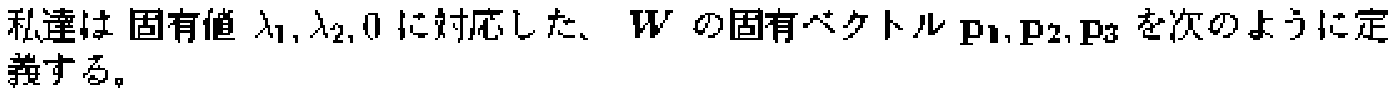}

 We define \( \vecW \)'s eigen vector \( { \bf p_1,p_2,p_3 } \)
corresponding to its eigen value \( \lambda_1,\lambda_2,0 \) as follows.
\begin{equation}
  \vecP^{-1} = \left( { \bf p_1 \ p_2 \ p_3 } \right) = 
 \left( \begin{array}{ccc}
  \xi_1 & \xi_2 & 1 \\ \eta_1 & \eta_2 & 1 \\ \zeta_1 & \zeta_2 & 1
 \end{array} \right)
= 
 \left( \begin{array}{ccc}
  \xi_1 & \xi_2 & 1 \ \\ \eta_1 & \eta_2 & 1 \ \\ -\xi_1-\eta_1 &
   -\xi_2-\eta_2 & 1 \
 \end{array} \right)
\end{equation}
\noindent\includegraphics[scale=1.0]{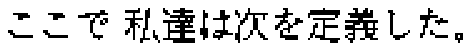}

 Here we difined
\begin{equation}
 \begin{array}{ll}
  \xi_1 = \omega_{12}^2 \omega_{23}^2 - \omega_{13}^2 (\omega_{13}^2
   -{\mit \Delta} \omega^2) \quad & \xi_2 = \omega_{12}^2 \omega_{23}^2 - \omega_{13}^2 (\omega_{13}^2
   +{\mit \Delta} \omega^2) \\
  \eta_1 = \omega_{12}^2 \omega_{13}^2 - \omega_{23}^2 (\omega_{23}^2
   -{\mit \Delta} \omega^2) \quad & \eta_2 = \omega_{12}^2 \omega_{13}^2 - \omega_{23}^2 (\omega_{23}^2
   +{\mit \Delta} \omega^2) \\
  \zeta_1 = -\xi_1 -\eta_1 \quad & \zeta_2 = -\xi_2 -\eta_2
 \end{array}
\end{equation}
\noindent\includegraphics[scale=1.0]{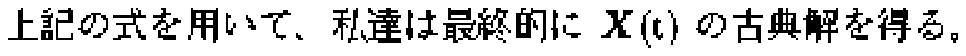}

 Using formulae above, we can get the classical solution of \( \vecX
 \)(t) finally.
\begin{equation}
 \vecX_{(t)} = 
  \left( \begin{array}{c} 
   x_{1(t)} \\ x_{2(t)} \\ x_{3(t)}
  \end{array} \right) =
  \left( \begin{array}{c} 
    \xi_1 z_{1(t)} +\xi_2 z_{2(t)} +z_{3(t)} \\
    \eta_1 z_{1(t)} +\eta_2 z_{2(t)} +z_{3(t)} \\
    -(\xi_1+\eta_1) z_{1(t)} -(\xi_2+\eta_2) z_{2(t)} +z_{3(t)} \\
  \end{array} \right)  \label{eqn025}
\end{equation}

\noindent\includegraphics[scale=1.0]{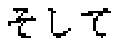}

Using 
\begin{equation}
  \Delta \equiv \eta_2 \xi_1 -\eta_1 \xi_2 
\end{equation}
\noindent\includegraphics[scale=1.0]{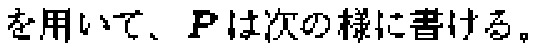}

, you can write \( \vecP \)
\begin{equation}
  \vecP = \frac{1}{3\Delta} 
   \left( \begin{array}{ccc}
    2\eta_2 +\xi_2 & -\eta_2 -2\xi_2 & -\eta_2 +\xi_2 \\
    -2\eta_1 -\xi_1 & \eta_1 +2\xi_1 & \eta_1 -\xi_1 \\
    \Delta & \Delta & \Delta
   \end{array} \right)
\end{equation}
\noindent\includegraphics[scale=1.0]{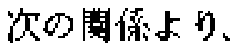}

 From the relation
\begin{equation}
  \vecZ_{(t)} = \vecP \vecX_{(t)}
\end{equation}
\noindent\includegraphics[scale=1.0]{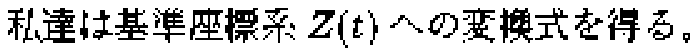}

 , we can get the formula for transformation to the normal coordinate \( \vecZ (t) \).
\begin{equation}
  \vecZ_{(t)} = 
  \left( \begin{array}{c} 
   z_{1(t)} \\ z_{2(t)} \\ z_{3(t)}
  \end{array} \right) = \frac{1}{3\Delta}
  \left( \begin{array}{c} 
    (2\eta_2 +\xi_2) x_{1(t)} +(-\eta_2 -2\xi_2) x_{2(t)} +(-\eta_2 +\xi_2) x_{3(t)} \\
    (-2\eta_1 -\xi_1) x_{1(t)} +(\eta_1 +2\xi_1) x_{2(t)} +(\eta_1 -\xi_1)x_{3(t)} \\
    \Delta x_{1(t)} +\Delta x_{2(t)} +\Delta x_{3(t)} \\
  \end{array} \right) \label{eqn028}
\end{equation}
\noindent\includegraphics[scale=1.0]{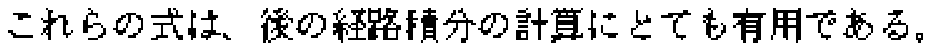}

 These formurae are very useful for evaluation of path integrals later.

 \subsection{Derivation of a propagator}
\noindent\includegraphics[scale=1.0]{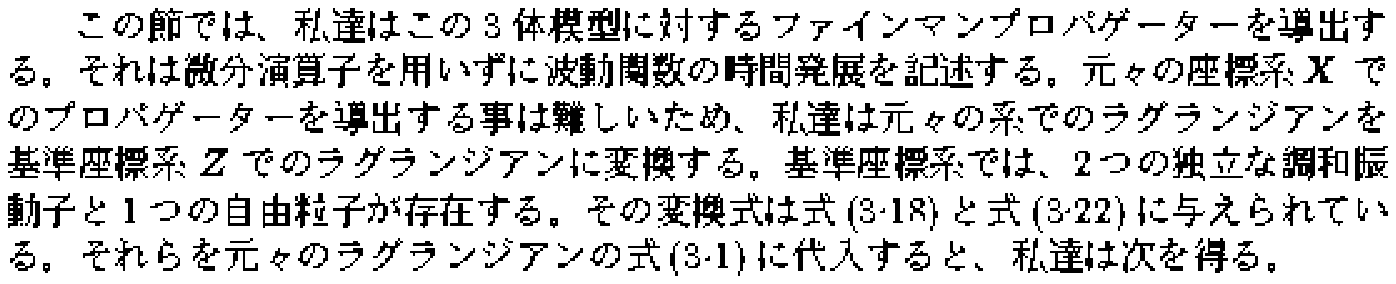}

  In this section, we derive the Feynman propagator for this 3 body model. It describes
 evolution of wave functions without differential operators. It is difficult to
 derive the propagator in the original coordinates \( \vecX \), therefore we transform the
 Lagrangean in the original coordinates into the Lagrangean in the
 normal coordinates \( \vecZ \), where there are two
 uncoupled harmonic oscillators and a free particle. Its transformation
 formulae are given in eq.(\ref{eqn025}) and eq.(\ref{eqn028}). We substitute them for the original
 Lagrangean eq.(\(\ref{eqn007}\)), then we get

\begin{equation}
  L = \frac{m_1}{2} \dot{z}^2_{1(t)} +\frac{m_2}{2} \dot{z}^2_{2(t)}
   +\frac{m_3}{2} \dot{z}^2_{3(t)} -\frac{m_1}{2} \omega_1^2 z^2_{1(t)}
   -\frac{m_2}{2} \omega_2^2 z^2_{2(t)}
\end{equation}
\noindent\includegraphics[scale=1.0]{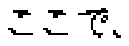}

 where 
\begin{equation}
  m_1 \equiv 2m(\xi_1^2 +\xi_1\eta_1 +\eta_1^2), 
  \quad m_2 \equiv 2m(\xi_2^2 +\xi_2\eta_2 +\eta_2^2),
  \quad m_3 \equiv 3m
\end{equation}
\begin{equation} \left\{
 \begin{array}{c}
  \omega_1^2 \equiv  \frac{m}{m_1} \{ \
   w_1^2 (2\xi_1^2 -\xi_1\eta_1 -\eta_1^2)
   +w_2^2(-\xi_1^2 -\xi_1\eta_1 +2\eta_1^2)
   +w_3^2(2\xi_1^2 +5\xi_1\eta_1 +2\eta_1^2) \ \} \\
  \omega_2^2 \equiv  \frac{m}{m_2} \{ \
   w_1^2 (2\xi_2^2 -\xi_2\eta_2 -\eta_2^2)
   +w_2^2(-\xi_2^2 -\xi_2\eta_2 +2\eta_2^2)
   +w_3^2(2\xi_2^2 +5\xi_2\eta_2 +2\eta_2^2) \ \} \\
  ( \ w_1^2 = \omega_{12}^2+\omega_{13}^2, \quad  w_2^2 =
   \omega_{12}^2+\omega_{23}^2, \quad  w_3^2 =
   \omega_{13}^2+\omega_{23}^2 \ )
  \end{array} \right.
\end{equation}
\noindent\includegraphics[scale=1.0]{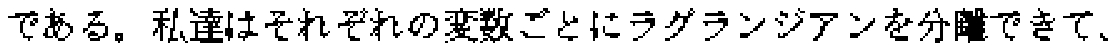}

  , we can decouple the Lagrangean with each variable. 
\begin{eqnarray}
  &&L_1(z_{1(t)},t) \equiv \frac{m_1}{2} \dot{z}^2_{1(t)} -\frac{m_1}{2} \omega_1^2
   z^2_{1(t)}, \
  L_2(z_{2(t)},t) \equiv \frac{m_2}{2} \dot{z}^2_{2(t)} -\frac{m_2}{2} \omega_2^2
  z^2_{2(t)}, \ \nonumber \\
  && \quad L_3(z_{3(t)},t) \equiv \frac{m_3}{2} \dot{z}^2_{3(t)}  \quad
   : L = L_1 +L_2 + L_3 
\end{eqnarray}
\noindent\includegraphics[scale=1.0]{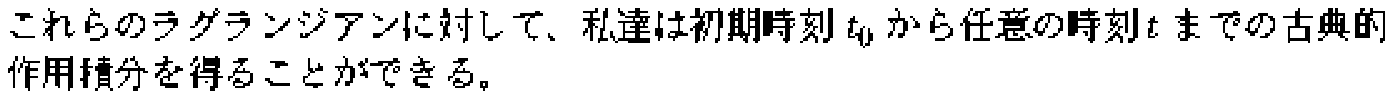}

 For these Lagrangeans, we can get the classical action integrals summed up
 from an initial time \(t_0\) to an arbitrary time \(t\).
\begin{eqnarray}
 &\mbox{ }& S^{(cl)}(\vecZ_{(t)},t:\vecZ_{(t_0)},t_0) 
  = \int_{t_0}^t L_{1(\tau)} d\tau +\int_{t_0}^t L_{2(\tau)} d\tau
  +\int_{t_0}^t L_{3(\tau)} d\tau \nonumber \\
 &\mbox{ }& \quad\quad \equiv S^{(cl)}_1(z_{1(t)},t:z_{1(t_0)},t_0)
  +S^{(cl)}_2(z_{2(t)},t:z_{2(t_0)},t_0)
  +S^{(cl)}_3(z_{3(t)},t:z_{3(t_0)},t_0) \nonumber \\
 && \
\end{eqnarray}
\noindent\includegraphics[scale=1.0]{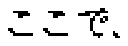}

 where 
\begin{eqnarray}
  && S^{(cl)}_1 = \frac{m_1 \omega_1}{2\sin{\omega_1(t-t_0)}} 
       \left\{  \cos{\omega_1(t-t_0)} ( z^2_{1(t)} +z^2_{1(t_0)} )
	-2z_{1(t)}z_{1(t_0)}  \right\} \\
  && S^{(cl)}_2 = \frac{m_2 \omega_2}{2\sin{\omega_2(t-t_0)}} 
       \left\{  \cos{\omega_2(t-t_0)} ( z^2_{2(t)} +z^2_{2(t_0)} )
	-2z_{2(t)}z_{2(t_0)}  \right\} \\
  && S^{(cl)}_3 = \frac{m_3 ( z_{3(t)} -z_{3(t_0)} )^2}{2(t-t_0)}
\end{eqnarray}
\noindent\includegraphics[scale=1.0]{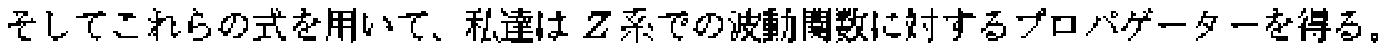}

 and using these formulae, we get the propagator for wave function
 in the system \( \vecZ \). 
\begin{eqnarray}
  U(\vecZ,t:\vecZ_0,t_0) &=& \int_{\vecZ(t_0)=\vecZ_0}^{\vecZ(t)=\vecZ}
   D\vecZ(\tau) \exp \left\{ \frac{i}{\hbar}
		      S(\vecZ,t:\vecZ(\tau),\tau:\vecZ_0,t_0)   \right\}   \\
  &\propto& \exp \left\{\frac{i}{\hbar} S^{(cl)}(\vecZ,t:\vecZ_0,t_0)
		 \right\} \label{eqn038}
\end{eqnarray}
\noindent\includegraphics[scale=1.0]{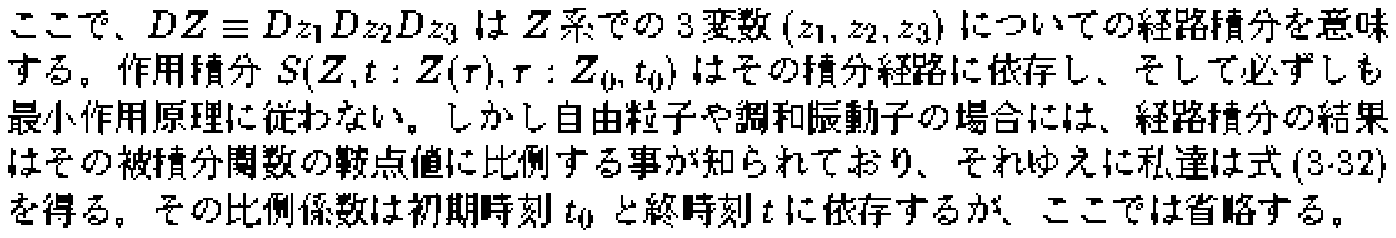}

 Here \( D\vecZ \equiv Dz_1 Dz_2 Dz_3 \) means path integrals about three
 variables (\(z_1,z_2,z_3\)) in system \( \vecZ \). The action integral \(
 S(\vecZ,t:\vecZ(\tau),\tau:\vecZ_0,t_0) \) depends on its integral
 paths and does not always follow the principle of minimum action. But it is known that the
 result of path integrals is in proportion to the value of saddle point
 of their integrands for free particles and for harmonic oscillators,
 therefore we get equation (\ref{eqn038}). Its
 proportional factor depends on the initial time \(t_0\) and the final time
 \(t\), but here we omit the factor.

\noindent\includegraphics[scale=1.0]{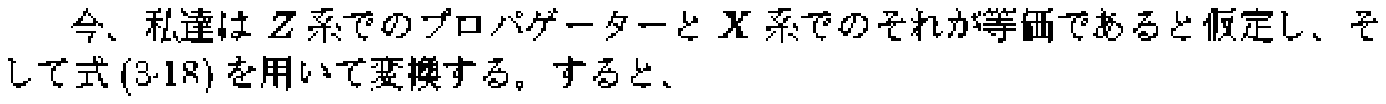}

 Now we assume that the propagator in \( \vecZ \) is equivalent to the
 one in original system \( \vecX \), and transform it using equation
 (\ref{eqn025}) . Then we get
\begin{eqnarray}
 &\mbox{ }& U(\vecX,t:\vecX_0,t_0) = U(\vecZ,t:\vecZ_0,t_0) \nonumber \\
  &\mbox{ }& \qquad \propto \exp \biggl[ \frac{i}{\hbar} \left\{ A_1 x_{1(0)}^2 +A_2 x_{2(0)}^2 
  +A_3 x_{3(0)}^2 +B_{12} x_{1(0)} x_{2(0)} \right. \biggr.\nonumber \\
   &\mbox{ }& \qquad\qquad \biggl. \left. +B_{23} x_{2(0)} x_{3(0)} +B_{13} x_{1(0)} x_{3(0)} +C_1 x_{1(0)}
  +C_2 x_{2(0)} +C_3 x_{3(0)} +D \ \right\} \biggr] \label{eqn039}
   \nonumber \\
  && \
\end{eqnarray}
\noindent\includegraphics[scale=1.0]{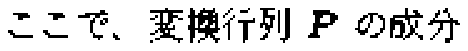}

 Here using elements of the transformation matrix \( \vecP \) 
\begin{equation}
 \begin{array}{lll}
   a_1 = \frac{1}{3\Delta}(2\eta_2+\xi_2) & 
   a_2 = \frac{1}{3\Delta}(-\eta_2-2\xi_2) & 
   a_3 = \frac{1}{3\Delta}(-\eta_2+\xi_2) \\
   b_1 = \frac{1}{3\Delta}(-2\eta_1-\xi_1) & 
   b_2 = \frac{1}{3\Delta}(\eta_1+2\xi_1) & 
   b_3 = \frac{1}{3\Delta}(\eta_1-\xi_1) \\
   c_1 = \frac{1}{3} & c_2 = \frac{1}{3} & c_3 = \frac{1}{3}
 \end{array}
\end{equation}
\noindent\includegraphics[scale=1.0]{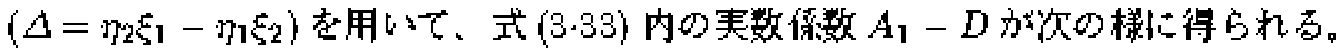}

 (\( \Delta = \eta_2\xi_1-\eta_1\xi_2 \)), we get the real coefficients in
 (\(\ref{eqn039}\)) \( A_1 - D \) as follows. 
\begin{eqnarray}
  &\mbox{ }& A_i = \frac{m_1\omega_1}{2}\cot[\omega_1(t-t_0)]a_i^2
   +\frac{m_2\omega_2}{2}\cot[\omega_2(t-t_0)]b_i^2
   +\frac{m_3}{2(t-t_0)} c_i^2 \\
  &\mbox{ }& B_{ij} = m_1\omega_1 \cot [\omega_1 (t-t_0)]a_i a_j +m_2\omega_2 \cot
   [\omega_2 (t-t_0)]b_i b_j +\frac{m_3}{(t-t_0)} c_i c_j  \\
  &\mbox{ }& C_i(\vecX) = -\frac{m_1 \omega_1}{\sin[\omega_1(t-t_0)]}
     \left( a_1 x_1 + a_2 x_2 + a_3 x_3 \right) a_i \nonumber \\
  &\mbox{ }& \qquad\qquad\qquad\qquad\qquad -\frac{m_2 \omega_2}{\sin[\omega_2(t-t_0)]}
     \left( b_1 x_1 + b_2 x_2 + b_3 x_3 \right) b_i \nonumber \\
  &\mbox{ }& \qquad\qquad\qquad\qquad\qquad\qquad\qquad\qquad -\frac{m_3}{(t-t_0)}
     \left( c_1 x_1 + c_2 x_2 + c_3 x_3 \right) c_i \\
  &\mbox{ }& D(\vecX) = \frac{m_1 \omega_1}{2} \cot[\omega_1(t-t_0)](a_1 x_1 +
  a_2 x_2 + a_3 x_3)^2  \nonumber \\
  &\mbox{ }& \qquad\qquad\qquad\qquad\qquad +\frac{m_2 \omega_2}{2} \cot[\omega_2(t-t_0)](b_1 x_1 + b_2 x_2 + b_3
   x_3)^2  \nonumber \\
  &\mbox{ }& \qquad\qquad\qquad\qquad\qquad\qquad\qquad\qquad\quad +\frac{m_3}{2(t-t_0)}(c_1 x_1 + c_2 x_2 + c_3 x_3)^2
\end{eqnarray}
\noindent\includegraphics[scale=1.0]{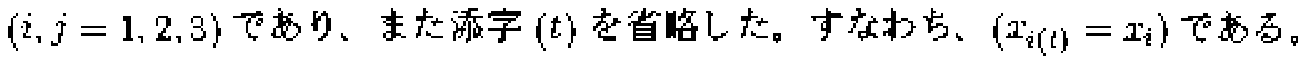}

 \( (i,j = 1,2,3 )\), and we omitted index \((t)\) , that is
 \((x_{i(t)}=x_i)\).

 \subsection{Derivation of wave function and numerical calculation
  of reduced density function.}
\noindent\includegraphics[scale=1.0]{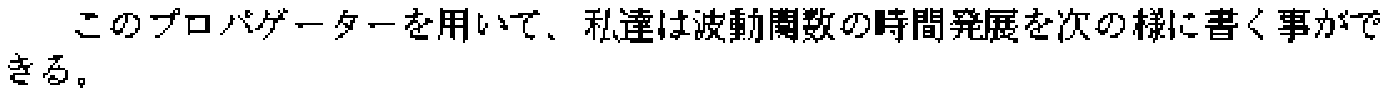}

 Using this propagator, we can write the development of wave
  function as follows. 
\begin{equation}
  \psi(\vecX,t) = \int_{-\infty}^{\infty} d\vecX_0 \ U(\vecX,t:\vecX_0,t_0)
   \ \psi(\vecX_0,t_0) \label{eqn045}
\end{equation}
\noindent\includegraphics[scale=1.0]{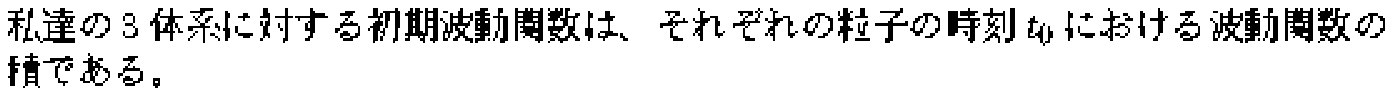}

 The initial wave function for our 3 body system is a product of wave functions of
 each particle at the time \(t_0\).

\begin{equation}
 \psi(\vecX_0,t_0) = \psi_1(x_{1(0)},t_0) \ \psi_2(x_{2(0)},t_0) \
  \psi_3(x_{3(0)},t_0) \label{eqn046}
\end{equation}
\noindent\includegraphics[scale=1.0]{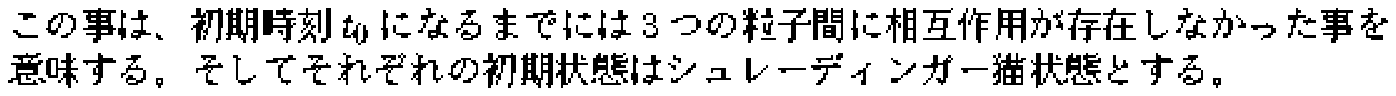}

This means that there has been no interaction among those 3 particles
 until the initial time \(t_0\).
 And the each initial state is the Schr\"odinger cat state
\begin{equation}
 \psi_1(x_{1(0)},t_0) = \tilde{ N_1 } \left[
				       \exp\left\{-\frac{x_{1(0)}^2}{4\sigma_1^2}\right\}+\exp\left\{-\frac{(x_{1(0)}-d_1)^2}{4\sigma_1^2}\right\}
				      \right]  \quad \mbox{etc...} \label{eqn047}
\end{equation}
\noindent\includegraphics[scale=1.0]{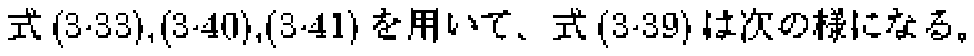}

 With equations (\(\ref{eqn039}\)),(\(\ref{eqn046}\)),(\(\ref{eqn047}\)), equation
 (\(\ref{eqn045}\)) comes to  

\begin{eqnarray}
 && \qquad\ \psi(\vecX,t) \propto \int_{-\infty}^{\infty} dx_{1(0)}
  \int_{-\infty}^{\infty} dx_{2(0)} \int_{-\infty}^{\infty} dx_{3(0)}
  \nonumber \\
 &&  \times \exp \left[ \frac{i}{\hbar} \bigl\{  A_1 x_{1(0)}^2 +A_2 x_{2(0)}^2
			       +A_3 x_{3(0)}^2  
 +B_{12} x_{1(0)} x_{2(0)} +B_{23} x_{2(0)} x_{3(0)}  \bigr. \right. \nonumber \\ 
 && \qquad\qquad\qquad\qquad\qquad\qquad\qquad
  \biggl. \bigl. +B_{13} x_{1(0)} x_{3(0)} +C_1 x_{1(0)} +C_2 x_{2(0)} +C_3 x_{3(0)} +D  \bigr\}
  \biggr] \nonumber \\
 && \times \biggl[ \exp \left\{
			 -\frac{x_{1(0)}^2}{4\sigma_1^2}-\frac{x_{2(0)}^2}{4\sigma_2^2}-\frac{x_{3(0)}^2}{4\sigma_3^2}  \right\} +\exp \left\{ -\frac{(x_{1(0)}-d_1)^2}{4\sigma_1^2}-\frac{x_{2(0)}^2}{4\sigma_2^2}-\frac{x_{3(0)}^2}{4\sigma_3^2}  \right\} + \cdots \nonumber \\
 && \qquad\qquad \cdots +\exp \left\{
			       -\frac{(x_{1(0)}-d_1)^2}{4\sigma_1^2}-\frac{(x_{2(0)}-d_2)^2}{4\sigma_2^2}-\frac{(x_{3(0)}-d_3)^2}{4\sigma_3^2}  \right\} \biggr] \label{eqnnew342}
\end{eqnarray}
\noindent\includegraphics[scale=1.0]{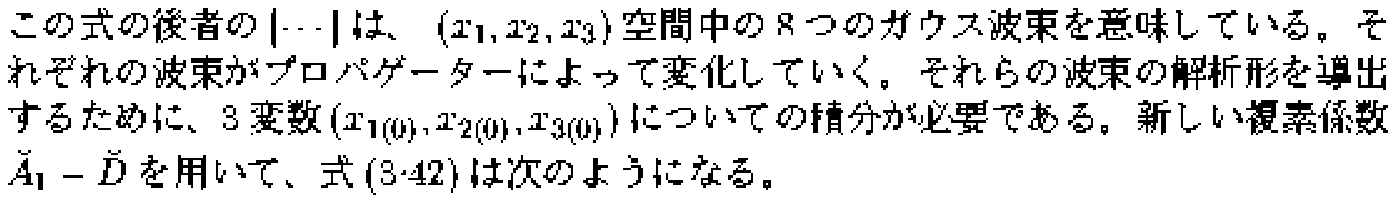}

 The latter \([ \cdots ]\) of this formula means 8 Gaussian packet in
 (\( x_1,x_2,x_3 \)) space. Each packet changes by propagator. For
 evaluating analytic forms of those packets, integrations with three variables (\( x_{1(0)},x_{2(0)},x_{3(0)} \))
 are needed. With the new complex coefficients \(\breve{A}_1 - \breve{D}
 \), eq.(\ref{eqnnew342}) turns into
\begin{eqnarray}
 && \qquad\ \psi(\vecX,t) = \sum_{k=0-7} \psi^{(k)} (\vecX,t) \\
 && \qquad\ \psi^{(k)}(\vecX,t) \propto \int_{-\infty}^{\infty} dx_{1(0)}
  \int_{-\infty}^{\infty} dx_{2(0)} \int_{-\infty}^{\infty} dx_{3(0)}
  \nonumber \\
 &&  \times \exp \left[   -\breve{A}_1 x_{1(0)}^2 -\breve{A}_2 x_{2(0)}^2
			       -\breve{A}_3 x_{3(0)}^2  
 +\breve{B}_{12} x_{1(0)} x_{2(0)} +\breve{B}_{23} x_{2(0)} x_{3(0)}  \right. \nonumber \\ 
 && \qquad\qquad\qquad\qquad\qquad \biggl. +\breve{B}_{13} x_{1(0)}
  x_{3(0)} +\breve{C}^{(k)}_1 x_{1(0)} +\breve{C}^{(k)}_2 x_{2(0)}
  +\breve{C}^{(k)}_3 x_{3(0)} +\breve{D}^{(k)}  \biggr] \nonumber \\
 && \
\end{eqnarray}
\noindent\includegraphics[scale=1.0]{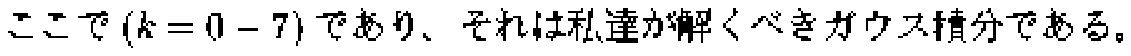}

 (\( k=0-7 \)), which is the Gaussian integrals we have to solve .
\begin{eqnarray}
 && \breve{A}_1 = \frac{1}{4\sigma_1^2} - \frac{i}{\hbar}A_1, \quad 
  \breve{B}_{12} = \frac{i}{\hbar}B_{12}, \quad  \breve{C}^{(k)}_1(\vecX) =
  \frac{d^{(k)}_1}{2\sigma_1^2} +\frac{i}{\hbar}C_1(\vecX)
\quad \mbox{etc...} \nonumber \\
 && \qquad\qquad \breve{D}^{(k)}(\vecX) = -\frac{d^{(k)2}_1}{4\sigma_1^2}
  -\frac{d^{(k)2}_2}{4\sigma_2^2} -\frac{d^{(k)2}_3}{4\sigma_3^2} +\frac{i}{\hbar}D(\vecX)
\end{eqnarray}
\noindent\includegraphics[scale=1.0]{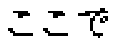}

 where 
\begin{equation} \left(
 \begin{array}{rrcccl}
   \quad & [& d^{(k)}_1 & d^{(k)}_2 & d^{(k)}_3 &] \\
   k = 0 & [& \ 0 & \ 0 & \ 0 &] \\
       1 & [& d_1 & \ 0 & \ 0 &] \\
       2 & [& \ 0 & d_2 & \ 0 &] \\
       3 & [& \ 0 & \ 0 & d_3 &] 
 \end{array}
 \quad
 \begin{array}{rrcccl}
       \ & \ & \quad \ & \quad \ & \quad \  & \ \\
       4 & [& d_1 & d_2 & \ 0 &] \\
       5 & [& d_1 & \ 0 & d_3 &] \\
       6 & [& \ 0 & d_2 & d_3 &] \\
       7 & [& d_1 & d_2 & d_3 &] 
 \end{array} \right) \label{eqn052}
\end{equation}
\noindent\includegraphics[scale=1.0]{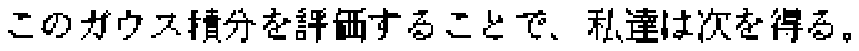}

 Evaluating this Gaussian integrals, we get
\begin{equation}
  \psi^{(k)}(\vecX,t) \propto \sqrt{\frac{\pi^3}{\Delta}} \exp \left[
 \frac{1}{16\Delta}\Phi^{(k)}(\vecX,t) +\breve{D}^{(k)}(\vecX,t) \right] \label{eqn053}
\end{equation}
\noindent\includegraphics[scale=1.0]{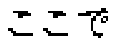}

 where
\begin{eqnarray}
  && \Delta(t) = \breve{A}_1 \breve{A}_2 \breve{A}_3
   -\frac{1}{4}( \breve{A}_2 \breve{B}^2_{13} +\breve{A}_3
   \breve{B}^2_{12} +\breve{A}_1 \breve{B}^2_{23} )
   -\frac{1}{4}\breve{B}_{12} \breve{B}_{13} \breve{B}_{23} \\[5pt]
  && \qquad \ \equiv \Re e \Delta(t) +i \ \Im m \Delta(t) \\[5pt]
  && \quad \Re e \Delta(t) = \frac{1}{4^3 \sigma_1^2 \sigma_2^2
   \sigma_3^2}-\frac{1}{4\hbar^2}\left( \frac{A_2 A_3}{\sigma_1^2}
  +\frac{A_3 A_1}{\sigma_2^2} +\frac{A_1 A_2}{\sigma_3^2} \right)
   \nonumber \\
  && \qquad\qquad\qquad\qquad  +\frac{1}{16\hbar^2} \left(
   \frac{B_{23}^2}{\sigma_1^2} +\frac{B_{13}^2}{\sigma_2^2}
   +\frac{B_{12}^2}{\sigma_3^2} \right) \\
  && \quad \Im m \Delta(t) = \frac{1}{\hbar^3}A_1 A_2 A_3
   -\frac{1}{16\hbar}\left( \frac{A_3}{\sigma_1^2 \sigma_2^2}
    +\frac{A_1}{\sigma_2^2 \sigma_3^2} +\frac{A_2}{\sigma_3^2
    \sigma_1^2} \right) \nonumber \\
  && \qquad\qquad\qquad\qquad -\frac{1}{4\hbar^3}(A_1 B_{23}^2 +A_2 B_{13}^2
    +A_3 B_{12}^2) +\frac{1}{4\hbar^3}B_{12}B_{13}B_{23} 
\end{eqnarray}
\noindent\includegraphics[scale=1.0]{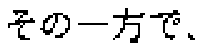}

 while
\begin{eqnarray}
  && \Phi^{(k)}(\vecX,t) = 4( \ \breve{A}_2 \breve{A}_3 \breve{C}^{2}_1(\vecX)
   +\breve{A}_1 \breve{A}_3 \breve{C}^{2}_2(\vecX) +\breve{A}_1 \breve{A}_2
   \breve{C}^{2}_3(\vecX) \ ) \nonumber \\
  && \qquad\qquad\qquad -\breve{B}_{23}^2 \breve{C}_1^2(\vecX) -\breve{B}_{13}^2 \breve{C}_2^2(\vecX) -\breve{B}_{12}^2
   \breve{C}_3^2(\vecX) \nonumber \\
  &&\qquad +2( \ \breve{B}_{13}\breve{B}_{23}\breve{C}_1(\vecX)\breve{C}_2(\vecX)
   +\breve{B}_{12}\breve{B}_{13}\breve{C}_2(\vecX)\breve{C}_3(\vecX)
  +\breve{B}_{12}\breve{B}_{23}\breve{C}_1(\vecX)\breve{C}_3(\vecX) \ ) \nonumber \\
  &&\qquad\qquad +4( \ \breve{A}_1\breve{B}_{23}\breve{C}_2(\vecX)\breve{C}_3(\vecX)
   +\breve{A}_2\breve{B}_{13}\breve{C}_1(\vecX)\breve{C}_3(\vecX)
   +\breve{A}_3\breve{B}_{12}\breve{C}_1(\vecX)\breve{C}_2(\vecX) \ )
   \nonumber \\
 && \
\end{eqnarray}
\noindent\includegraphics[scale=1.0]{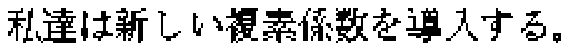}

 We introduce new complex coefficients
\begin{equation}
 (4\breve{A}_2\breve{A}_3-\breve{B}_{23}^2) \equiv \lambda_1 \
  \mbox{etc..} \ , \
  (2\breve{B}_{13}\breve{B}_{23}+4\breve{A}_3\breve{B}_{12}) \equiv
  \mu_{12} \ \mbox{etc..}
\end{equation}
\noindent\includegraphics[scale=1.0]{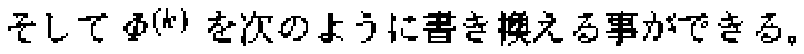}

 , then we can rewrite \(\Phi^{(k)}\) as follows.
\begin{equation}
  \Phi^{(k)}(\vecX,t) = \sum_{i=1}^3 \lambda_i \breve{C}_i^2(\vecX) \ + \ \sum_{
   \stackrel{ \scriptstyle (i,j)=(1,2) \mbox{ \small or}}{ \scriptstyle
   (2,3) \mbox{ \small or } (3,1)} } \mu_{ij} \breve{C}_i(\vecX) \breve{C}_j(\vecX) \qquad\qquad\quad
\end{equation}
\noindent\includegraphics[scale=1.0]{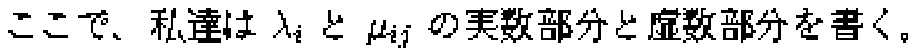}

 Here, we write the real part and the imaginary part of \(\lambda_i\)
 and \(\mu_{ij}\). 
\begin{eqnarray}
  && \lambda_i \equiv \Re e \lambda_i + i \ \Im m \lambda_i \\[5pt]
      && \quad \Re e \lambda_i = -\frac{1}{\hbar^2} \biggl\{ \ m_1 \omega_1
	\cot[\omega_1(t-t_0)] \ m_2 \omega_2 \cot[\omega_2(t-t_0)] \
	(a_jb_k-a_kb_j)^2  \biggr. \nonumber \\
    && \qquad\qquad\qquad\qquad\qquad + m_2 \omega_2 \cot[\omega_2(t-t_0)] \ \frac{m_3}{(t-t_0)} \
	(b_jc_k-b_kc_j)^2 \nonumber \\
    && \qquad\qquad\qquad\qquad \biggl. + \frac{m_3}{(t-t_0)} \ m_1 \omega_1 \cot[\omega_1(t-t_0)] \
	(c_ja_k-c_ka_j)^2 \ \biggr\} +\frac{1}{4 \sigma_j^2 \sigma_k^2} \\[5pt]
      && \quad \Im m \lambda_i = -\frac{1}{2\hbar} \biggl\{ \ m_1 \omega_1 \cot[\omega_1(t-t_0)]\left( \frac{a_j^2}{\sigma_k^2}+\frac{a_k^2}{\sigma_j^2} \right) \nonumber \\
      &&  \qquad\qquad\qquad\qquad\quad +m_2
       \omega_2\cot[\omega_2(t-t_0)]\left(
				     \frac{b_j^2}{\sigma_k^2}+\frac{b_k^2}{\sigma_j^2} \right)  +\frac{m_3}{(t-t_0)}\left( \frac{c_j^2}{\sigma_k^2}+\frac{c_k^2}{\sigma_j^2}  \right) \ \biggr\} \nonumber \\
     && \
\end{eqnarray}
\noindent\includegraphics[scale=1.0]{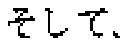}

 and 
\begin{eqnarray}
 && \mu_{ij} \equiv \Re e \mu_{ij} +i \ \Im m \mu_{ij} \\[5pt]
  && \  \Re e \mu_{ij} = -\frac{2}{\hbar^2} \biggl\{
  m_1\omega_1\cot[\omega_1(t-t_0)] \ m_2\omega_2\cot[\omega_2(t-t_0)] \
  (a_jb_k-a_kb_j)(a_kb_i-a_ib_k) \biggr. \nonumber \\
  && \qquad\qquad\qquad\qquad \ +m_2 \omega_2 \cot[\omega_2(t-t_0)] \
   \frac{m_3}{(t-t_0)} \ (b_jc_k-b_kc_j)(b_kc_i-b_ic_k) \nonumber \\
  && \qquad\qquad\qquad\qquad\quad \  \biggl. + \frac{m_3}{(t-t_0)} \ m_1
   \omega_1 \cot[\omega_1(t-t_0)] \ (c_ja_k-c_ka_j)(c_ka_i-c_ia_k) \
   \biggr\} \nonumber \\
  && \ \\
  && \ \Im m \mu_{ij} = \frac{1}{\sigma_k^2 \hbar} \biggl\{
  m_1\omega_1\cot[\omega_1(t-t_0)]a_ia_j \biggr. \nonumber \\
  && \qquad\qquad\qquad\qquad\qquad\qquad\qquad\quad \
   \biggl. +m_2\omega_2\cot[\omega_2(t-t_0)]b_ib_j \
   +\frac{m_3}{(t-t_0)}c_ic_j \ \biggr\} \nonumber \\
  && \ \\
  && \qquad\qquad\qquad\qquad \bigl( \  (i,j,k) = (1,2,3)
  \ \mbox{or} \ (2,3,1) \ \mbox{or} \ (3,1,2) \ \bigr)\nonumber
\end{eqnarray}
\noindent\includegraphics[scale=1.0]{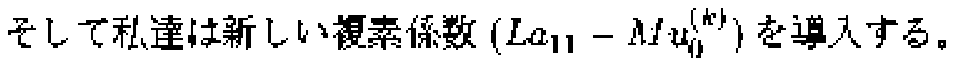}

And we introduce new complex factors (\( La_{11}-Mu^{(k)}_0 \)).
\begin{eqnarray}
  && \quad La_{dd} = \sum_{i=1}^3 \lambda_i \alpha_d^{(i)2} ,\quad
  Mu_{dd} = \sum_{ \stackrel{ \scriptstyle (i,j)=(1,2) \mbox{ \small or}}{
   \scriptstyle (2,3) \mbox{ \small or } (3,1)} } \mu_{ij}
   \alpha_d^{(i)} \alpha_d^{(j)} \quad ( \ d=1,2,3 \ ) \\
  && La_{df} = 2 \sum_{i=1}^3 \lambda_i \alpha_d^{(i)} \alpha_f^{(i)} , \quad
  Mu_{df} = \sum_{(i,j)} \mu_{ij}
   \left( \alpha_d^{(i)} \alpha_f^{(j)} + \alpha_d^{(j)} \alpha_f^{(i)}
   \right) \\
  && \qquad\qquad\qquad \bigl( \ (d,f)=(1,2) \ \mbox{or} \ (2,3) \
   \mbox{or} \ (3,1)  \ \bigr) \nonumber \\
  && La_d^{(k)} = \sum_{i=1}^3 \lambda_i \frac{d_i^{(k)}}{\sigma_i^2}
  \alpha_d^{(i)} , \quad 
  Mu_d^{(k)} = 0.5  \sum_{(i,j)} \mu_{ij} 
   \left( \frac{d_i^{(k)}}{\sigma_i^2} \alpha_d^{(j)} +
    \frac{d_j^{(k)}}{\sigma_j^2} \alpha_d^{(i)}  \right) \\
  && \quad La_0^{(k)} = 0.25 \sum_{i=1}^3 \lambda_i \frac{d_i^{(k)2}}{\sigma_i^4},
  \quad 
  Mu_0^{(k)} = 0.25 \sum_{(i,j)} \mu_{ij}
   \frac{d_i^{(k)}}{\sigma_i^2} \frac{d_j^{(k)}}{\sigma_j^2}
\end{eqnarray}
\noindent\includegraphics[scale=1.0]{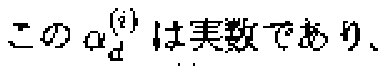}

 this \(\alpha_d^{(i)}\) are real,
\begin{eqnarray}
 && \quad C_i(\vecX) = - \alpha_1^{(i)} x_1 -\alpha_2^{(i)} x_2 -\alpha_3^{(i)}
	 x_3  \\[5pt]
 && \alpha_d^{(i)} \equiv 
	\left(\frac{m_1\omega_1}{\sin[\omega_1(t-t_0)]}a_i\right)a_d
	+\left(\frac{m_2\omega_2}{\sin[\omega_2(t-t_0)]}b_i\right)b_d
        +\left(\frac{m_3}{(t-t_0)}c_i\right)c_d \nonumber \\
 && \
\end{eqnarray}
\noindent\includegraphics[scale=1.0]{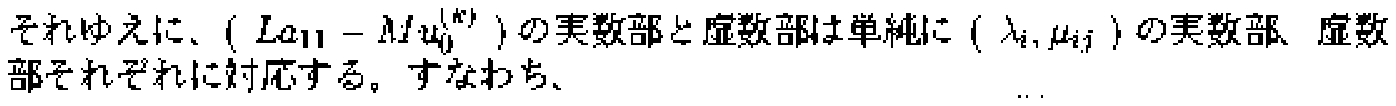}

 therefore the real and the imaginary part of ( \(La_{11}-Mu^{(k)}_0\) )
 simply correspond to the real part and the imaginary part of ( \(\lambda_{i},\mu_{ij}\) ) respectively, that is \\
\begin{equation}
 \quad \Re e La_{dd} = \sum_{i=1}^3 \Re e \lambda_i \alpha_d^{(i)2}
  \quad \mbox{etc..}
\end{equation}
\noindent\includegraphics[scale=1.0]{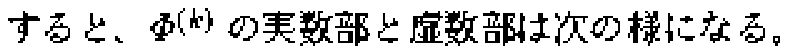}

Then the real and the imaginary part of \(\Phi^{(k)}\) are 
\begin{equation}
  \Phi^{(k)}(\vecX,t) \equiv \Re e \Phi^{(k)}(\vecX,t) +i \ \Im m \Phi^{(k)}(\vecX,t) 
\end{equation}
\noindent\includegraphics[scale=1.0]{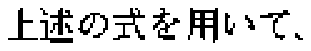}

 With the help of the formulae above,
\begin{eqnarray}
  && \quad \Re e \Phi^{(k)}(\vecX,t) = -\frac{1}{\hbar^2}(\Re e
   La_{11}+\Re e Mu_{11})x_1^2  -\frac{1}{\hbar^2}(\Re e La_{22}+\Re e
   Mu_{22})x_2^2 \nonumber \\
  && \qquad\qquad\qquad\qquad \  -\frac{1}{\hbar^2}(\Re e La_{33}+\Re e Mu_{33})x_3^2
    -\frac{1}{\hbar^2}(\Re e  La_{12}+\Re e Mu_{12})x_1 x_2 \nonumber \\
  && \qquad\qquad\qquad\qquad  -\frac{1}{\hbar^2}(\Re e  La_{23}+\Re e Mu_{23})x_2 x_3 
   -\frac{1}{\hbar^2}(\Re e  La_{31}+\Re e Mu_{31})x_3 x_1  \nonumber \\
  && \qquad\qquad\qquad\qquad \  +\frac{1}{\hbar}(\Im m La^{(k)}_1 +\Im m
   Mu^{(k)}_1 ) x_1 +\frac{1}{\hbar}(\Im m La^{(k)}_2 +\Im m
   Mu^{(k)}_2 ) x_2 \nonumber \\
  && \qquad\qquad\qquad\qquad\quad +\frac{1}{\hbar}(\Im m La^{(k)}_3 +\Im m
   Mu^{(k)}_3 ) x_3 + \Re e La^{(k)}_0 +\Re e Mu^{(k)}_0 \\ [5pt]
  && \quad \Im m \Phi^{(k)}(\vecX,t) = -\frac{1}{\hbar^2}(\Im m
   La_{11}+\Im m Mu_{11})x_1^2  -\frac{1}{\hbar^2}(\Im m La_{22}+\Im m
   Mu_{22})x_2^2 \nonumber \\
  && \qquad\qquad\qquad\qquad \ \ -\frac{1}{\hbar^2}(\Im m La_{33}+\Im m Mu_{33})x_3^2
    -\frac{1}{\hbar^2}(\Im m  La_{12}+\Im m Mu_{12})x_1 x_2 \nonumber \\
  && \qquad\qquad\qquad\qquad  -\frac{1}{\hbar^2}(\Im m  La_{23}+\Im m Mu_{23})x_2 x_3 
   -\frac{1}{\hbar^2}(\Im m  La_{31}+\Im m Mu_{31})x_3 x_1  \nonumber \\
  && \qquad\qquad\qquad\qquad \ -\frac{1}{\hbar}(\Re e La^{(k)}_1 +\Re e
   Mu^{(k)}_1 ) x_1 -\frac{1}{\hbar}(\Re e La^{(k)}_2 +\Re e
   Mu^{(k)}_2 ) x_2 \nonumber \\
  && \qquad\qquad\qquad\qquad\quad -\frac{1}{\hbar}(\Re e La^{(k)}_3 +\Re e
   Mu^{(k)}_3 ) x_3 + \Im m La^{(k)}_0 +\Im m Mu^{(k)}_0
\end{eqnarray}
\noindent\includegraphics[scale=1.0]{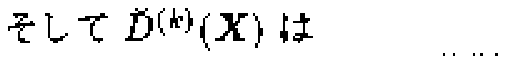}

 And another formula \( \breve{D}^{(k)}(\vecX) \) is
\begin{eqnarray}
  && \breve{D}^{(k)}(\vecX) \equiv \Re e \breve{D}^{(k)} + i \ \Im m
   \breve{D}(\vecX) \\[5pt]
  && \qquad \Re e \breve{D}^{(k)} =
   -\frac{d_1^{(k)2}}{4\sigma_1^2}-\frac{d_2^{(k)2}}{4\sigma_2^2}-\frac{d_3^{(k)2}}{4\sigma_3^2}
   \\
  && \qquad \Im m \breve{D}(\vecX) = D(\vecX)/\hbar
\end{eqnarray}
\noindent\includegraphics[scale=1.0]{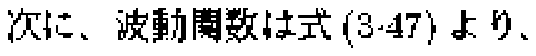}

 Next, the wave function is from eq.(\ref{eqn053}), 
\begin{eqnarray}
  && \psi^{(k)}(\vecX,t) \propto \sqrt{ \frac{\pi^3\Delta^{*} }{
   |\Delta|^2} } \ \exp \left[ \ \frac{\Delta^{*}}{16|\Delta|^2}
		       \Phi^{(k)}(\vecX,t) + \breve{D}^{(k)}(\vecX,t)
			\ \right] \\[5pt]
  && \qquad\qquad\quad \equiv \ \breve{Q} \ \exp [ \ \Theta^{(k)}(\vecX,t) \ ]
\end{eqnarray}
\noindent\includegraphics[scale=1.0]{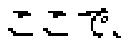}

 where 
\begin{eqnarray}
  &&  \qquad\qquad\qquad\qquad\qquad \breve{Q} \equiv \Re e \breve{Q} + i \ \Im m \breve{Q} \\[5pt]
   &&  \Re e \breve{Q} = \sqrt{\frac{\pi^3}{|\Delta|}} \
       \cos\frac{\phi}{2} \ , \quad \Im m \breve{Q} =
       \sqrt{\frac{\pi^3}{|\Delta|}} \ \sin\frac{\phi}{2} \ \quad :
       \phi = \arctan\left( \Im m \Delta/ \Re e \Delta \right) \nonumber
       \\
   && \
\end{eqnarray}
\noindent\includegraphics[scale=1.0]{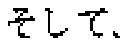}

 and 
\begin{eqnarray}
  && \qquad\qquad\qquad \Theta^{(k)}(\vecX,t) \equiv \Re e \Theta^{(k)}(\vecX,t) +i \ \Im m \Theta^{(k)}(\vecX,t)
   \\[5pt]  && \Re e \Theta^{(k)}(\vecX,t) = \frac{1}{16|\Delta|^2} ( \ \Re e \Delta \cdot \Re e \Phi^{(k)} +\Im m \Delta \cdot \Im m \Phi^{(k)} \ )
   + \Re e \breve{D}^{(k)} \\
  && \Im m \Theta^{(k)}(\vecX,t) = \frac{1}{16|\Delta|^2} ( \ \Re e
   \Delta \cdot \Im m \Phi^{(k)} -\Im m \Delta \cdot \Re e \Phi^{(k)} \ )
   + \Im m \breve{D}(\vecX)
\end{eqnarray}
\noindent\includegraphics[scale=1.0]{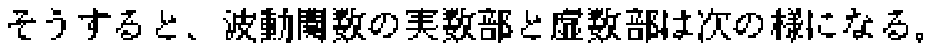}

 Then, the real part and the imaginary part of the wave function are as
 follows.
\begin{eqnarray}
 && \quad\qquad\quad \ \psi^{(k)}(\vecX,t) \equiv \Re e \psi^{(k)}(\vecX,t) +i \ \Im m
  \psi^{(k)}(\vecX,t) \\[5pt]
 && \Re e \psi^{(k)}(\vecX,t) \propto \exp[ \ \Re e \Theta^{(k)} \ ]
 \left( \ \Re e \breve{Q} \cdot \cos[\Im m \Theta^{(k)}] -\Im m \breve{Q}
  \cdot \sin[\Im m \Theta^{(k)}] \ \right) \nonumber \\
 && \qquad\qquad\qquad = \sqrt{\frac{\pi^3}{|\Delta|}} \ \exp[ \ \Re e
  \Theta^{(k)} \ ]\cdot\cos\left[ \ \Im m \Theta^{(k)} + \frac{\phi}{2} \ \right] \\
 && \Im m \psi^{(k)}(\vecX,t) \propto \exp[ \ \Re e \Theta^{(k)} \ ]
 \left( \ \Re e \breve{Q} \cdot \sin[\Im m \Theta^{(k)}] +\Im m \breve{Q}
  \cdot \cos[\Im m \Theta^{(k)}] \ \right) \nonumber \\
 && \qquad\qquad\qquad = \sqrt{\frac{\pi^3}{|\Delta|}} \ \exp[ \ \Re e
 \Theta^{(k)} \ ]\cdot\sin\left[ \ \Im m \Theta^{(k)} + \frac{\phi}{2} \ \right]
\end{eqnarray}
\noindent\includegraphics[scale=1.0]{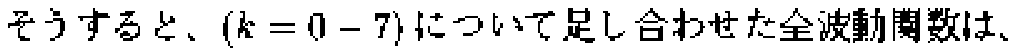}

 Therefore the total wave function summed with (\( k=0-7 \)) is 
\begin{eqnarray}
  && \psi^{(total)}(\vecX,t) \equiv \Re e \psi^{(total)}(\vecX,t) +i \  \Im
   m \psi^{(total)}(\vecX,t) \\[5pt]
  && \qquad \Re e \psi^{(total)} = C \sqrt{\frac{\pi^3}{|\Delta|}} \ \sum_{k=0}^7 \exp[ \ \Re e
  \Theta^{(k)} \ ]\cdot\cos\left[ \ \Im m \Theta^{(k)} + \frac{\phi}{2}
			    \ \right] \\
  && \qquad \Im m \psi^{(total)} = C \sqrt{\frac{\pi^3}{|\Delta|}} \ \sum_{k=0}^7 \exp[ \ \Re e
  \Theta^{(k)} \ ]\cdot\sin\left[ \ \Im m \Theta^{(k)} + \frac{\phi}{2} \ \right]
\end{eqnarray}
\noindent\includegraphics[scale=1.0]{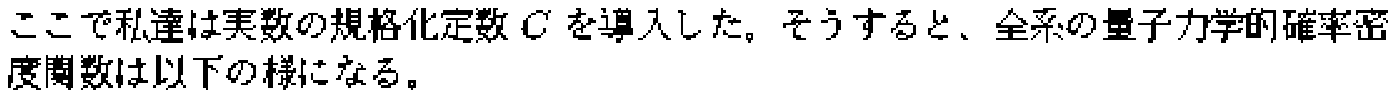}

 Here we introduced a real normalization constant \(C\).  Then the quantum
 mechanical probability density function of total system become as follows.
\begin{eqnarray}
  && \rho^{(total)}(\vecX,t) = \Re e^2 \psi^{(total)} + \Im m^2
   \psi^{(total)} \nonumber \\[5pt]
  && \qquad   = C^2 \frac{\pi^3}{|\Delta|} \ \sum_{k=0}^7 \sum_{l=0}^7
		  \exp[ \ \Re e \Theta^{(k)} +\Re e \Theta^{(l)} \
		  ]\cdot\cos[ \ \Im m \Theta^{(k)} -\Im m
			     \Theta^{(l)} \ ]  \nonumber \\
  && \qquad = C^2 \frac{\pi^3}{|\Delta|} \biggl( \sum_{k=0}^7 
		  \exp[ \ 2\Re e \Theta^{(k)}\ ]  \nonumber \\
  && \qquad\qquad\qquad\qquad +2\sum_{k<l}^{0-7} \exp [ \ \Re e
   \Theta^{(k)}+ \Re e \Theta^{(l)} \ ]\cdot\cos[ \ \Im m
   \Theta^{(k)}-\Im m \Theta^{(l)} \ ] \biggr) \label{eqn097} \nonumber
   \\
  &&
\end{eqnarray}
\noindent\includegraphics[scale=1.0]{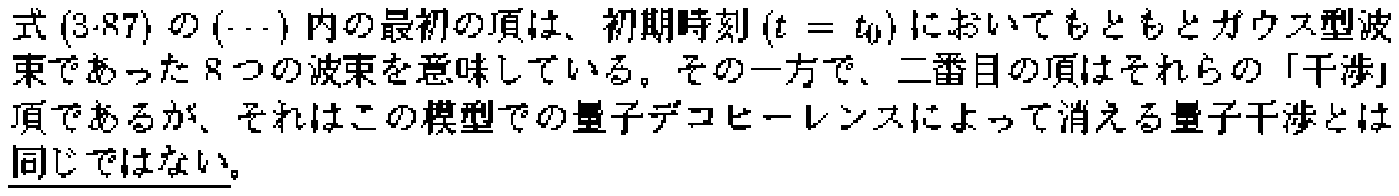}

 The first term in \(( \cdots )\) of equation (\(\ref{eqn097}\)) means
 the eight wave packets which originally are the Gaussian packets at the
 initial time
 (\(t=t_0\)). 
While the second term is their
 ``interference'' term, which is \underline{not} the same as the quantum
 interference vanishes by quantum decoherence in this model.

\noindent\includegraphics[scale=1.0]{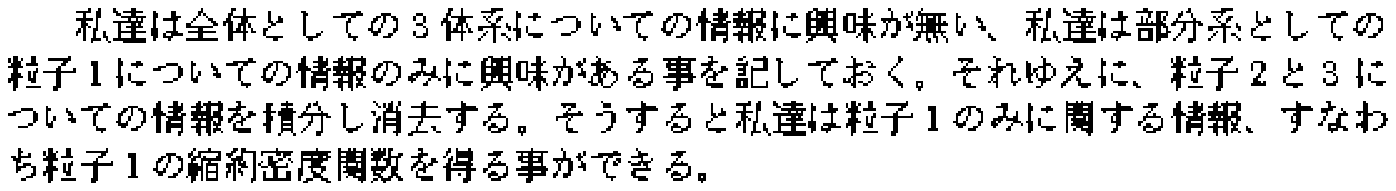}

 Note that we are \underline{not} interested in informations about
 total 3-body system,
 we are interested in only the information about particle-1 as a
 sub-system. Therefore we should integrate out
 informations about particle-2 and -3. Then we can get the information
 about particle-1 only, that is, the reduced density function for particle-1.

\begin{equation}
  \tilde{\rho}_1^{(reduced)}(x_1,t) \equiv \int_{-\infty}^{\infty}
   \int_{-\infty}^{\infty}dx_2 \ dx_3 \ \rho^{(total)}(\vecX,t) \label{eqn098}
\end{equation}
\noindent\includegraphics[scale=1.0]{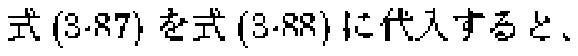}

We substitute eq.(\(\ref{eqn097}\)) into eq.(\(\ref{eqn098}\)),
\begin{eqnarray}
  && \tilde{\rho}_1^{(reduced)}(x_1,t) = C^2 \frac{\pi^3}{|\Delta|}
   \biggl( \sum_{k=0}^7 \int_{-\infty}^{\infty}
   \int_{-\infty}^{\infty}dx_2 \ dx_3 \ \exp[ \ 2\Re e \Theta^{(k)}\ ]  \nonumber \\
  && \qquad +2\sum_{k<l}^{0-7} \int_{-\infty}^{\infty}
   \int_{-\infty}^{\infty}dx_2 \ dx_3 \ \exp [ \ \Re e
   \Theta^{(k)}+ \Re e \Theta^{(l)} \ ]\cdot\cos[ \ \Im m
   \Theta^{(k)}-\Im m \Theta^{(l)} \ ] \biggr) \nonumber \\
  && \
\end{eqnarray}
\noindent\includegraphics[scale=1.0]{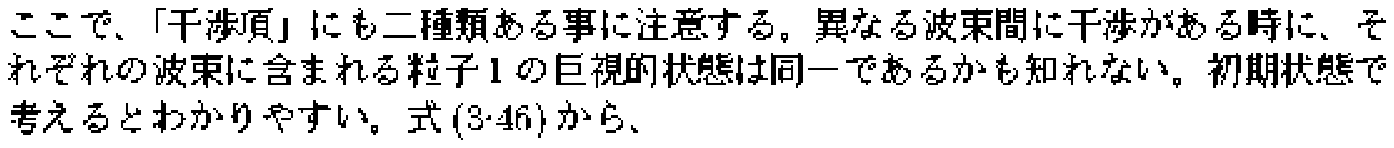}

Here we notice that there are two kinds of ``interference term''. When there is
a interference between different packets, macroscopic states of
particle-1 included in each packet may be the same. It is simple that we think the
initial states. From eq.(\(\ref{eqn052}\)),
\begin{equation}
  \begin{array}{cl}
    k=0,2,3,6 & \mbox{: \ The packet around (\(x_1=0\)) at initial time (\(t=t_0\)).} \\
    k=1,4,5,7 & \mbox{: \ The packet around (\(x_1=d_1\)) at initial time. }
  \end{array}
\end{equation}
\noindent\includegraphics[scale=1.0]{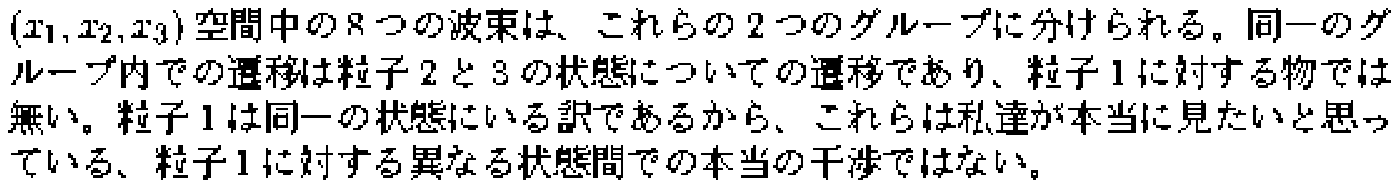}

The 8 packets in the (\(x_1,x_2,x_3\)) space are separated into these
two groups. The interference between packets in the same group means the
transition between the states for particle-2 and 3, not for particle-1. 
 Because the particle-1 is in the same its own state, these are
 not the true interferences between different states for particle-1
 which really we want to see.  

\noindent\includegraphics[scale=1.0]{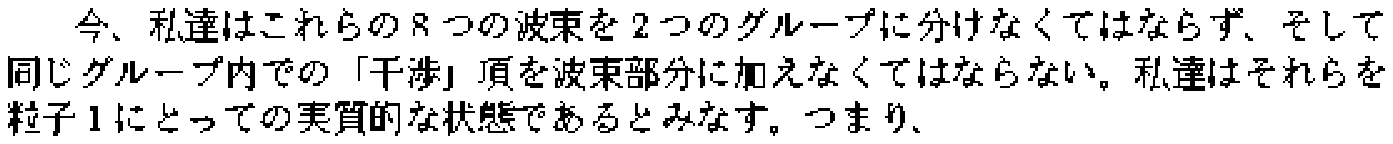}

Now we have to separate these 8 packets into two groups
above, and we have to add the ``interference'' terms among the packets in the
same group to the packet terms. We regard them as the effective
states for the particle-1.  That is, \\[5pt]
\(\bigcirc\) The packet around (\(x_1=0\)) at initial time \(t_0\).¡§
\begin{eqnarray}
  && \tilde{\rho}_{1\_0}^{\mbox{ \tiny eff}}(x_1,t) \equiv C^2 \frac{\pi^3}{|\Delta|}
    \sum_{k=0,2,3,6} \biggl( \int_{-\infty}^{\infty}
   \int_{-\infty}^{\infty}dx_2 \ dx_3 \ \exp[ \ 2\Re e \Theta^{(k)}\ ]
   \nonumber  \\
  && \quad +2\sum_{l=0,2,3,6}^{k<l} \int_{-\infty}^{\infty}
   \int_{-\infty}^{\infty}dx_2 \ dx_3 \ \exp [ \ \Re e
   \Theta^{(k)}+ \Re e \Theta^{(l)} \ ]\cdot\cos[ \ \Im m
   \Theta^{(k)}-\Im m \Theta^{(l)} \ ] \biggr) \nonumber \\
  && \
\end{eqnarray}
\(\bigcirc\) The packet around (\(x_1=d_1\)) at initial time \(t_0\).¡§
\begin{eqnarray}
  && \tilde{\rho}_{1\_d}^{\mbox{ \tiny eff}}(x_1,t) \equiv C^2 \frac{\pi^3}{|\Delta|}
    \sum_{k=1,4,5,7} \biggl( \int_{-\infty}^{\infty}
   \int_{-\infty}^{\infty}dx_2 \ dx_3 \ \exp[ \ 2\Re e \Theta^{(k)}\ ]
   \nonumber  \\
  && \quad +2\sum_{l=1,4,5,7}^{k<l} \int_{-\infty}^{\infty}
   \int_{-\infty}^{\infty}dx_2 \ dx_3 \ \exp [ \ \Re e
   \Theta^{(k)}+ \Re e \Theta^{(l)} \ ]\cdot\cos[ \ \Im m
   \Theta^{(k)}-\Im m \Theta^{(l)} \ ] \biggr) \nonumber \\
  && \
\end{eqnarray}
\(\bigcirc\) Their interference term.¡§
\begin{equation}
   \tilde{\rho}_{1\_int}^{\mbox{ \tiny eff}}(x_1,t) \equiv 4C^2
    \frac{\pi^3}{|\Delta|} \sum_{  \stackrel{ \scriptstyle k=0,2,3,6}{
    \scriptstyle l=1,4,5,7} }^{k<l} \int_{-\infty}^{\infty}
   \int_{-\infty}^{\infty}dx_2 \ dx_3 \ \exp [ \ \Re e
   \Theta^{(k)}+ \Re e \Theta^{(l)} \ ]\cdot\cos[ \ \Im m
   \Theta^{(k)}-\Im m \Theta^{(l)} \ ] 
\end{equation}
\noindent\includegraphics[scale=1.0]{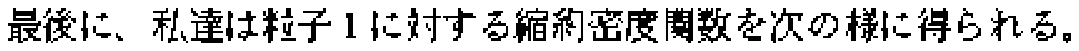}

Finally, we get the reduced density for particle-1 as follows.
\begin{equation}
  \tilde{\rho}_1(x_1,t) = \tilde{\rho}_{1\_0}^{\mbox{ \tiny
   eff}}(x_1,t) + \tilde{\rho}_{1\_d}^{\mbox{ \tiny eff}}(x_1,t) + \tilde{\rho}_{1\_int}^{\mbox{ \tiny eff}}(x_1,t)
\end{equation}
\noindent\includegraphics[scale=1.0]{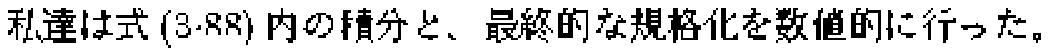}

We used numerical calculation for integrations in eq.(\ref{eqn098}) and
final normalization.

 \subsection{Simulation Result}

\noindent\includegraphics[scale=1.0]{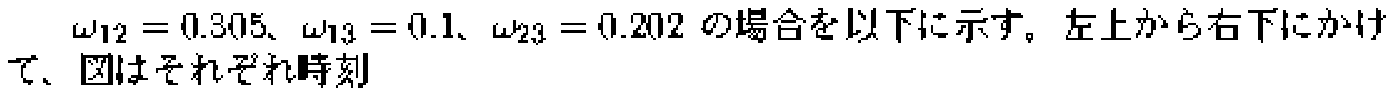}\\
\noindent\includegraphics[scale=1.0]{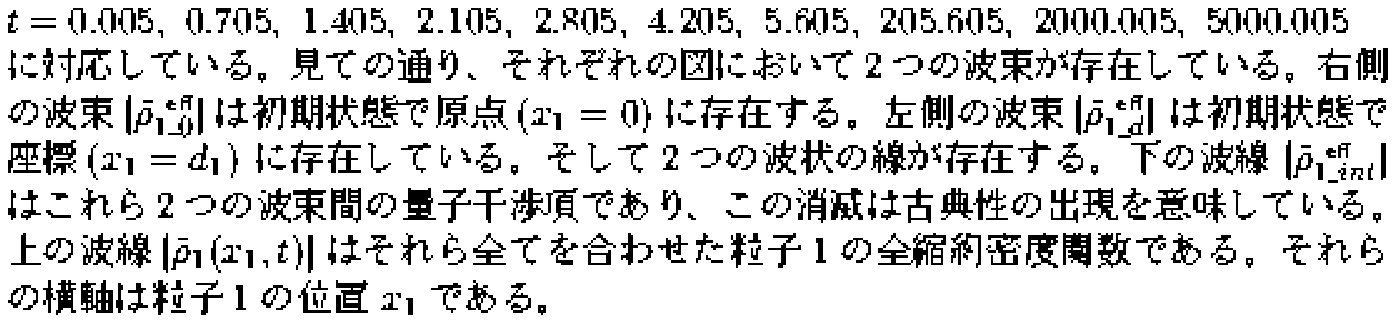}

The case \( \omega_{12}=0.305 \), \( \omega_{13}=0.1 \), \(
\omega_{23}=0.202 \) is showed as follows. From upper left to bottom
right, the figures are corresponding to time\\
 \( t=0.005, \ 0.705, \
1.405, \ 2.105, \ 2.805, \ 4.205, \ 5.605, \ 205.605, \ 2000.005, \
5000.005 \)\\
, respectively. As you can see, there are 2 packets in each figure. The right packet is
[\(\tilde{\rho}_{1\_0}^{\mbox{ \tiny eff}}\)] which is in the origin
(\(x_1 =0\)) initially. The left packet is 
[\(\tilde{\rho}_{1\_d}^{\mbox{ \tiny eff}}\)] which is in (\(x_1 =
d_1\)) initially.  And there are 2 wave-like lines. {\underline{The lower wave line is the quantum} \underline{interference term}}
[\(\tilde{\rho}_{1\_int}^{\mbox{ \tiny eff}}\)] between these 2 packets,
whose disappearance means the emergence of classicality.
The upper wave line is their total reduced density for particle 1,
[\(\tilde{\rho}_1(x_1,t)\)].
 Their horizontal axes are particle-1's position, \(x_1\).

\begin{figure}
\includegraphics[scale=.57]{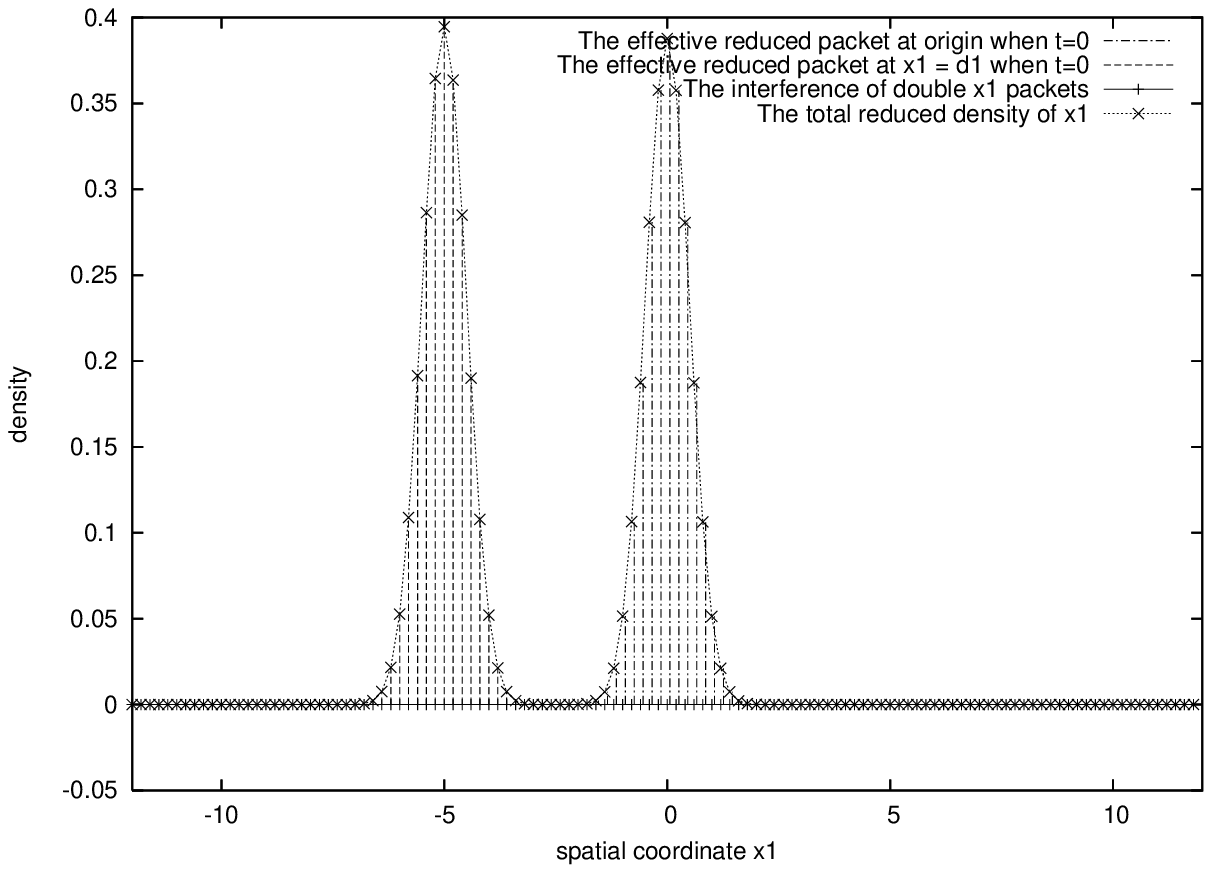}
\includegraphics[scale=.57]{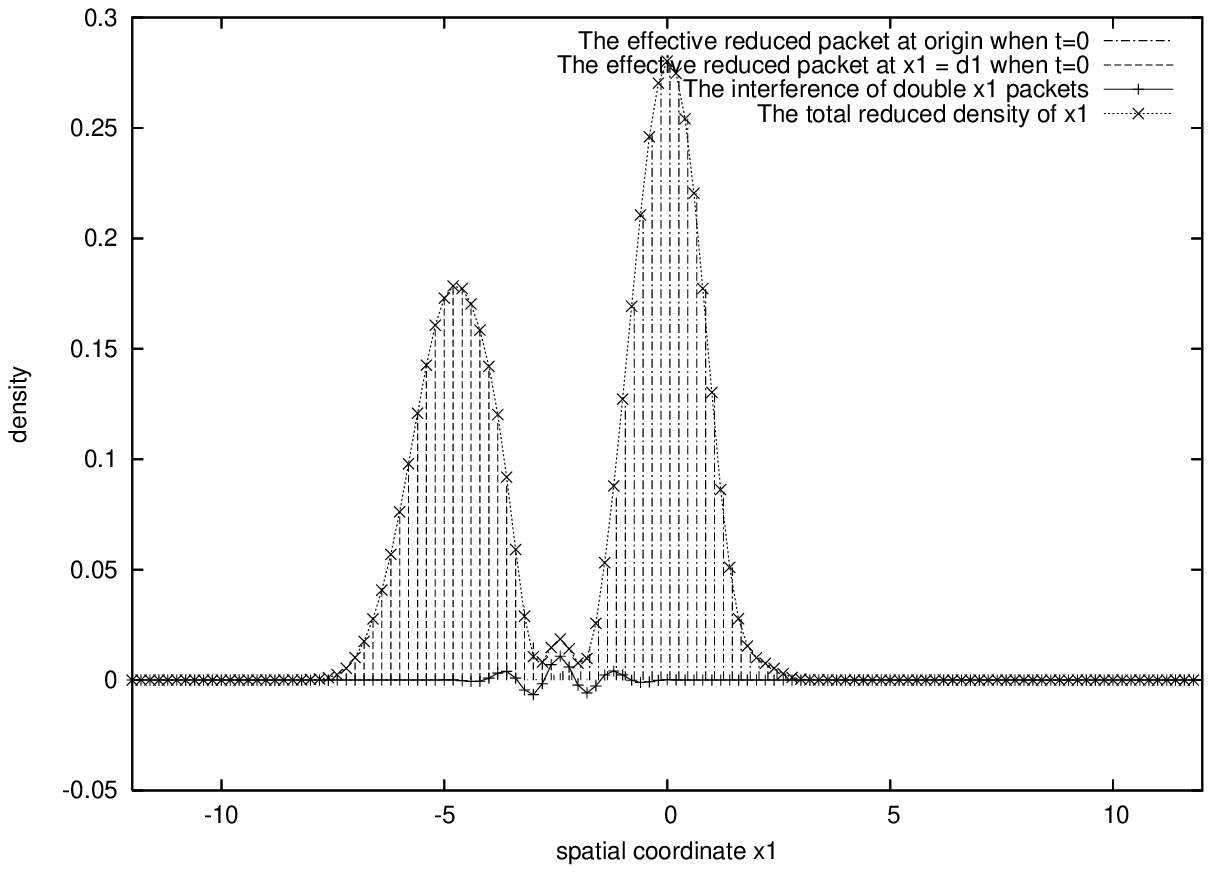} 
\caption{ \(\Uparrow\) (Left) t=0.005, (Right) t=0.705}
\end{figure}

\begin{figure}
\includegraphics[scale=.57]{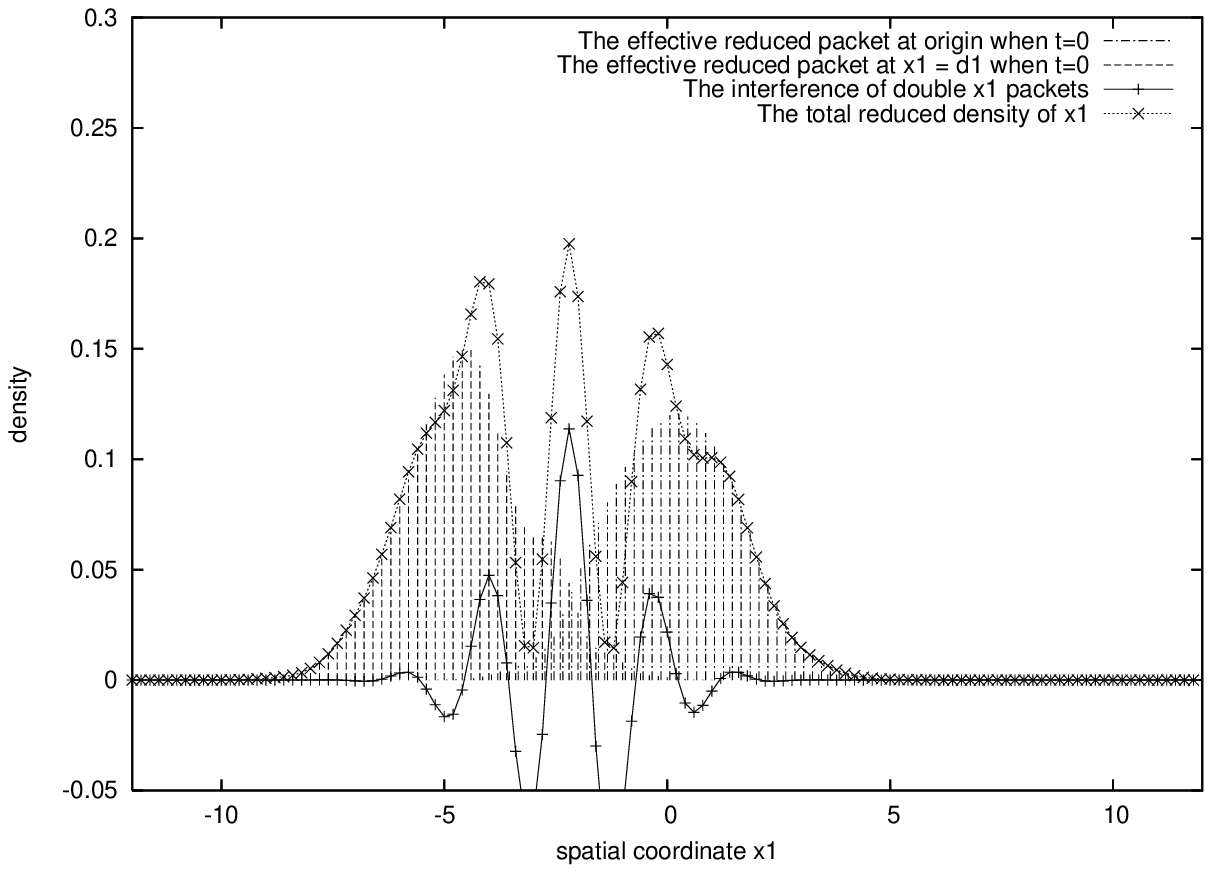}
\includegraphics[scale=.57]{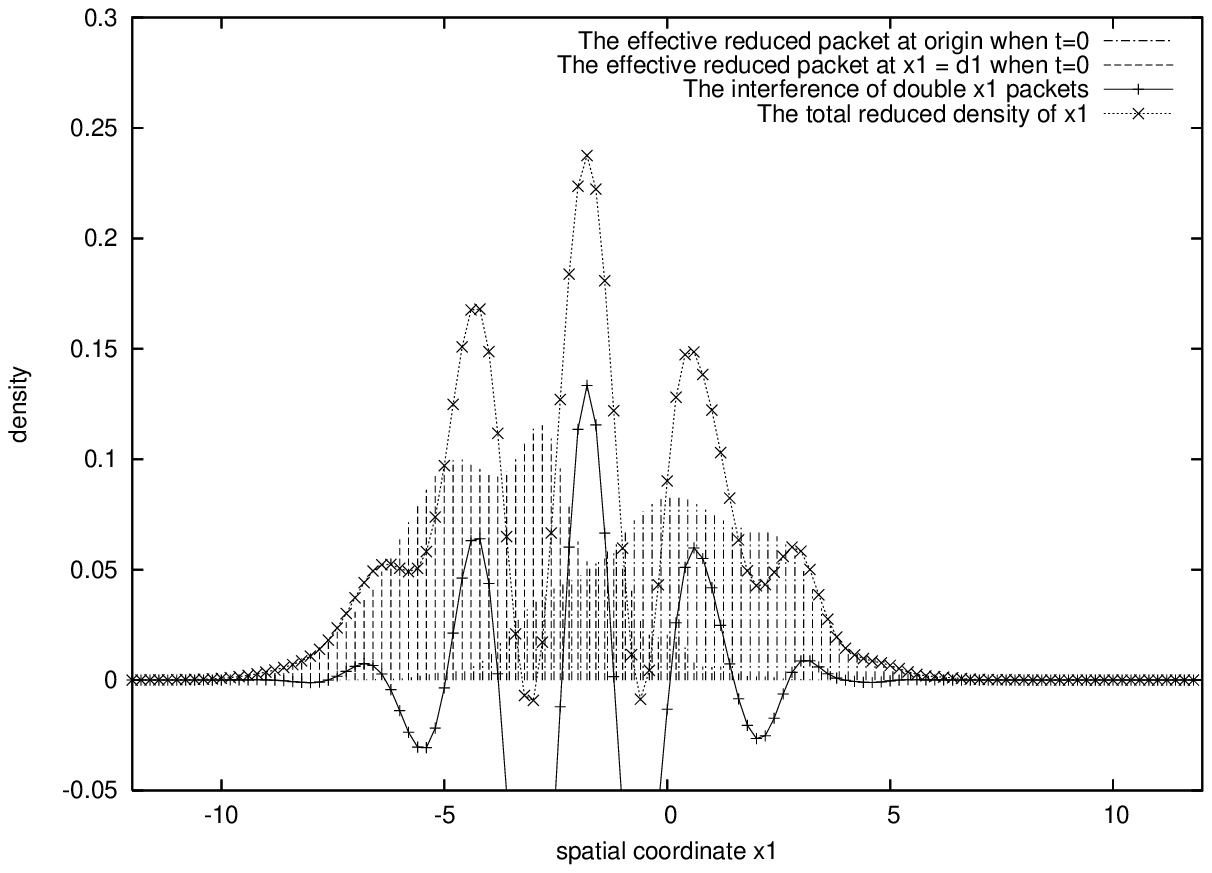} 
\caption{ \(\Uparrow\) (Left) t=1.405, (Right) t=2.105}
\end{figure}

\begin{figure}
\includegraphics[scale=.57]{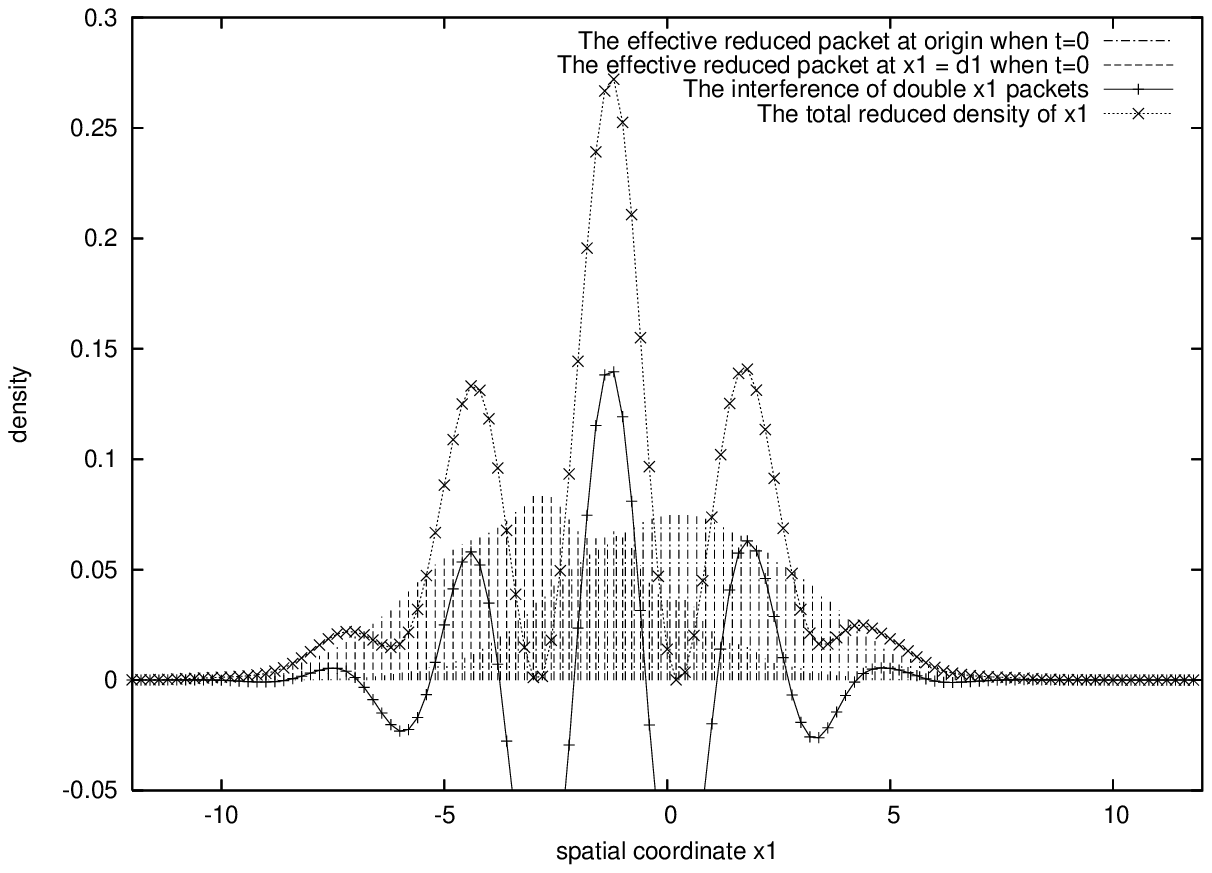} 
\includegraphics[scale=.57]{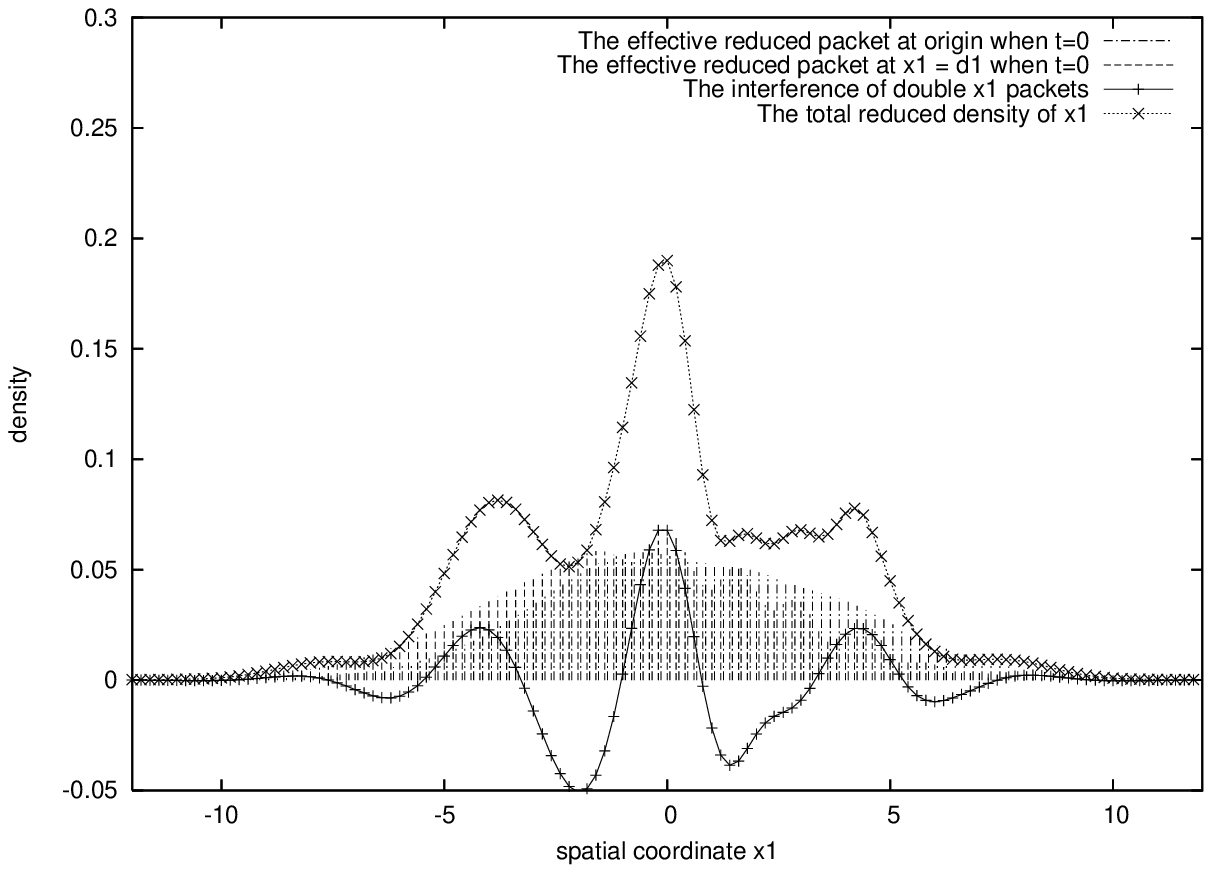} 
\caption{ \(\Uparrow\) (Left) t=2.805, (Right) t=4.205}
\end{figure}

\begin{figure}
\includegraphics[scale=.57]{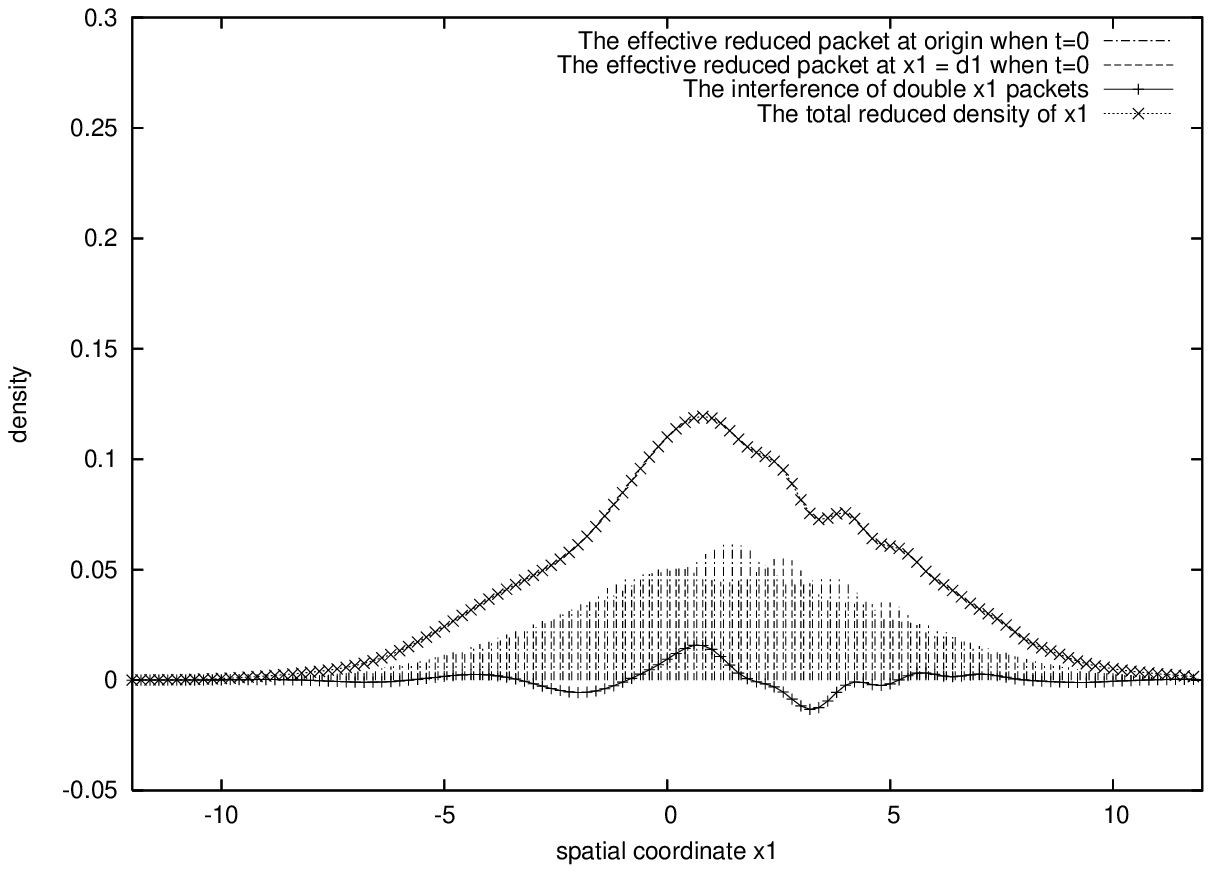}
\includegraphics[scale=.57]{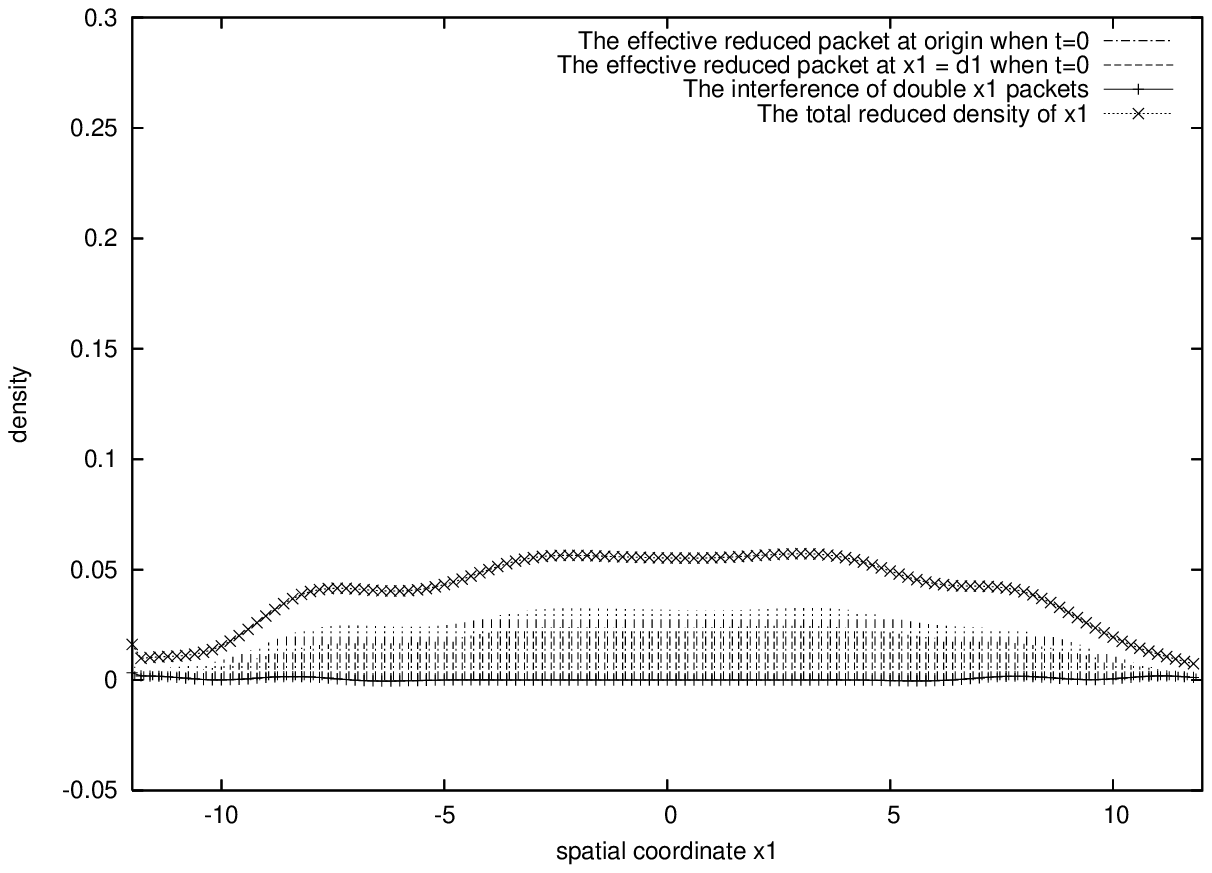} 
\caption{ \(\Uparrow\) (Left) t=5.605, (Right) t=205.605}
\end{figure}

\begin{figure}
\includegraphics[scale=.57]{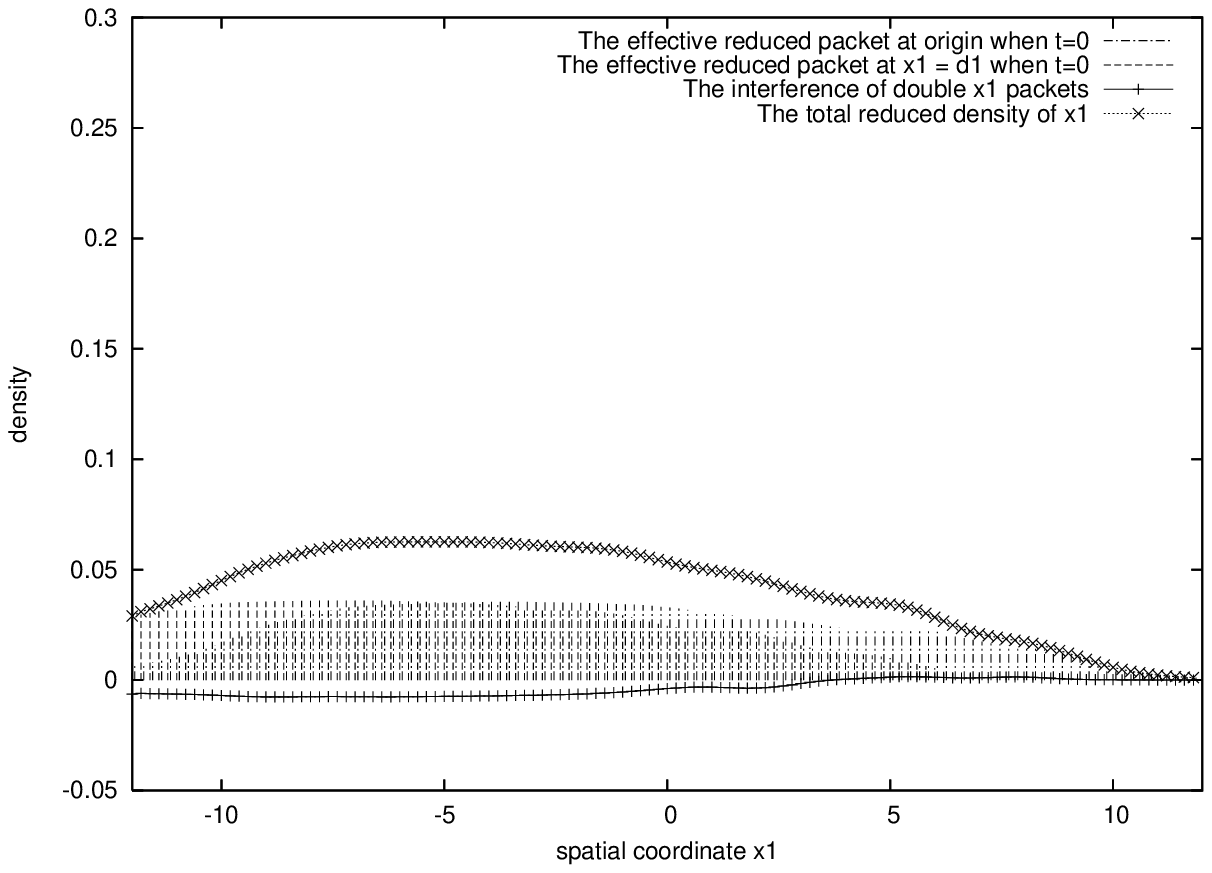}
\includegraphics[scale=.57]{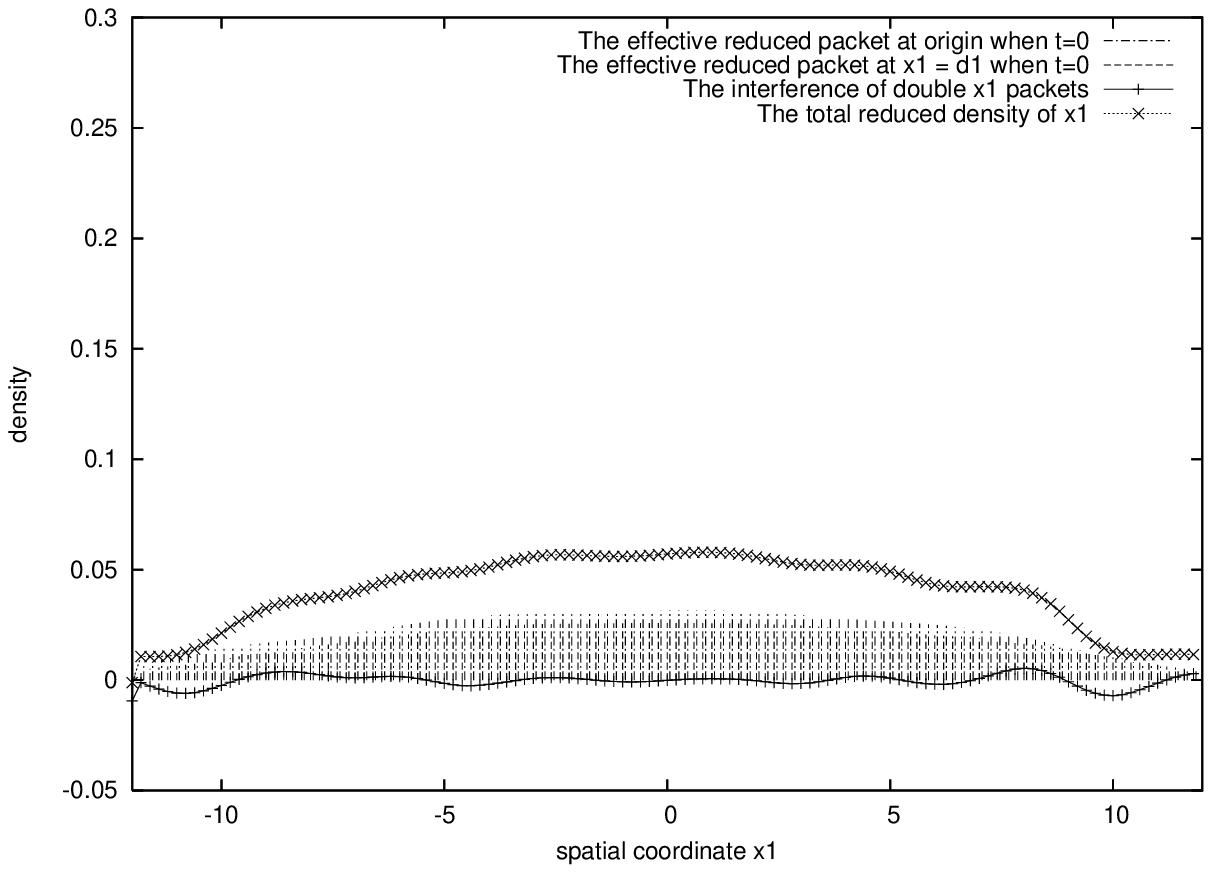} 
\caption{ \(\Uparrow\) (Left) t=2000.005, (Right) t=5000.005}
\end{figure}

\noindent\includegraphics[scale=1.0]{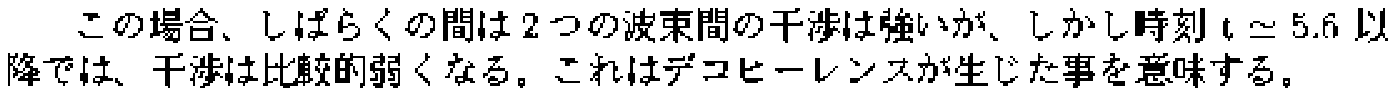}

In this case, momentarily the interference between two packets are
strong, but after the time t \(\simeq\) 5.6, the interferences are
comparatively weakened. It means the quantum decoherence arises. 

\noindent\includegraphics[scale=1.0]{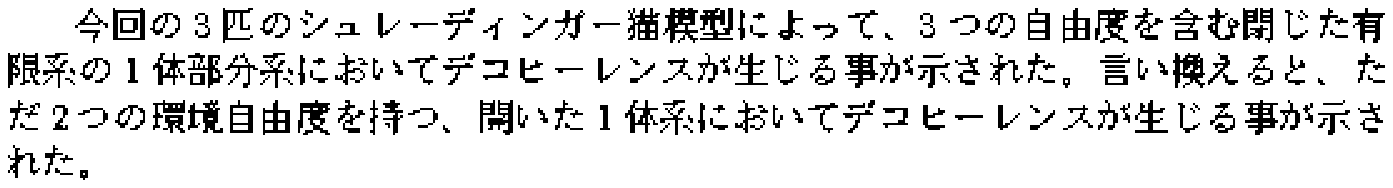}

By this Three Schr\"odinger cats model, it was showed that the quantum decoherence would arise in a
1-body sub-system of a closed finite system which consists of three degrees of
freedom. In other words, it was showed that decoherence would arise in an
opened 1-body system with only two envoronmental degrees
of freedom.  \(\diamondsuit\)
\newpage

\section{Discussion}

\noindent\includegraphics[scale=1.0]{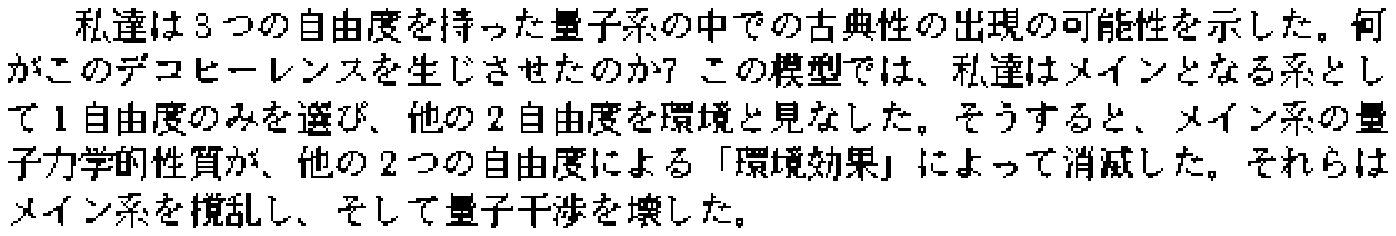}

We showed the possibility of emergence of classicality in a quantum
mechanical system with
3 degrees of freedom. What did make this quantum decoherence? In this
model, we selected only 1 degree as main system,
and regarded other 2 degrees of freedom as environments. Then, the quantum
mechanical property of the main system vanished because of the
``environmental effects'' of other two
degrees of freedom. They disturbed the main system and destroyed its
quantum interference. 

\noindent\includegraphics[scale=1.0]{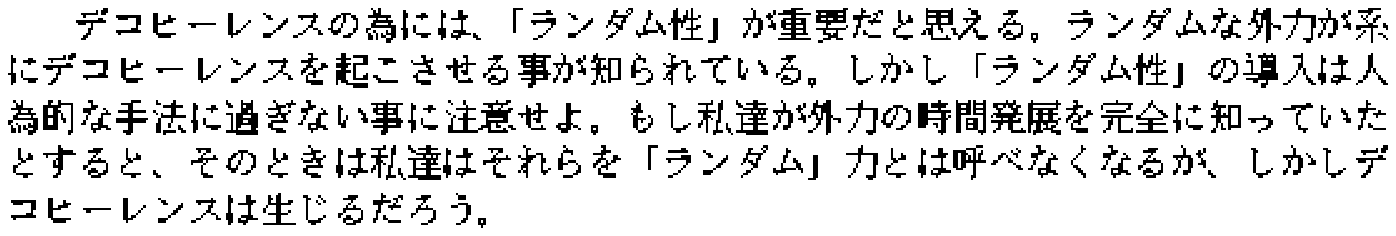}

For decoherence, it seems that ``randomness'' is important. It is known
that external random forces make a system decoherence. But remember that the introduction of
``randomness'' is only an artificial procedure. If we know the time
evolution of external forces perfectly, then we can not say they are ``random''
forces, but the quantum decoherence will arise.

\noindent\includegraphics[scale=1.0]{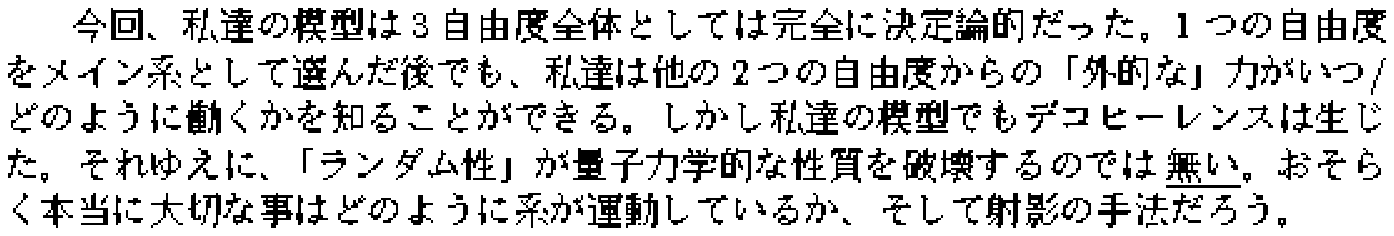}

This time, our model is fully deterministic for 3 degrees of freedom as
a whole. After we select a degrees of freedom as a main system, we can
know when/how the ``external'' forces from other 2 degrees of freedom work. 
But the quantum decoherence occured in our model. Therefore the
``randomness'' does \underline{not} destroy the quantum mechanical
nature. Maybe, the truly important thing is how the system drives, and
the projection procedure. 

\noindent\includegraphics[scale=1.0]{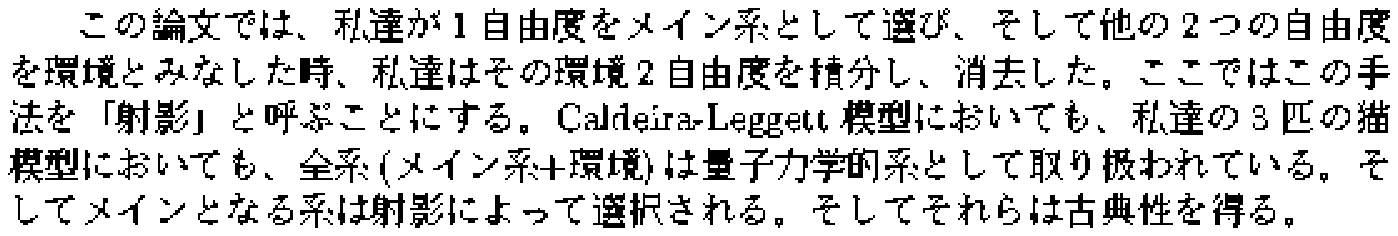}

In this paper, when we select 1 degree of freedom as a main system and regard other 2 degrees
of freedom as a envoronment, we integrate out the environmental 2
degrees of freedom. We
call the procedure ``projection'' here. Both in the Caldeira-Leggett
Model and our 3-Schr\"odinger cats model, the total (main system +
environment) system is treated as a quantum mechanical system. And main
system is selected by projection, and they get classicality.

\noindent\includegraphics[scale=1.0]{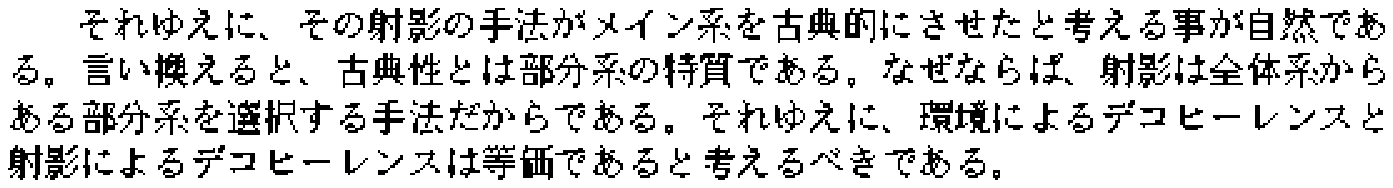}

Therefore it is natural that we should think that the projection procedure
makes the main system classical.
In other words, classicality is the property of subsystems. Because the
projection is a method to select some subsystem from a whole
system. Therefore we should assume that the decoherence by environmental
effects and the one by projection are equivalent.

\noindent\includegraphics[scale=1.0]{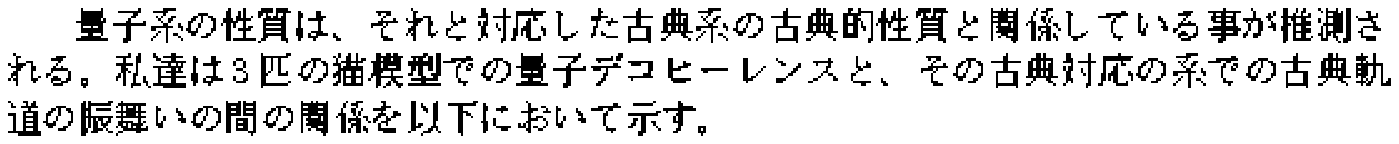}

It is expected that the quamtum mechanical property of the system relates to
the classical machanical property of its equivalent classical system. We
will show a relationship between the quantum decoherence in our
3-Schr\"odinger cats model and a behavior of classical orbits in its
classical equivalent model as follows. \(\diamondsuit\)

\begin{figure}
\includegraphics[scale=.24]{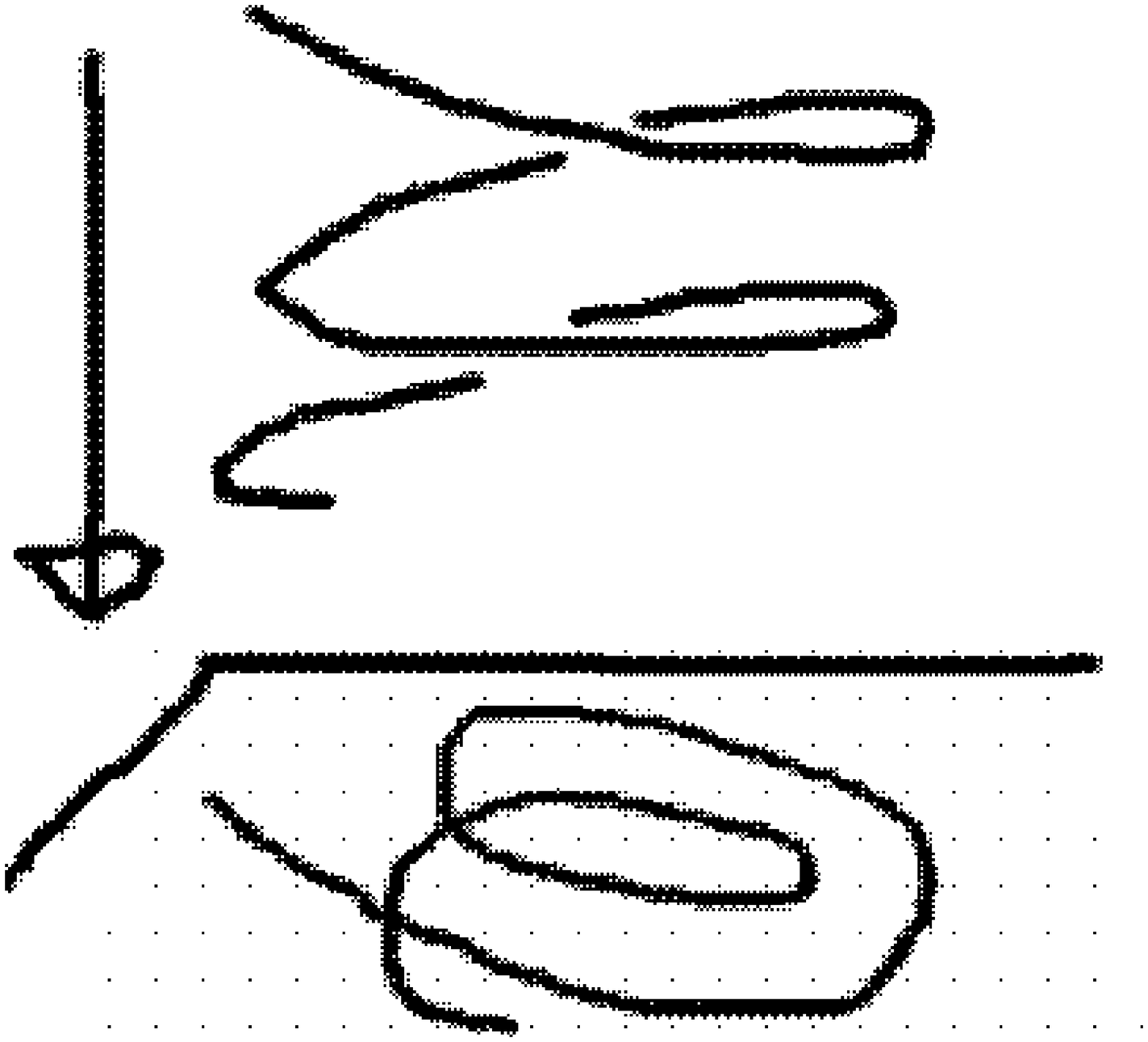}
\includegraphics[scale=.45]{nekosc.eps}
\caption{ Projection makes orbits crossing.}
\end{figure}

\section{Quantum Decoherence/Irreversibility}

\noindent\includegraphics[scale=1.0]{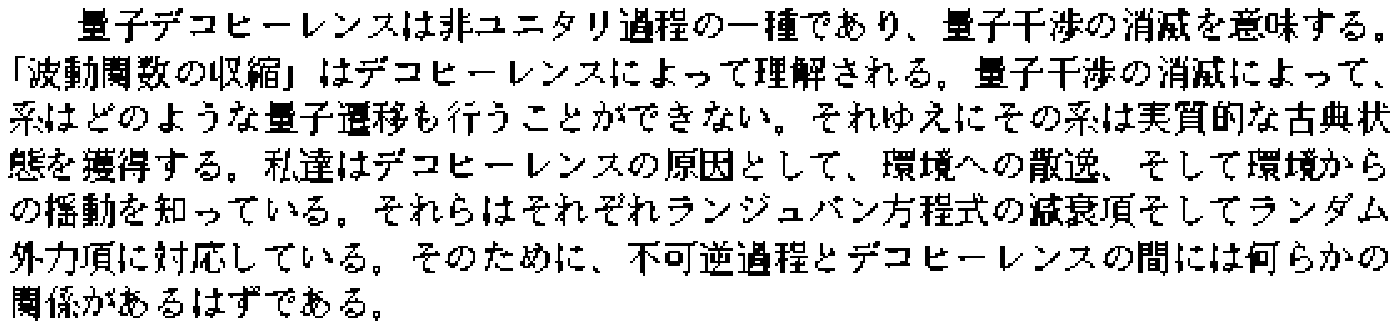}

The Quantum Decoherence is a kind of non-unitary process, which means
disappearance of quantum interference. ``The collapse of the wave
function'' is understood by decoherence. Because of the disappearance of
quantum interference, a system can not do any quantum transitions. Therefore the system gets effective classicality. We know the cause of decoherence such as the dissipation to environment, and the fluctuation from environment. They are corresponding to the damping and the random forces of Langevin equation respectively. So there must be any relationship between the irreversible process and decoherence.

\noindent\includegraphics[scale=1.0]{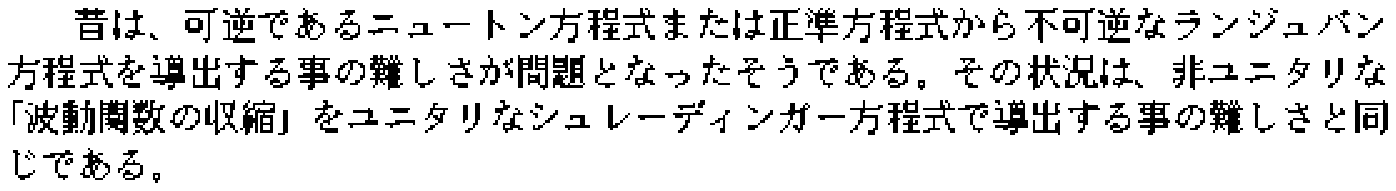}

In old days, there was a problem of difficulty to derive the
irreversible Langevin equation from the reversible Newton equation or
the canonical equations. The situation is the same as the problem of
difficulty to derive non-unitary ``the collapse of the wave function''
by the unitary Schr\"odinger equation. 

\noindent\includegraphics[scale=1.0]{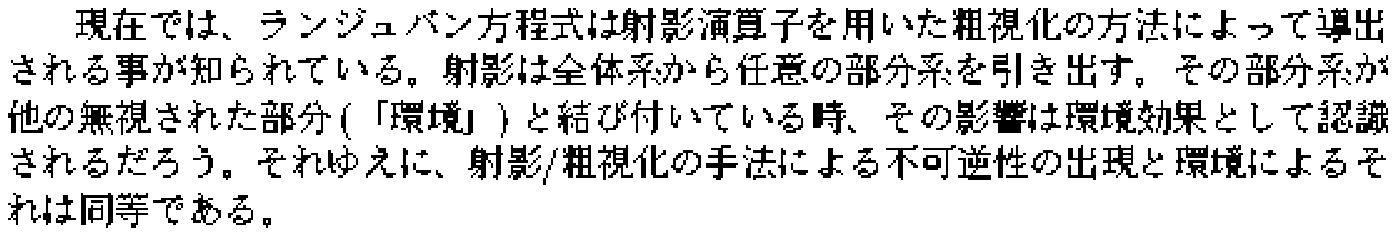}

Now, it is known that the Langevin equation is derived by the coarse
graining procedures using the projection operators. The projection
extracts the arbitrary subsystem from the total system. When the subsystem couples with the other neglected part(``environment''), the effect would be considered as the environmental effects. Therefore, the appearance of the irreversibility by projection/coarse graining procedure and that by the environment are equivalent.   

\noindent\includegraphics[scale=1.0]{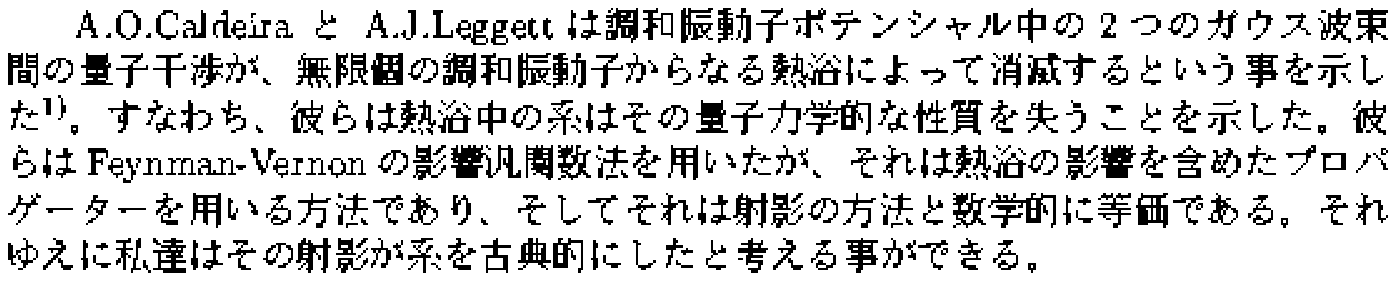}

A.O.Caldeira and A.J.Leggett showed the disappearance of quantum
interference between two Gaussian packets in a harmonic oscillator
potential by the heat bath which consists of infinite numbers of
harmonic oscillators{\cite{calleg}}. That is to say, they showed that
the system in a heat bath lost its quantum property. They used the
Feynman-Vernon's influence functional method, which is the way of using
the propagator with effects of heat bath, and is mathematically
equivalent to the projection method. Therefore we can understand the projection make the system classical.

\noindent\includegraphics[scale=1.0]{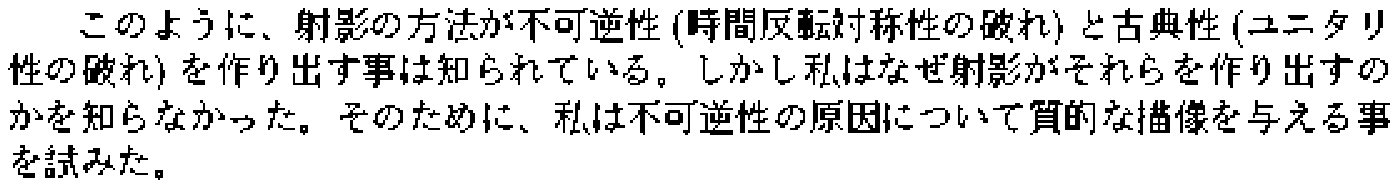}

Thus, it is known that the projection procedures make an
irreversibility(the break down of time reversal symmetry) and a
classicality(the break down of unitarity). But I didn't know why a
projection makes them. So I have tried to make a qualitative picture for
an origin of irreversibility. 

\noindent\includegraphics[scale=1.0]{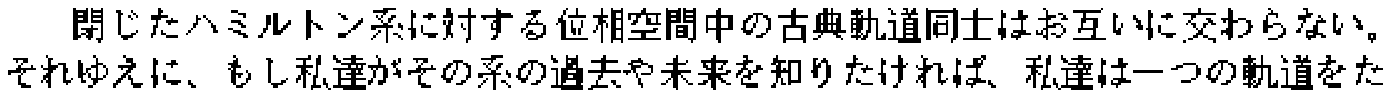}
\noindent\includegraphics[scale=1.0]{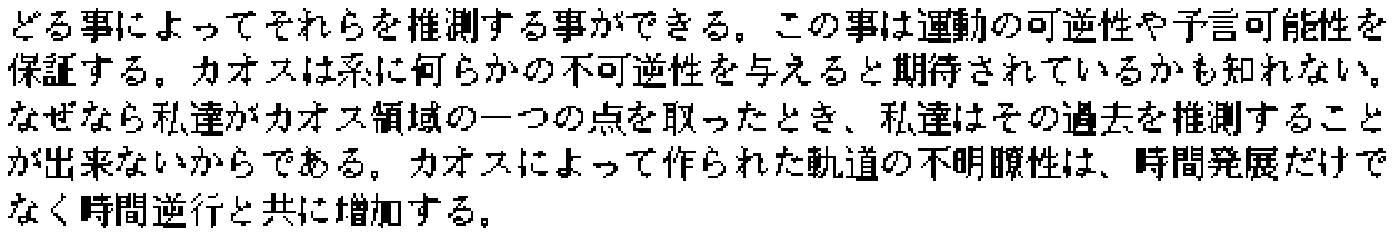}

Classical orbits within a the phase space
for a closed Hamiltonian system do not cross each other. Therefore, if
we want to know the past or the future of the system, we can guess them by
following an orbit. This guarantees the reversibility and the
predictability of motion. Chaos may be expected to make any
irreversibility for system. Because when we take a point in phase space
with chaos region, we can not guess its past. The indefiniteness of
orbits made by chaos increases not only with the time evolution but also
with the time reversal.

\noindent\includegraphics[scale=1.0]{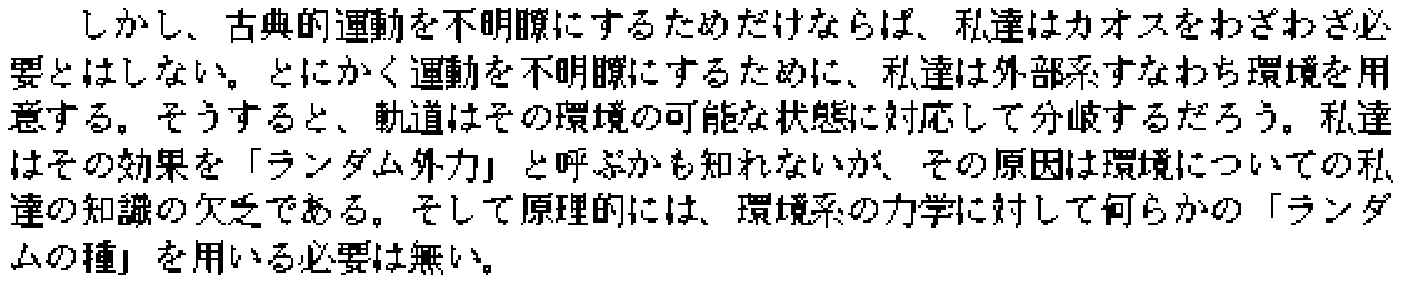}

But, we do not bother to need chaos only for making the classical motion
obscure. Anyway to make any indefiniteness of motion, we prepare an
external system or an environment. Then the orbit will branch according
to probable states of the environment. We may call the effect ``the
random force'', its origin is the lack of our knowledge of the
environment. And in principle there is no need to use any ``random
seed'' for mechanics of the environment. 

\noindent\includegraphics[scale=1.0]{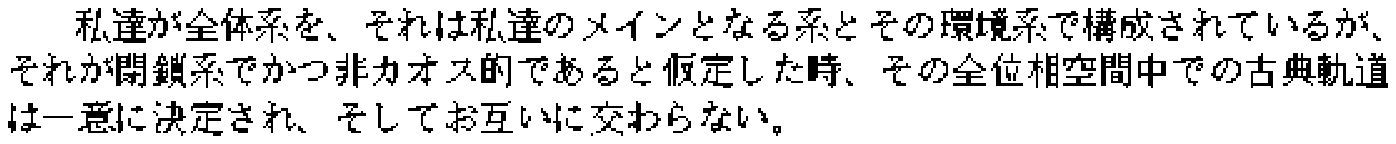}

When we assume the total system, which consists of our main system and
the environment, is closed and non-chaotic, then its classical orbits in
the total phase space have to be defined uniquely and do not cross each
other.

\noindent\includegraphics[scale=1.0]{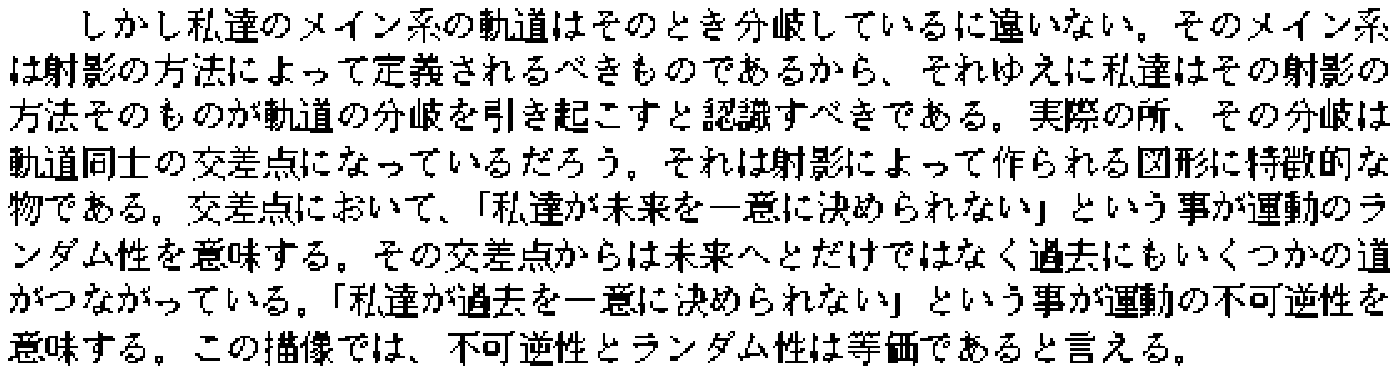}

But our main system's orbits must be branching then. The main system should be
defined by a projection procedure, therefore we should realize that the
projection procedure itself would make
the branch of orbits. Really the branch would be an intersection of
orbits, which is characteristic of figures made by projection. At the
intersection, the thing ''We can not decide the future uniquely.'' means
the randomness of motion.  From the
crossing point, there are the some ways not only to the futures but also
to the
pasts. The thing ``We can not decide the past uniquely.'' means the
irreversibility of motion. In this image, the irreversibility and the
randomness are the equivalent.\(\diamondsuit\)\\

\begin{figure}
\includegraphics[scale=.11]{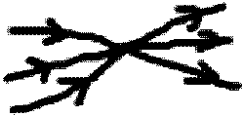}
\caption{Crossing makes the system irreversible?}
\end{figure}

\section{Crossing of Classical Orbits  and Quantum Decoherence}

\noindent\includegraphics[scale=1.0]{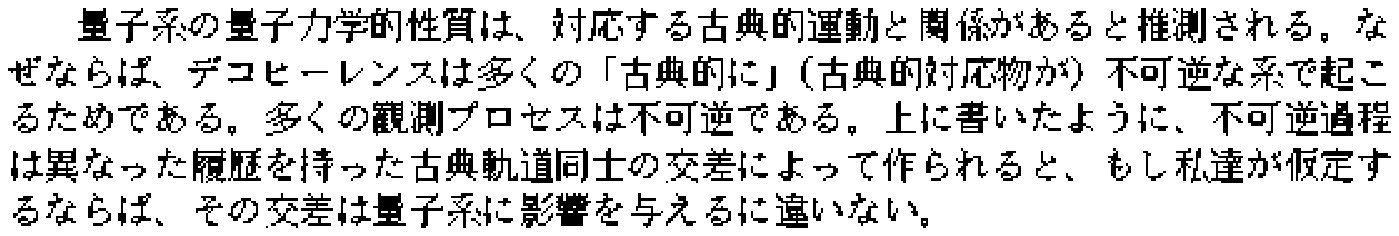}

The quantum mechanical behavior of quantum system is supposed to relate with the corresponding classical motion. Because decoherence occurs in many ``classically''(classical equivalent) irreversible systems. Most observation processes are irreversible. As I noted above, if we assume the irreversibility is made by crossing of classical orbits with different histories, that crossing must affect the quantum systems.

\noindent\includegraphics[scale=1.0]{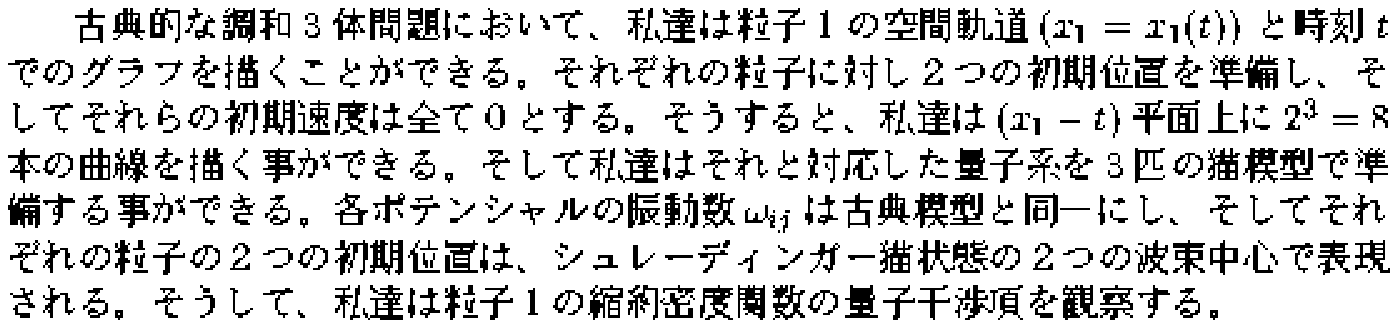}

In a classical harmonic three body problem, we can draw the spatial orbit of particle-1 (\(x_1=x_1(t)\)) versus time \(t\). Each particle has 2 initial positions and their all initial velocities are set 0. Then we can draw \( 2^3 = 8 \) lines on (\(x_1-t\)) plane. And we can simulate its equivalent quantum system by the three cats model. Each frequency of potentials \(\omega_{ij}\) is same as the classical model, and 2 initial positions of each particle are expressed by centers of packets of the Schr\"odinger cat state. Then we will observe the quantum interference term of the reduced density of particle-1.       

\begin{figure}
 \includegraphics[scale=.8]{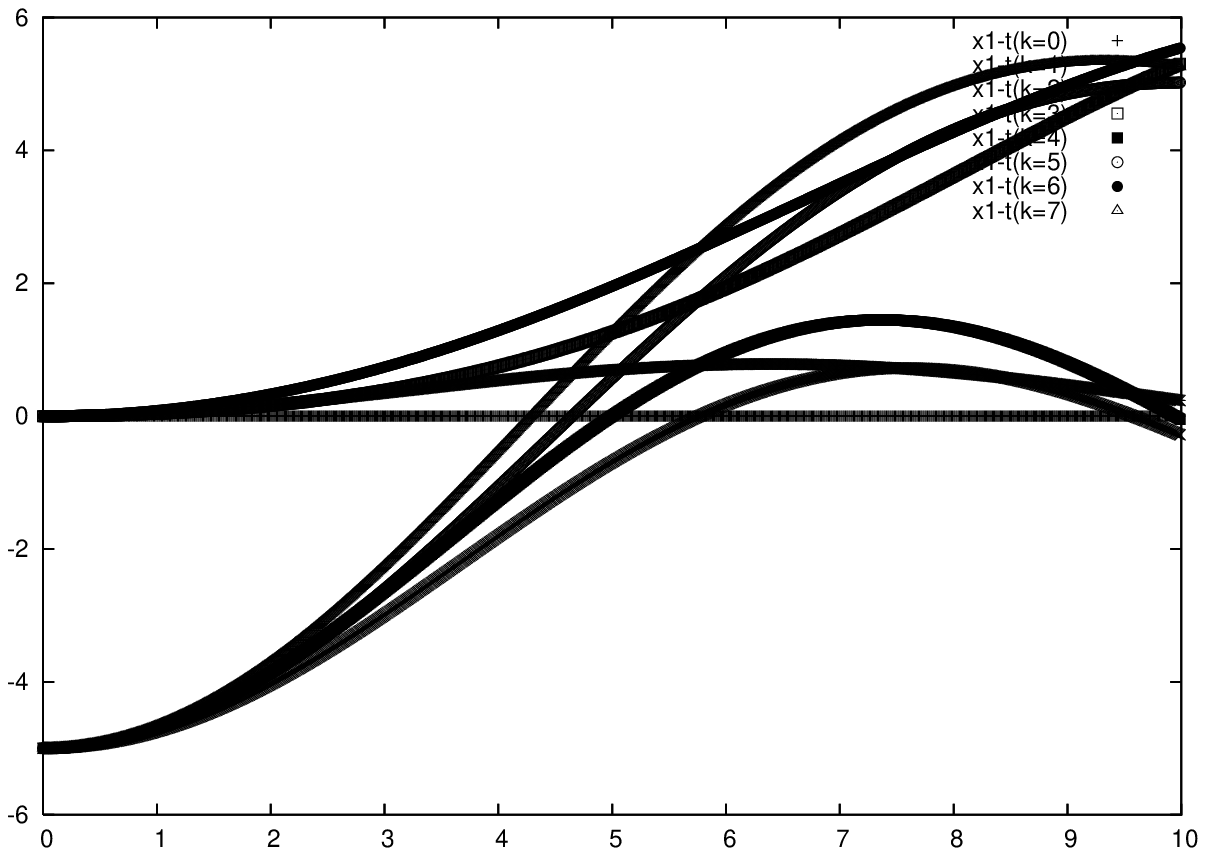}
 \caption{The crossing of classical orbits for particle-1. The
 horizontal axis is time \(t\), and the vertical axis is particle-1's
 position, \(x_1(t)\). Orbits are crossing at time \(t\)=4.0-6.0.}
 \includegraphics[scale=.3]{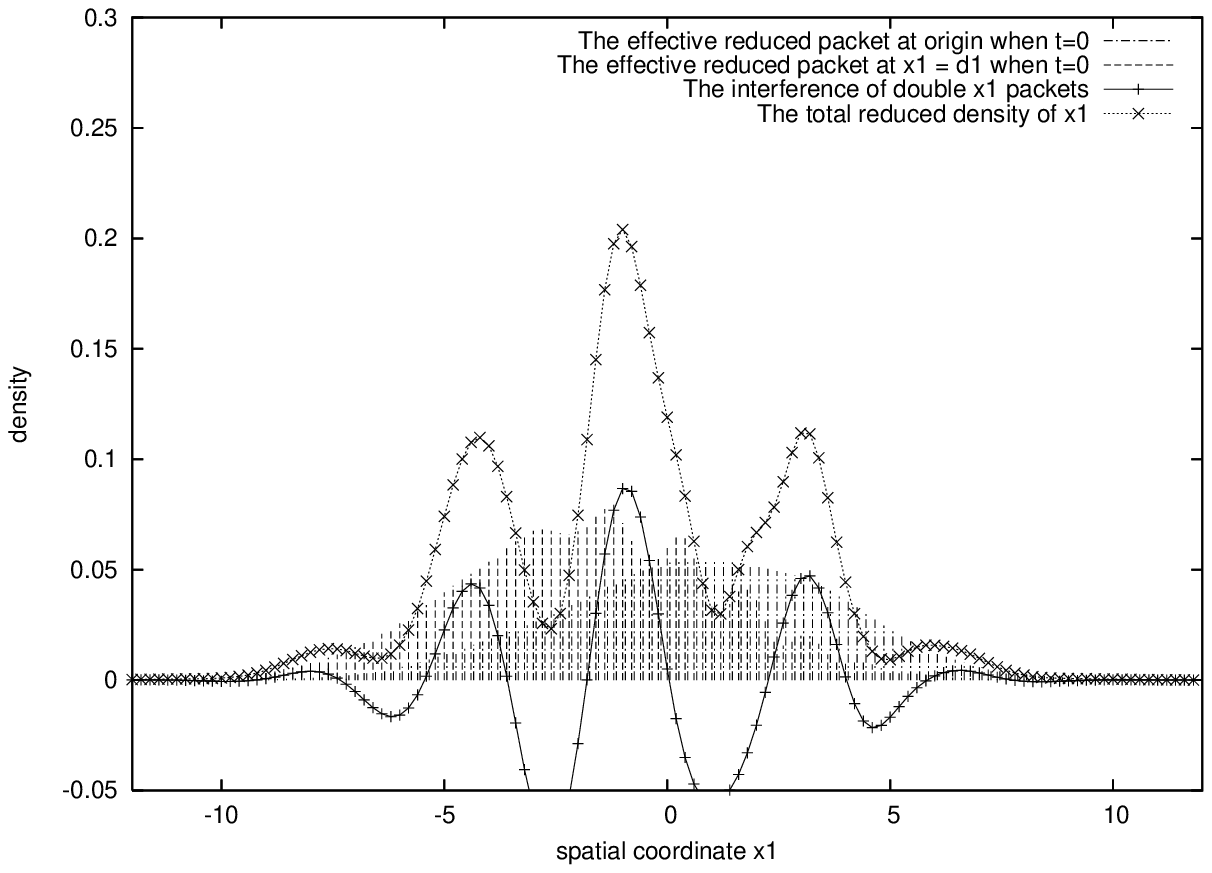}
 \includegraphics[scale=.3]{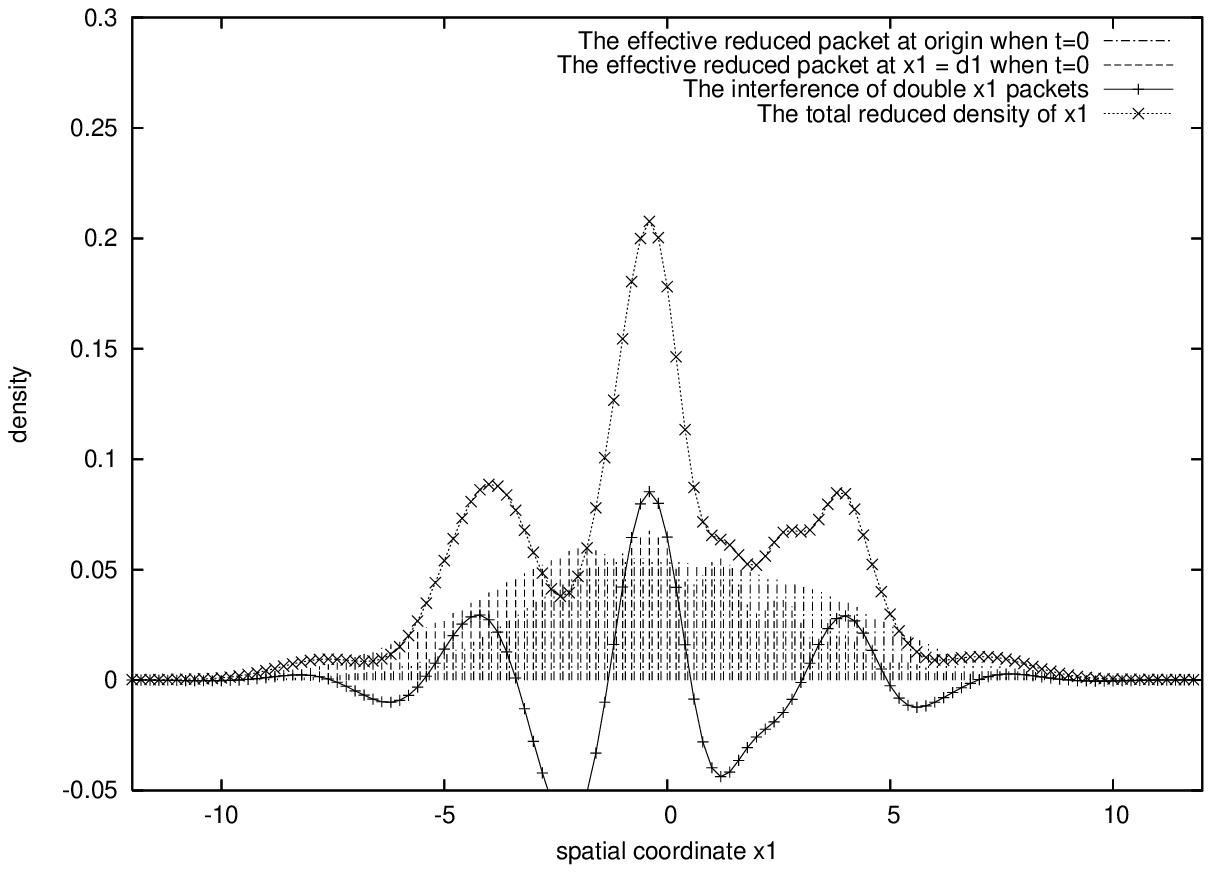}
 \includegraphics[scale=.3]{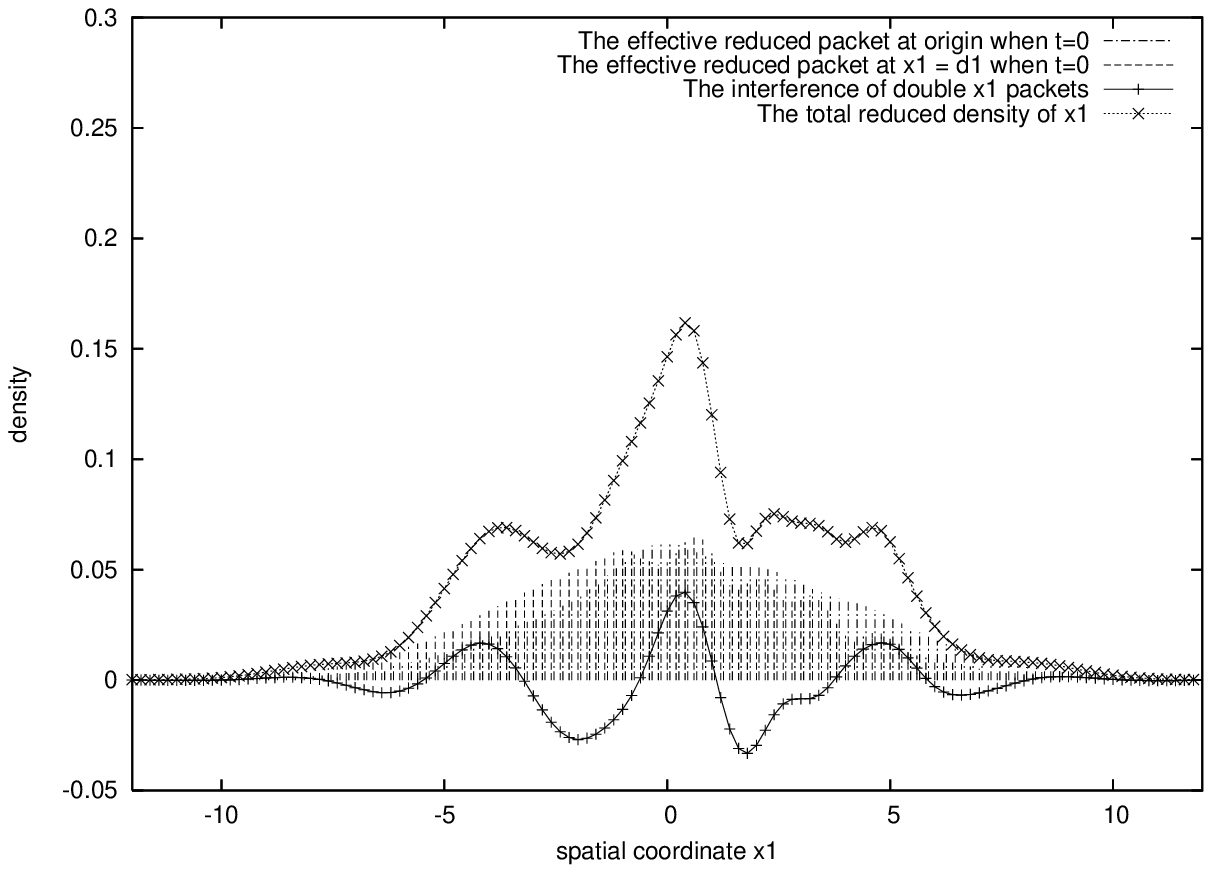}
 \includegraphics[scale=.3]{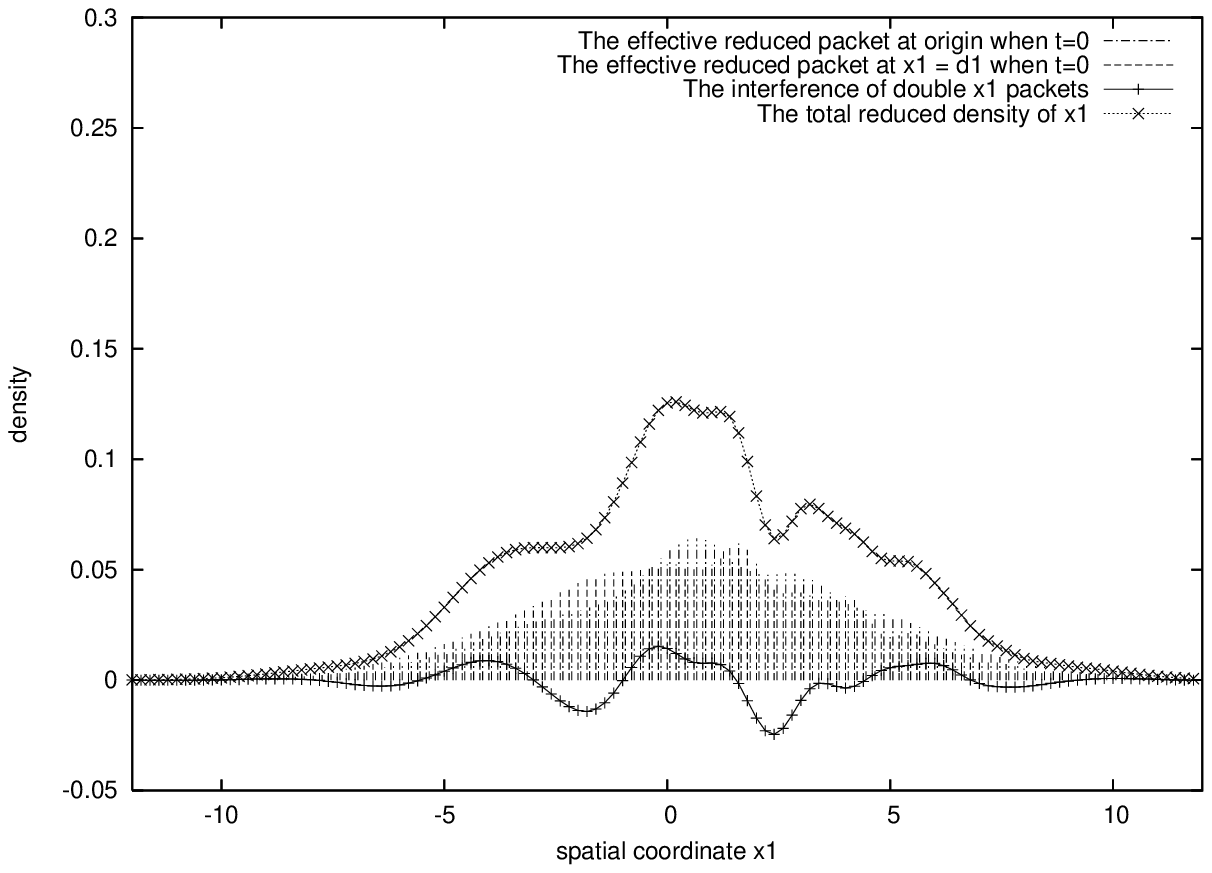}
\caption{The disappearance of quantum interference for particle-1 at three
 Schr\"odinger cats model. From left, the time \(t\)=3.5, \(t\)=4.0, \(t\)=4.5, \(t\)=5.0
 . The horizontal axis is particle 1's position \(x_1(t)\), and the 
 vertical axis is the reduced probability density for particle-1, \(\tilde{\rho}_1(x_1)\). The interference is damped at time \(t\)=4.0-5.0. }
\end{figure}

\noindent\includegraphics[scale=1.0]{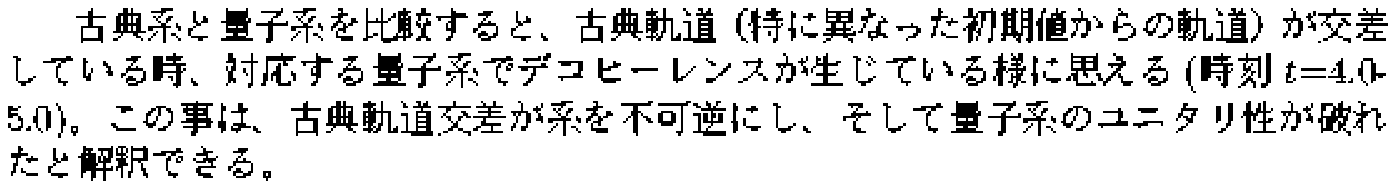}

Comparing the quantum system with the classical system, when classical
orbits (especially the ones from the different initial points) are crossing,
decoherence seems to arise in corresponding quantum system(time \(t\)=4.0-5.0). This can be understood that the classical crossing make system irreversible, then unitarity of the quantum system breaks down.

\noindent\includegraphics[scale=1.0]{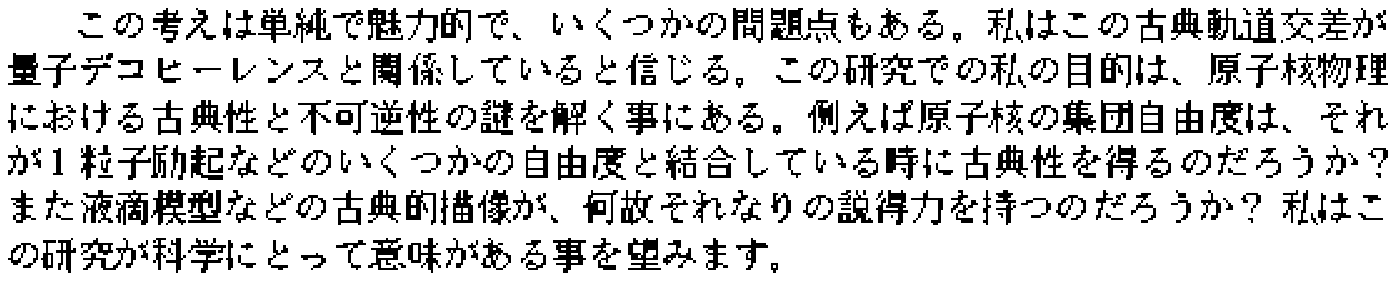}

This thought seems to be simple and tempting, and there are some
difficulties. I believe that this classical orbits' crossing
relates to quantum decoherence. My goal of this study is to reveal the mystery of classicality and irreversibility
in nuclear physics. For example, does a nuclear collective degree of freedom get
classicality when it is
coupled with some degrees of freedom such as single particle excitation? And why is it valid to use any classical pictures for nuclei such as the liquid drop model? I hope this study is meaningful for science.
\(\diamondsuit\)\\[2pt]

\section*{Acknowledgements}
I would like to thank Prof. Fumihiko SAKATA. He taught me nuclear
physics, introduced
Caldeira-Leggett's paper to me and gave me a lot of worthy advices and
severe judgments. I would like to thank people in the nuclear summer
school. It is a very valuable memory for me. I would like to thank people in
my life in Ibaraki. Finally, I would like to thank my family.\\[2pt]

%

\begin{figure}
\includegraphics[scale=.45]{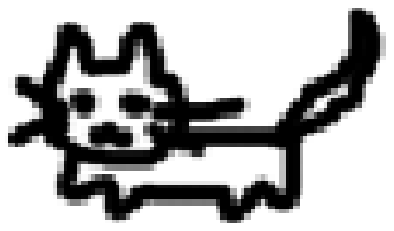}
\end{figure}


\begin{thebibliography}{99}
  
\bibitem{calleg} A.O.Caldeira and A.J.Leggett. Phys.Rev.A 31 (1985), 1059-1066.
\bibitem{calleg2} A.O.Caldeira and A.J.Leggett. Physica A 121 (1983), 587-616.
\bibitem{zurek} Wojciech H.Zurek. Los Alamos Science 27 (2002), 2-25.


\end{thebibliography}
\end{document}